\documentclass[numberedappendix]{emulateapj}
\usepackage{epsfig,graphics}
\usepackage[colorlinks=true,  citecolor=blue,  pagecolor=blue,
linkcolor=blue,  menucolor=blue,  urlcolor=blue,
linkbordercolor={0 0 1},  frenchlinks=True,  breaklinks]{hyperref}
\begin{document}
\submitted{Received 2010 August 31; accepted 2011 April 26; published 2011 June 21}
\newcommand{\Ha}{H$\alpha$}
\newcommand{\OIII}{[\ion{O}{3}]}
\newcommand{\NII}{[\ion{N}{2}]}
\newcommand{\OII}{[\ion{O}{2}]}
\newcommand{\SIII}{[\ion{S}{3}]}
\newcommand{\SII}{[\ion{S}{2}]} 
\newcommand{\zphotf}{z_{\rm phot}}
\newcommand{\zspecf}{z_{\rm spec}}
\newcommand{\zphot}{$\zphotf$}
\newcommand{\zspec}{$\zspecf$}
\newcommand{\Rcf}{R_{\rm C}}
\newcommand{\Rc}{$\Rcf$}

\newcommand{\tcolor}{(A color version of this figure is available in the online journal.)}

\newcommand{\dz}{$\Delta z_{68}$}
\newcommand{\Udrop}{$U$-dropout}
\newcommand{\Udrops}{$U$-dropouts}
\newcommand{\galex}{{\it GALEX}}
\newcommand{\spitzer}{{\it Spitzer}}
\newcommand{\nuv}{{\it NUV}}
\newcommand{\fuv}{{\it FUV}}
\newcommand{\EBV}{E(B-V)}

\newcommand{\Un}{U_n}
\newcommand{\Rs}{\mathcal{R}}
\newcommand{\UBV}{U - BV}
\newcommand{\BVRi}{BV - \Rcf i\arcmin}

\newcommand{\Lya}{Ly$\alpha$}
\newcommand{\mm}{$\mu$m}
\newcommand{\Msun}{$M_{\sun}$}

\newcommand\aper[1]{#1\arcsec\ $\phi$}
\newcommand{\iyr}{yr$^{-1}$}
\newcommand{\vMpc}{Mpc$^{-3}$}
\newcommand{\sbzk}{sBzK}

\newcommand{\fBK}{5BK3$\sigma$}
\newcommand{\fBf}{5B5$\sigma$}

\newcommand{\LyRcutn}{25.5}
\newcommand{\LyRcut}{$\Rcf\leq\LyRcutn$}

\newcommand{\Nabs}{$\sim$53,000} 
\newcommand{\Ninta}{6\%}  
\newcommand{\Nintb}{8\%}  
\newcommand{\Nintc}{20\%} 

\newcommand{\absa}{81\%--90\%} 

%
\newcommand{\nBXs}{6280}    
\newcommand{\nBMs}{6776}    
\newcommand{\nsBzKs}{8715}  
\newcommand{\npBzKs}{228}   
\newcommand{\nzULBGs}{2578} 
\newcommand{\nzNLBGs}{8574} 
\newcommand{\nzFLBGs}{1067} 
%
\newcommand{\nBX}{5585}      
\newcommand{\nBM}{6408}      
\newcommand{\nsBzK}{\nsBzKs} 
\newcommand{\npBzK}{214}     
\newcommand{\nzULBG}{2375}   
\newcommand{\nzNLBG}{7878}   
\newcommand{\nzFLBG}{1042}   
\newcommand{\ntot}{32217}    
%
\newcommand{\nBXz}{5070}    
\newcommand{\nBMz}{6097}    
\newcommand{\nsBzKz}{7866}  
\newcommand{\npBzKz}{208}   
\newcommand{\nzULBGz}{2060} 
\newcommand{\nzNLBGz}{7663} 
\newcommand{\nzFLBGz}{1039} 
\newcommand{\ntotz}{30003}  
%
\newcommand{\nBXsb}{8699}     
\newcommand{\nBMsb}{8952}     
\newcommand{\nzULBGsb}{4026}  
\newcommand{\nzNLBGsb}{10441} 
%
\newcommand{\nBXb}{8004}    
\newcommand{\nBMb}{8584}    
\newcommand{\nzULBGb}{3823} 
\newcommand{\nzNLBGb}{9745} 
\newcommand{\ntotb}{40127}  
%
\newcommand{\nBXbz}{6887}    
\newcommand{\nBMbz}{7771}    
\newcommand{\nzULBGbz}{3265} 
\newcommand{\nzNLBGbz}{9397} 
\newcommand{\ntotbz}{36433}  
%
\newcommand{\nUa}{20687}  
\newcommand{\nUaz}{18830} 
\newcommand{\nUb}{26411}  
\newcommand{\nUbz}{23279} 

\newcommand{\nUaround}{$\approx$21,000}  
\newcommand{\nUazround}{$\approx$19,000} 

\newcommand{\nBXi}{1235}  
\newcommand{\nBMi}{778}  
\newcommand{\nBXif}{24.4} 
\newcommand{\nBMif}{12.8} 
\newcommand{\nUif}{13.4} 
%
\newcommand{\nBXbif}{19.3} 
\newcommand{\nBMbif}{11.1} 

\newcommand{\Nspec}{495}
\newcommand{\accur}{1.2\%} 
\newcommand{\accurd}{0.012}

\newcommand{\deltaall}{53217} 
\newcommand{\deltaK}{21768}   
\newcommand{\deltab}{52984}   

\newcommand{\nhighz}{4079}   
\newcommand{\nlowz}{21564}   
\newcommand{\ncensus}{27574} 

\newcommand{\nNLBGi}{25.4\%} 

\newcommand{\lsfr}{66\%}      
\newcommand{\lsfrBXBM}{70\%}  

\newcommand{\stataa}{96.6\%} 
\newcommand{\statab}{76.6\%} 
\newcommand{\statac}{67.8\%} 
\newcommand{\statad}{75.3\%} 
\newcommand{\statact}{two-thirds} 

\newcommand{\stataza}{75.9\%} 
\newcommand{\statazb}{68.5\%} 
\newcommand{\statazc}{59.6\%} 
\newcommand{\statazd}{73.5\%} 

\newcommand{\statsa}{32.2\%} 
\newcommand{\statsb}{36.5\%} 
\newcommand{\statsc}{9.0\%}  
\newcommand{\statsd}{77.7\%} 

\newcommand{\statNa}{32.4\%} 
\newcommand{\statNb}{41.4\%} 
\newcommand{\statNc}{2.4\%}  
\newcommand{\statNd}{76.2\%} 

\newcommand{\statca}{79.7\%} 
\newcommand{\statcb}{87.0\%} 
\newcommand{\statcc}{77.2\%} 

\newcommand{\nprange}{75\%--96\%} 

\newcommand{\ncensusC}{17566} 
\newcommand{\censuscomp}{$\approx$90\%} 

\newcommand{\sbzkgood}{85\%} 

\newcommand{\zBX}{74.7\%}   
\newcommand{\zBM}{84.5\%}   
\newcommand{\zU}{78.83\%}   
\newcommand{\zsbzk}{93.9\%} 
\newcommand{\zNUV}{91.9\%}  
%
\newcommand{\zNUVint}{17.4\%} 

\newcommand{\peakval}{60\%--80\%} 
\newcommand{\allz}{52\%--65\%} 

\newcommand{\VeffK}{$1.18\times10^6$ Mpc$^3$}  
\newcommand{\VeffR}{$1.24\times10^6$ Mpc$^3$}  
\newcommand{\sfrKn}{0.18$\pm$0.03}             
\newcommand{\sfrK}{\sfrKn\ \Msun\ \iyr\ \vMpc} %
\newcommand{\EBVlimit}{0.25}                   
\newcommand{\pEBV}{65\%}                       
\newcommand{\pEBVb}{$\approx$64\%}             
\newcommand{\NEBV}{one-fourth}                 
\newcommand{\NEBVb}{three-quarters}            
\newcommand{\pEBVl}{35\%}                      
%
\newcommand{\highR}{67\%}                      
\newcommand{\sfrKNUVn}{0.16$\pm$0.03}          
\newcommand{\sfrKNUV}{\sfrKNUVn\ \Msun\ \iyr\ \vMpc}
\newcommand{\sfrKNUVp}{10\%} 
\newcommand{\highRb}{$\sim$50\%}               

\newcommand{\BB}{\tablenotemark{b}}
\newcommand{\BC}{\tablenotemark{c}}
\newcommand{\CD}{\tablenotemark{d}}
\newcommand{\E}{\tablenotemark{e}}

\title{A Census of Star-Forming Galaxies at $z=1$--3 in the Subaru Deep Field}
\author{Chun Ly,\altaffilmark{1,9,10} Matthew A. Malkan,\altaffilmark{1} Masao Hayashi,\altaffilmark{2}
  Kentaro Motohara,\altaffilmark{3} Nobunari Kashikawa,\altaffilmark{4,5} Kazuhiro Shimasaku,\altaffilmark{2,6}
  Tohru Nagao,\altaffilmark{7} and Celestine Grady\altaffilmark{8}}
\shorttitle{A Census of Galaxies at $z=1$--3}
\shortauthors{Ly et al.}
\affil{\altaffilmark{1} Department of Physics and Astronomy, UCLA, Box 951547, Los Angeles, CA, USA; \email{chunly@stsci.edu}}
\affil{\altaffilmark{2} Department of Astronomy, School of Science, University of Tokyo, Bunkyo, Tokyo, Japan}
\affil{\altaffilmark{3} Institute of Astronomy, University of Tokyo, Mitaka, Tokyo, Japan}
\affil{\altaffilmark{4} Optical and Infrared Astronomy Division, National Astronomical Observatory, Mitaka,
  Toky, Japan}
\affil{\altaffilmark{5} Department of Astronomy, School of Science, Graduate University for Advanced Studies,
  Mitaka, Tokyo, Japan}
\affil{\altaffilmark{6} Research Center for the Early Universe, School of Science, University of Tokyo, Tokyo,
  Japan}
\affil{\altaffilmark{7} The Hakubi Project, Kyoto University, Kyoto, Japan}
\affil{\altaffilmark{8} Department of Applied Science, UC Davis, Davis, CA, USA}
\altaffiltext{9}{Current Address: Space Telescope Science Institute, Baltimore, MD, USA}
\altaffiltext{10}{Giacconi Fellow.}

\begin{abstract}
  Several UV and near-infrared color selection methods have identified galaxies at $z=1$--3. Since each
  method suffers from selection biases, we have applied three leading techniques (Lyman break, BX/BM, and BzK
  selection) simultaneously in the Subaru Deep Field. This field has reliable ($\Delta z/(1+z)=0.02$--0.09)
  photometric redshifts for \Nabs\ galaxies from 20 bands (1500 \AA--2.2 \mm). The BzK, LBG, and BX/BM
  samples suffer contamination from $z<1$ interlopers of \Ninta, \Nintb, and \Nintc, respectively.
  Around the redshifts where it is most sensitive ($z\sim1.9$ for star-forming BzK, $z\sim1.8$ for
  $z\sim2$ LBGs, $z\sim1.6$ for BM, and $z\sim2.3$ for BX), each technique finds \peakval\ of
  the census of the three methods. In addition, each of the color techniques shares \nprange\ of its galaxies
  with another method, which is consistent with previous studies that adopt identical criteria on magnitudes
  and colors. Combining the three samples gives a comprehensive census that includes \censuscomp\ of
  $\zphotf=1$--3 galaxies, using standard magnitude limits similar to previous studies. In fact,
  we find that among $z=1$--2.5 galaxies in the color selection census, \absa\ of them can be selected
  by just combining the BzK selection with one of the UV techniques ($z\sim2$ LBG or BX and BM).
  The average galaxy stellar mass, reddening, and star formation rates (SFRs) all decrease systematically
  from the \sbzk\ population to the LBGs, and to the BX/BMs. The combined color selections yield a total
  cosmic SFR density of \sfrK\ for $K_{\rm AB} \lesssim 24$. We find that \pEBV\ of the star formation is
  in galaxies with $\EBV>\EBVlimit$ mag, even though they are only \NEBV\ of the census by number.
\end{abstract}


\section{INTRODUCTION}\label{0}
Several color techniques have succeeded in identifying large samples (thousands) of galaxies in various
windows of high redshift. They work by using deep wide-field imaging in only a few broad-band filters. In
the simplest cases, only two colors are needed to isolate a spectral break. For imaging surveys limited to
optical observations, the Balmer/4000 \AA\ break is measurable up to $z\approx1$, and the Lyman continuum
break is detectable starting at redshifts of 3 and higher \citep[e.g.,][]{steidel99,bouwens06,yoshida06}.
But at intermediate redshifts, CCDs are only sensitive to the spectral region between these two strong
features. The resulting inability to identify galaxies at $z\approx1$--3, has lead to this range being called
the ``redshift desert.'' Finding large samples of galaxies in the redshift desert requires detections of the
Balmer break with near-infrared photometry, or the Lyman break with ultraviolet imaging.

\defcitealias{ly09}{L09}
The extension of the Lyman break technique to $z < 2.6$ requires deep near-ultraviolet (NUV) imaging, which
has only been recently available with the \galex\ satellite \citep[hereafter L09]{ly09} and {\it Hubble} WFC3/UVIS
\citep{hathi10}. Prior to this, different techniques were developed, to select ``BX'' and ``BM'' galaxies
\citep{adelberger04,steidel04}, BzK galaxies \citep{daddi04}, and ``distant red galaxies''
\citep[DRGs;][]{franx03,dokkum04}. However, each technique suffers from its own selection biases. For example,
UV selections (e.g., LBG, BX) tend to identify young star-forming galaxies with low dust extinction, while
IR techniques (e.g., BzK, DRG) select more massive, dusty galaxies.

In this paper, we apply several color selection techniques to our panchromatic photometry of the Subaru Deep
Field (SDF), to identify \nUaround\ galaxies at $z=1$--3 for a census of optical and near-infrared (NIR)
selected star-forming galaxies. Throughout this manuscript, the term ``census'' will be repeatedly
use to refer to the union of the different color selection methods that identify LBGs, BX/BMs, and
\sbzk\ galaxies down to specific magnitude depths that are typical of past and current $z\sim2$ census studies.
Our multi-technique survey, compared to previous work 
\citep[hereafter R05, Q07, and G07, respectively]{reddy05,quadri07,grazian07}, has significantly
more galaxies because our deep imaging covers 2--10 times more area. Since we also have a wide range of
photometry spanning many filters, we can then directly compare these simple color techniques to a sample
derived from photometric redshift \citep[][hereafter photo-$z$ or \zphot]{baum62}. 
\defcitealias{reddy05}{R05}\defcitealias{quadri07}{Q07}\defcitealias{grazian07}{G07}

\newcommand{\pz}{\phm{0}}
\newcommand{\ps}{\phm{1}}
\newcommand{\pss}{\phm{11}}
\newcommand{\psss}{\phm{111}}
\begin{deluxetable*}{lccccccll}[!htc]
\tablewidth{0pt}
\tabletypesize{\scriptsize}
\tablecaption{Summary of SDF Multi-wavelength Data}
\tablehead{
  \colhead{Filter} &
  \colhead{$\lambda_{\rm cen}$\tablenotemark{a}} & 
  \colhead{$\lambda_{\rm FWHM}$\tablenotemark{a}} & 
  \colhead{FWHM (\arcsec)} & 
  \colhead{$m_{\rm lim}^{\rm AP}$\tablenotemark{b}} &
  \colhead{Ap. Corr.\tablenotemark{c}} & 
  \colhead{$m_{\rm lim}^{\rm TOT}$\tablenotemark{d}} & 
  \colhead{$N$($3\sigma$)\tablenotemark{e}} & 
  \colhead{$N$($5\sigma$)\tablenotemark{e}}
}
\startdata
  $FUV$           &\ps1533 & \pss209 &$\sim$4.5& 26.286 &  1.454 & 25.880 & 13895 \pz(16859) &\pz6592 \pz(8016)\\
  $NUV$           &\ps2284 & \pss697 &$\sim$5.0& 26.596 &  1.845 & 25.931 & 47845 \pz(57668) &  30271 (36470)  \\
  $U$             &\ps3634 & \pss750 &     1.49& 26.287 &  1.164 & 26.122 & 34926 \pz(42379) &  24124 (29257)  \\
  $B$             &\ps4438 & \pss687 &     1.01& 28.066 &  1.192 & 27.876 & 88703   (107935) &  74836 (91017)  \\
  $V$             &\ps5463 & \pss885 &     1.11& 27.338 &  1.178 & 27.161 & 75522 \pz(92182) &  62341 (76025)  \\
  \Rc             &\ps6515 & \ps1100 &     1.11& 27.528 &  1.173 & 27.355 & 90452   (110151) &  76773 (93466)  \\
  $i$\arcmin      &\ps7659 & \ps1419 &     1.11& 27.202 &  1.168 & 27.034 & 82308   (100245) &  69666 (84902)  \\
  $z$\arcmin      &\ps9020 & \pss956 &     0.96& 26.270 &  1.181 & 26.090 & 57163 \pz(69778) &  46767 (57017)  \\
  $z_b$           &\ps8842 & \pss620 &     0.91& 25.806 &  1.175 & 25.631 & 47067 \pz(57551) &  36289 (44405)  \\
  $z_r$           &\ps9841 & \ps1000 &     0.91& 24.946 &  1.274 & 24.683 & 28989 \pz(35276) &  21853 (26571)  \\
  $IA598$         &\ps6007 & \pss303 &     0.91& 26.276 &  1.262 & 26.023 & 47764 \pz(58322) &  34987 (42724)  \\
  $IA679$         &\ps6780 & \pss340 &     0.96& 26.870 &  1.214 & 26.659 & 72221 \pz(87969) &  57672 (70361)  \\
  $J$             &  12492 & \ps1799 &     1.27& 22.894 &  1.214 & 22.684 & 16694 \pz(19985) &\pz9637 (11713)  \\
  $H$             &  16186 & \ps5700 &     1.18& 22.328 &  1.208 & 22.123 & 10662 \pz(12691) &\pz5684 \pz(6949)\\
  $K$             &  22035 & \ps2995 &     0.81& 23.673 &  1.172 & 23.500 & 24788            & 17415           \\
  NB704           &\ps7046 & \pss100 &     0.91& 26.112 &  1.150 & 25.960 & 52193 \pz(63537) & 39438 (47951)   \\ 
  NB711           &\ps7111 & \psss72 &     0.91& 25.533 &  1.206 & 25.330 & 36893 \pz(44871) & 25580 (31205)   \\ 
  NB816           &\ps8150 & \pss120 &     0.91& 26.404 &  1.142 & 26.259 & 66728 \pz(81313) & 52492 (63960)   \\ 
  NB921           &\ps9196 & \pss132 &     0.96& 26.045 &  1.167 & 25.878 & 57461 \pz(70347) & 44847 (54809)   \\ 
  NB973           &\ps9755 & \pss200 &     0.91& 25.086 &  1.255 & 24.840 & 37001 \pz(44990) & 26350 (32067)   \\ 
  $[3.6]$ \aper{2}&  35634 & \ps7500 &  \ldots & 24.433 &  1.059 & 24.371 & 92534  (104282)  & 87628 (97728)   \\
  $[3.6]$ \aper{4}& \ldots &  \ldots &    2.25 & 23.192 &  1.424 & 22.809 & 90411  (101482)  & 83352 (92057)   \\
  $[4.5]$ \aper{2}&  45110 &   10000 &  \ldots & 24.014 &  1.230 & 23.790 & 87228 \pz(97890) & 81200 (89907)   \\
  $[4.5]$ \aper{4}& \ldots &  \ldots &    2.10 & 22.817 &  1.456 & 22.409 & 81915 \pz(90890) & 75896 (82998)   \\[-3mm]
\enddata
\label{table1}
\tablenotetext{1}{Filter's center and FWHM are in units of \AA.}
\tablenotetext{2}{Limiting magnitudes are defined as 5$\sigma$ with a \aper{3} ($U$, $J$, and $H$),
  6\farcs8 $\phi$ ($FUV$ and $NUV$), or \aper{2} (remaining wave bands) aperture. IRAC limits are provided
  with a 4\arcsec\ and \aper{2} for original and HiRes data products.}
\tablenotetext{3}{Corrections of aperture flux to total source flux assuming a point source. These
  are determined within the same aperture as $m_{\rm lim}^{\rm AP}$.}
\tablenotetext{4}{The 5$\sigma$ aperture-corrected limiting magnitudes.}
\tablenotetext{5}{First values exclude sources in the $K$-band low sensitivity regions.
  Those in parentheses include sources in these regions.}
\end{deluxetable*}

The observations spanning wavelengths from 1500 \AA\ to 4.5 \micron\ and their reductions are described in
Section~\ref{1}. Section~\ref{2} describes the merging of the multi-wavelength data together, which accounts
for differences in the spatial resolution across different wave bands. Section~\ref{3} discusses the
determination of photo-$z$ and the modeling of the spectral energy distribution (SED). In Section~\ref{4},
the different color selection techniques used and individual sample properties are presented. Section~\ref{5}
discusses the selection effects of one technique against another, and compares them with our \zphot-selected
sample. We also compare our results with previous surveys. In Section~\ref{6}, we present a Monte Carlo
simulation that aims to generate a mock census of $z\approx1$--3 galaxies to support many of our observed
relations and results. Conclusions from our $z=1$--3 census survey are summarized in Section~\ref{7}. 

Throughout this paper, a flat cosmology with [$\Omega_{\Lambda}$, $\Omega_M$, $h_{70}$] = [0.7, 0.3, 1.0] is
adopted. Magnitudes are reported on the AB system \citep{oke74}, and unless otherwise indicated, limiting
magnitudes are 5$\sigma$, corrected for the flux falling outside of the apertures assuming an
unresolved source (i.e., total limiting magnitudes). Diameter apertures sizes are denoted with a ``$\phi$.''

\section{Subaru Deep Field Multi-Wavelength Observations}\label{1}
\begin{figure} 
  \epsscale{1.10}
  \plotone{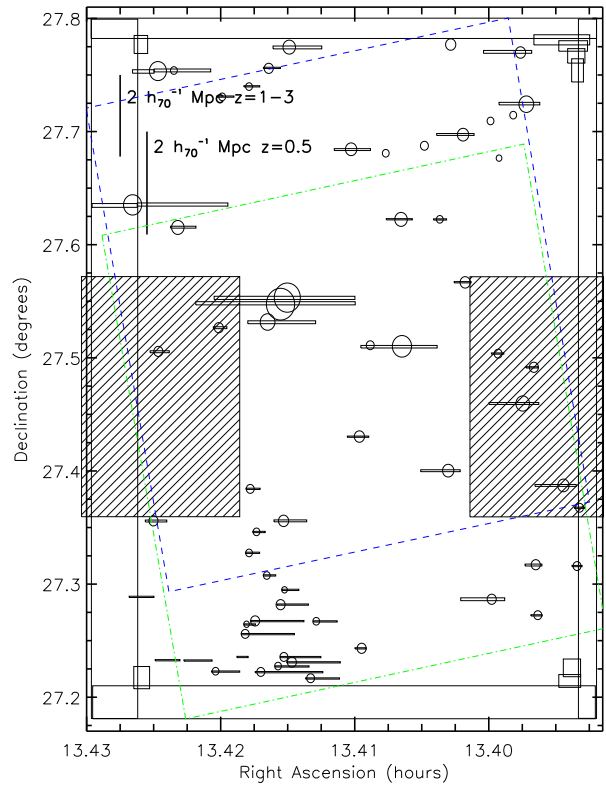}
  \caption{Sky coverage of SDF data described in Section~\ref{1}. Masked regions for the Suprime-Cam data
    are shown with the solid black rectangles and circles. Excluding these regions, the area coverage is 870.4
    arcmin$^2$. The shaded regions correspond to lower $K$-band sensitivity areas (718.7 arcmin$^2$ has deep
    $K$ coverage). \galex, NEWFIRM, and MOSAIC data have coverage over the entire SDF. \spitzer\ 1 ks IRAC
    coverage is illustrated with the dashed line (blue; [3.6] and [5.8]) and dot-dashed line (green; [4.5]
    and [8.0]). The angular size for 2 h$_{70}^{-1}$ Mpc is shown by the thick bars for $z=0.5$ and $z=1$--3.
    \tcolor}
  \label{cov}
\end{figure}

A summary of the depth, spatial resolution, and sample sizes for the imaging data is provided in 
Table~\ref{table1}, while the spatial coverage of the SDF at different wavebands is shown in Figure~\ref{cov}.
Although the reported depths illustrate the sensitivity of the data, we emphasize that all of the
sources actually analyzed later in this paper are detected well above these $5\sigma$ limits in almost all
of the wavebands. The only exceptions are the \fuv, \nuv, $J$, $H$, and $K$ bands.
Since most of these data are described in previous papers, only brief summaries are presented.
 
\subsection{Subaru/Suprime-Cam Data}\label{1.1}
The $BV\Rcf i\arcmin z\arcmin$ observations\footnote{These data are publicly available at
  \url{http://soaps.nao.ac.jp/SDF/v1/index.html}.} are the deepest data obtained to date from
Suprime-Cam \citep{miyazaki02}, and are described in \cite{kashik04}. The total limiting magnitudes were
redetermined with more rigorous masking of pixels affected by object flux and are $\approx0.1$--0.2 mag
deeper than those quoted in \cite{ly07}. They range from 26.1 to 27.9 mag.

In addition to these data, Suprime-Cam imaging in four intermediate-band filters, IA598, IA679, $z_b$, and
$z_r$ are also available. Studies \citep[e.g.,][]{combo02,ilbert09,dokkum09} using intermediate-band filters
have improved photo-$z$ estimates, as they are able to identify the location of spectral breaks more
accurately. The $z_b$ and $z_r$ observations were obtained in 2003 and 2004, and are described in
\cite{shima05}. The net integration times are 85 and 600 minutes for $z_b$ and $z_r$, which give total
limiting magnitudes of 25.6 and 24.7 mag, respectively. The IA598 and IA679 observations were obtained
in 2007 and are described in \cite{nagao08}. Their depths are 26.0 and 26.7 mag, respectively.

Finally, we include imaging in five narrow-band (NB) filters (NB704, NB711, NB816, NB921, and NB973).
These data are included to specifically help improve the \zphot's for emission-line galaxies, but we
find redshift accuracy improvements for galaxies at $z\lesssim1.6$. Their 5$\sigma$ total flux depths
vary between 24.9 and 26.2 mag. The narrow-band data reduction is further discussed in
\cite{ouchi03}, \cite{shima03}, \cite{shima04}, and \cite{kashik04}.

\subsection{UKIRT/WFCAM Data}\label{1.2}
$K$-band data for this survey were acquired with Wide-Field Camera \citep[WFCAM;][]{casali07} on the United
Kingdom Infrared Telescope (UKIRT) on 2005 April 14--15 and 2007 March 5--6. Because the field of view (FoV) of WFCAM is
split into four 13\farcm6$\times$13\farcm6 images separated by 12\farcm8, four pointings were required to
cover the whole SDF optical image. The total exposure time for each pointing is 294--300 minutes, except
for one pointing of 15 minutes. We ignore the less sensitive regions, so 80\% of the SDF optical coverage
has uniform $K$-band data down to the deepest sensitivity (see Figure~\ref{cov}). Previous results using a
subset of this data were presented in \cite{hayashi07}. The data were reduced following standard NIR
reduction procedures in IRAF. Photometric and astrometric calibrations were carried out against 
Two Micron All Sky Survey \citep[2MASS;][]{skrutskie06} stars. The final $K$-band data reach a
total $5\sigma$ depth of 23.5 mag. 

\subsection{Mayall/MOSAIC Data}\label{1.3}
$U$-band data were obtained from the KPNO Mayall 4 meter telescope with the Mosaic-1 Imager
\citep[FoV of 36\arcmin;][]{muller98} on 2007 Apr 18 and 19. Observing conditions were dark
with minimal cloud coverage. A series of 25 minute exposures was obtained to accumulate 47 ks for a 
5$\sigma$ total depth of 26.1 mag. A standard dithering pattern was followed to provide more uniform
imaging across the $\approx$10\arcsec\ CCD gaps. 
The MSCRED IRAF (version 2.12.2) package was used to reduce the data and produced final mosaic images with a
pixel scale of 0\farcs258 and an average-weighted seeing of $\approx$1\farcs5 FWHM. The reduction steps
followed closely the procedures outlined for the reduction of the MOSAIC data for the NOAO Deep Wide-Field Survey.

While several \cite{landolt92} standard star fields were observed, Landolt's $U$ filter differs significantly
from the MOSAIC/$U$, which makes it difficult to calibrate the photometry. One hundred and two Sloan Digital Sky Survey (SDSS)
stars distributed uniformly across the eight Mosaic-1 CCDs with $u\arcmin \lesssim 21$ mag were used. This
approach requires a transformation between SDSS $u$\arcmin\ and MOSAIC/$U$, which is obtained by convolving
the spectra of 175 Gunn--Stryker stars with the total system throughput at these wavebands. These stars span
$B-V = -0.2$--0.8 mag, and show that the $B-V$ color term is smaller than the scatter (1$\sigma$ = 0.06 mag)
in the predicted U magnitudes.

\subsection{\textit{GALEX}/\textit{NUV} and \textit{FUV} Data}\label{1.4}
The \galex\ space telescope \citep{martin05} provides simultaneous $FUV$ and $NUV$ imaging with a pixel scale
of 1\farcs5 and an FoV of 1\fdg2, thus covering the entire SDF. A total of 161225 s\footnote{The results
  of \citetalias{ly09} used a data set that is shallower by 20 ks.} (80851 s) was obtained in
the $NUV$ ($FUV$) band \citep{malkan04}. The data were processed through the \galex\ reduction pipeline and
stacked accordingly. The FWHM of unresolved images is 4\farcs5 in the $FUV$ and 5\farcs0 in the $NUV$.
The 5$\sigma$ total limiting magnitudes are both $\approx$25.9 mag.

\subsection{Mayall/NEWFIRM Data}\label{1.5}
$J$- and $H$-band observations were obtained from the NOAO Extremely Wide-field Infrared Imager
\citep[NEWFIRM;][]{probst04,probst08} on 2008 May 10--12. Exposures of 60 ($J$) and 20 or 30 ($H$)
s were taken with a simple five-point dithering pattern to image across the detector gaps
($\approx35\arcsec$ in size) and to avoid latent images of bright sources on the same source. A
total of 27.5 ks and 5.5ks were acquired in $J$ and $H$, respectively. While cloud coverage was
minimal, the conditions were non-photometric (e.g., the effective $J$ exposure time was approximately
9ks). These NEWFIRM data required a first-pass reduction to create a deep object mask. This mask is
then used to create flat-fields and sky images to enable proper subtraction of the sky background. Then
the sky-subtracted images were used for the second-pass processing to create the final mosaics.
The $J$ ($H$) limiting magnitude was determined to be 23.4 (22.8) mag in a \aper{3} aperture, and
the average seeing is $\approx$1\farcs3 (1\farcs2).

\subsection{\spitzer/IRAC Data}\label{1.6}
Imaging in the IRAC bands was conducted in 2006 February 11 and 14 \citep[Prop\#20229;][]{malkan05}. The observations
consisted of nearly uniform coverage in all four bands of 1 ks. The coverage is $26\arcmin\times25\farcm5$
for [3.6] and [5.8] or [4.5] and [8.0], and the overlapping region for all four bands is approximately
$19\arcmin\times25\farcm5$. Version S13.2 data products were used. Necessary pre-processing steps were taken
to remove detector artifacts (e.g., mux-bleeding, column pull-downs) with contributed software provided by
the IRAC community, and the data reduction followed standard procedures using the MOsaicker and Point
source EXtractor (MOPEX) package. The astrometric solutions for the mosaics were examined against 2MASS
stars and are accurate to with 0\farcs1.

Apertures were placed randomly on the images to determine the limiting depth. The depths are summarized in
Table~\ref{table1}. For the remainder of this paper, we only use the [3.6] and [4.5] data, since they
sample rest frame $\sim1$--2 \mm\ for $z=1$--3, which are unaffected by polycyclic aromatic hydrocarbon (PAH)
features.

The greatest limitation of the  IRAC data is the spatial resolution: measuring unresolved galaxies generally
requires at least a \aper{4} aperture. Recently, \cite{velusamy08} developed a high-resolution image
deconvolution code (HiRes) to handle \spitzer\ data. The code has been found to significantly improve the
quality of the images, allowing for smaller photometric apertures and a reduction of source confusion. HiRes
was used on these IRAC data, and the results show improvements in the aperture fluxes by a factor of 2--4
using the same aperture sizes. For example, a \aper{2} aperture was able to encompass 95\% and 81\% of the
total light in [3.6] and [4.5], respectively. We examined sources in both images and found that photometric
fluxes are conserved, indicating that these HiRes images can be used to attain reliable flux measurements
in these bands. This improvement allowed for the crucial detection of lower mass galaxies at $z\sim2$ at
rest 1 \mm.

\section{The SDF Multi-band Galaxy Sample}\label{2}
\subsection{Synthesizing Multi-band Data}\label{2.1}
{\it Methodology.}
A key requirement of this survey is to merge the full SDF multi-wavelength data. However, the spatial resolution
varies by a factor of a few for the different instruments. Simply degrading the best data ($\sim$1\arcsec) to
match the poorest resolution (4\arcsec--5\arcsec\ for \galex\ data) will hamper the identification of faint galaxies,
most of which are at high redshift.

A solution to this problem is a hybrid approach of (1) slightly degrading high-resolution images to match a
``baseline'' spatial resolution that is not significantly different and (2) using aperture corrections. This
method will not greatly compromise sensitivity, and the aperture corrections are reliable, since this work
mostly probes compact galaxies (they are faint and at high-$z$).

We begin by combining multiple broad- and intermediate-band images into a single ultra-deep frame.
This approach benefits from the inclusion of weak sources that are marginally detected in individual
wave bands, but are highly significant in the merged image. This ultra-deep image serves as a reference
image when running SExtractor \citep{bertin96} in ``dual image'' mode.
It yields a catalog of sources in the SDF, and later serves as the ``master object list'' for the
photo-$z$ catalogs and the color-selected samples.
The similarities of the spatial resolution for $BV\Rcf i\arcmin z\arcmin K$, IA598, and IA679 allow us to
degrade these images to a 1\farcs1 (hereafter ``psf-matched''). The inclusion of $K$ implies that this
ultra-deep mosaic is simultaneously sensitive to optically selected and NIR-selected galaxies.

To stack these images, we use the $\chi^2$-weighting scheme of \cite{szalay99}.
This method uses the $\chi^2$ of each pixel, imaged in $N$ bands, to estimate the probability that it is
sampling a source or the sky. First, for image $i$, the mean ($\mu_i$) is removed, and then the
image is scaled by the sky rms fluctuation ($\sigma_i$):
\begin{equation}
  g_i(x,y) = \frac{f_i(x,y) - \mu_i}{\sigma_i},
\end{equation}
where $f_i(x,y)$ is the measurement (in count rates, electrons, or fluxes) in image $i$, and $g_i(x,y)$
can be viewed as the signal-to-noise ratio (S/N) for a given pixel.
Then, we sum the squares of these 8 S/N images to yield the ultra-deep image:
$y = \displaystyle\sum_{i=1}^{8} g_i^2$.

We considered including additional wave bands into the ultra-deep image, but they degraded the results
due to their lower sensitivities and/or poorer spatial resolutions. For example, the $U$-, $J$-, and $H$-band
data have lower spatial resolution ($\sim1\farcs5$) and are less sensitive than the nearby $B$- and $K$-band
data. Similarly, including the narrow-band observations does not increase the sample sizes. These data are
shallower than the broad-band SDF observations, so almost all of the narrow-band selected sources are identified
in the broad-band images. While there are some galaxies that are only detected in the narrow-band observations
and thus missed by our multi-band $\chi^2$ image (e.g., Ly$\alpha$ emitting galaxies at $z\gtrsim5$), this
tiny galaxy population is less than 0.05\% of the full catalog, and will not affect the $z=1$--3 galaxy census
that we later derive. For these Ly$\alpha$ emitters, it is better to generate a separate sample using only
the narrow-band observations, as done in \citet{kashik06} and \cite{shima06}, for example.

{\it Source catalogs.}
We run SExtractor on the square root of the ultra-deep $\chi^2$ image to obtain a total of 243,964 sources
with a minimum of five connected pixels above a $\chi=3.5$ threshold, of which 211,594 sources are in the
unmasked regions shown in Figure~\ref{cov}. This sample is further reduced to 174,837 sources when the poorly
sampled $K$-band regions are excluded. This is the largest photometric catalog for the SDF.

One concern with the multi-band catalog is that it may systemically miss some galaxies, particularly the
bluest and reddest ones, because they are undetected in the longer or shorter bandpasses, respectively.  We
investigate this possibility by generating $BV\Rcf$ and $i\arcmin z\arcmin K$ stacked mosaics using the
$\chi^2$ weighting approach. We run SExtractor on these stacked images, and compare these catalogs against the
ultra-deep eight-band catalog. First, we find that the overlap is $\approx$90\% or larger at any given magnitude
in any of the individual bands. Second, since galaxies with extreme colors could be systematically missed, we
also examine the overlap of the bluest (reddest) quartile of $BV\Rcf$ ($i\arcmin z\arcmin K$) catalog against
our eight-band catalog. The blue sample consists of $\sim$20,000 galaxies with $B-V < 0.2$ mag while the red
sample contains $\sim$6000 galaxies with $z\arcmin - K > 0.4$ mag. Fortunately, the overwhelming majority
of these galaxies with extreme colors were already included in the original selection using the eight-band
ultra-deep image, 93\% and 89\% of the blue and red samples, respectively. We find that these fractions hold
regardless of magnitudes or colors. These tests and comparisons indicate that the multi-band synthesis
approach does not inherently miss any significant population of galaxies. We will discuss this issue further
in Section~\ref{6}.
\cite{szalay99} also find that the approach is optimal when generating source catalogs for the Hubble Deep Field.

{\it Measured fluxes.}
Generally, fluxes are measured in fixed circular apertures with diameter sizes comparable to
twice the FWHM of resolution, and are then corrected for aperture losses, which are listed in
Table~\ref{table1}.
Thus, for images with $\lesssim1\arcsec$, we measure photometric fluxes in \aper{2} aperture with SExtractor.
For the $U$, $J$, and $H$ bands, we use \aper{3} aperture measurements. The \galex\ and IRAC/HiRes bands
use a 6\farcs8 and \aper{2} apertures, respectively. For the \galex\ and IRAC/HiRes data, the images were
not matched to the Suprime-Cam pixel scale. Instead, we use DAOPHOT to measure the fluxes at known
positions, which follows the identical procedure discussed in \citetalias{ly09}. One advantage of DAOPHOT
is that additional noise is not introduced when these images are regridded to the reference SDF
Suprime-Cam image with a pixel scale of 0\farcs2.

{\it GALEX source confusion.}
One concern for GALEX  \fuv\ and \nuv\ measurements is source confusion---the overestimation of the
  flux for faint sources that happen to fall in the wings of brighter nearby sources. To correct for this,
we take a point-spread function (PSF)-fitting approach, which supersedes the baseline approach of fixed aperture measurements.
This consists of two stages: (1) fitting and removing relatively bright sources with an empirical PSF
constructed from a dozen unresolved bright sources, and (2) then for fainter sources, aperture photometry
can be done on the image with the relatively bright sources removed. The \galex\ fluxes for the bright and
faint sources are then combined together for a single catalog.

The PSF fitting is done using a few DAOPHOT tools. DAOPHOT assumes that sources are unresolved relative
to the PSF. Since 97\% (51\%) of the objects in our multi-band source catalogs are 3\arcsec\ (1\farcs1) FWHM
or smaller, the large spatial resolution of 4\arcsec--5\arcsec\ for \galex\ implies that modeling and removing
sources with a point-like PSF is a valid approach for almost all sources.
  
Prior inputs to fitting the GALEX observations are based on detections in the $U$ band. 
A total of 43,839 (31,084) sources were considered
for \nuv\ (\fuv) PSF fitting down to 26.75 mag (for \nuv) and 26.25 mag (for \fuv), which is much deeper
than any previous photometric studies with GALEX. Only 18,036 (14,341) sources were actually PSF-fitted,
since the remaining sources were either too faint or too close to another brighter object.

While PSF fitting is done for bright sources, we still determine the total fluxes using the fixed
aperture flux measurements and apply aperture corrections. This ensures that all fluxes are obtained in a
consistent manner. We note that we compare ``total'' flux measurements with those
obtained from PSF-fitting for bright sources, and find them to be consistent within the uncertainties. This
is expected, since the aperture corrections are derived from the empirical PSF, which is used to fit and
remove the flux from the wings of sources. 
One form of ``quality assurance" is to compare the ``noise" in the original GALEX image to that of
the ``cleaned" image, after PSF-fitting subtraction.   Visual inspection shows that the source-subtracted
image looks very much like pure white sky noise. We quantified this by measuring the 1$\sigma$
fluctuations in the same area with same clipping algorithm, and find that the rms is reduced by 50\%
after the sources are PSF-fitted and removed.  In fact, the rms of the cleaned NUV image, 19 photons
per pixel, is only slightly larger than the predicted rms based on Poisson count statistics ($\sim16$).
Similarly, the cleaned FUV image has an average sky background of 13 photons per pixel. The square root
of this is 3.6 photons, which is very close to our observed rms fluctuations of 3.9 photons. 
Thus the sky noise in our cleaned picture is nearly as small as the limit set by the photon statistics
of the sky background.

\subsection{The Identification of Stars}\label{2.2}
To identify foreground stars we use two techniques. First, stars can be easily identified in the NIR, since it
measures the Rayleigh--Jeans tail, and forms a stellar locus in color space distinct from galaxies. Studies
that have selected BzK galaxies also distinguished stars by their $z\arcmin-K$ colors
\citepalias[see e.g., ][]{quadri07}. For sources detected in the $K$ band, we classify 1603 stars with
$z\arcmin-K \leq 0.455(B-z\arcmin) - 0.773$ and $z\arcmin-K \leq 0.235(B-z\arcmin) -0.279$.

\begin{figure*}[htc] 
  \epsscale{0.45}
  \plotone{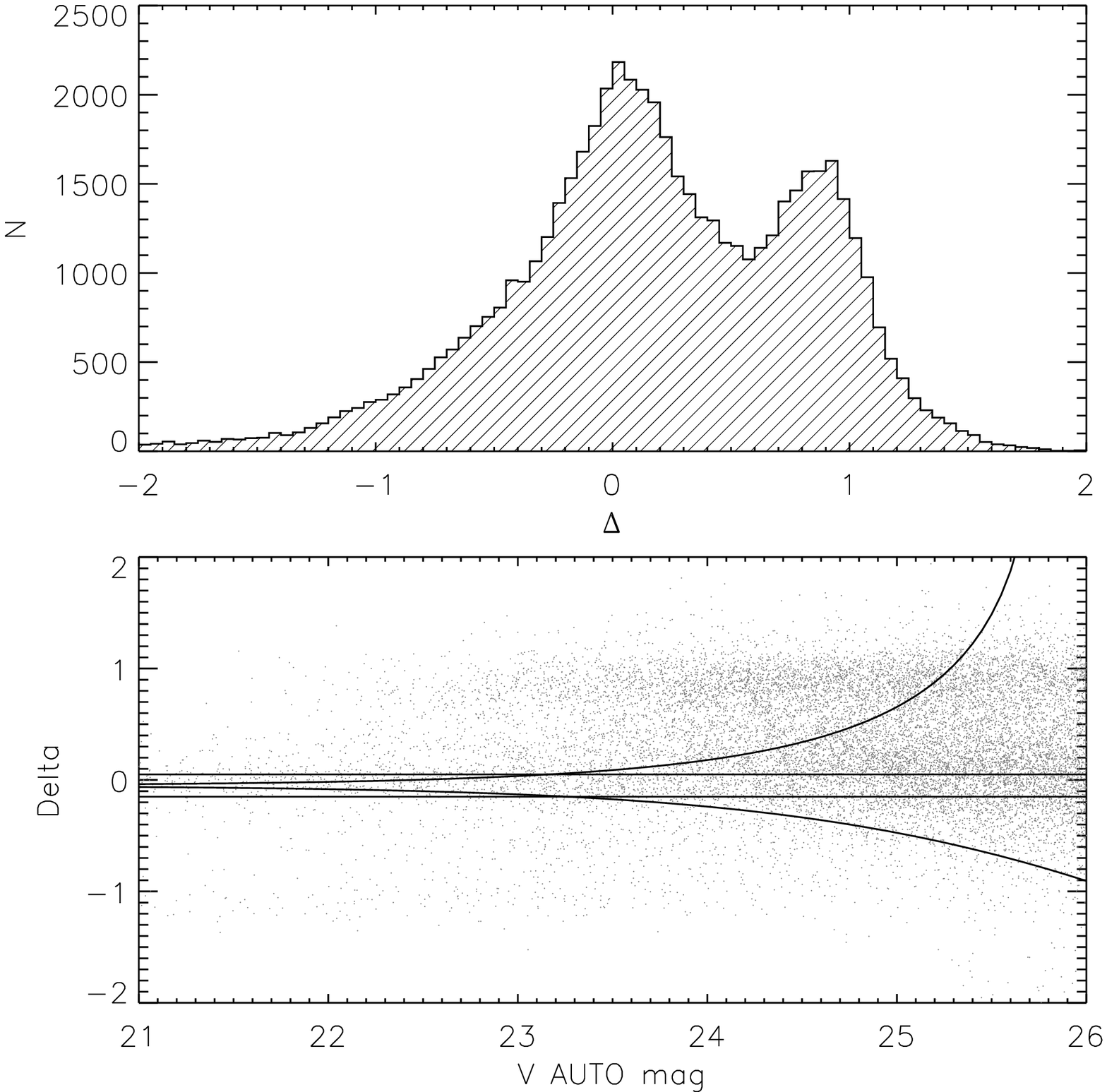} \plotone{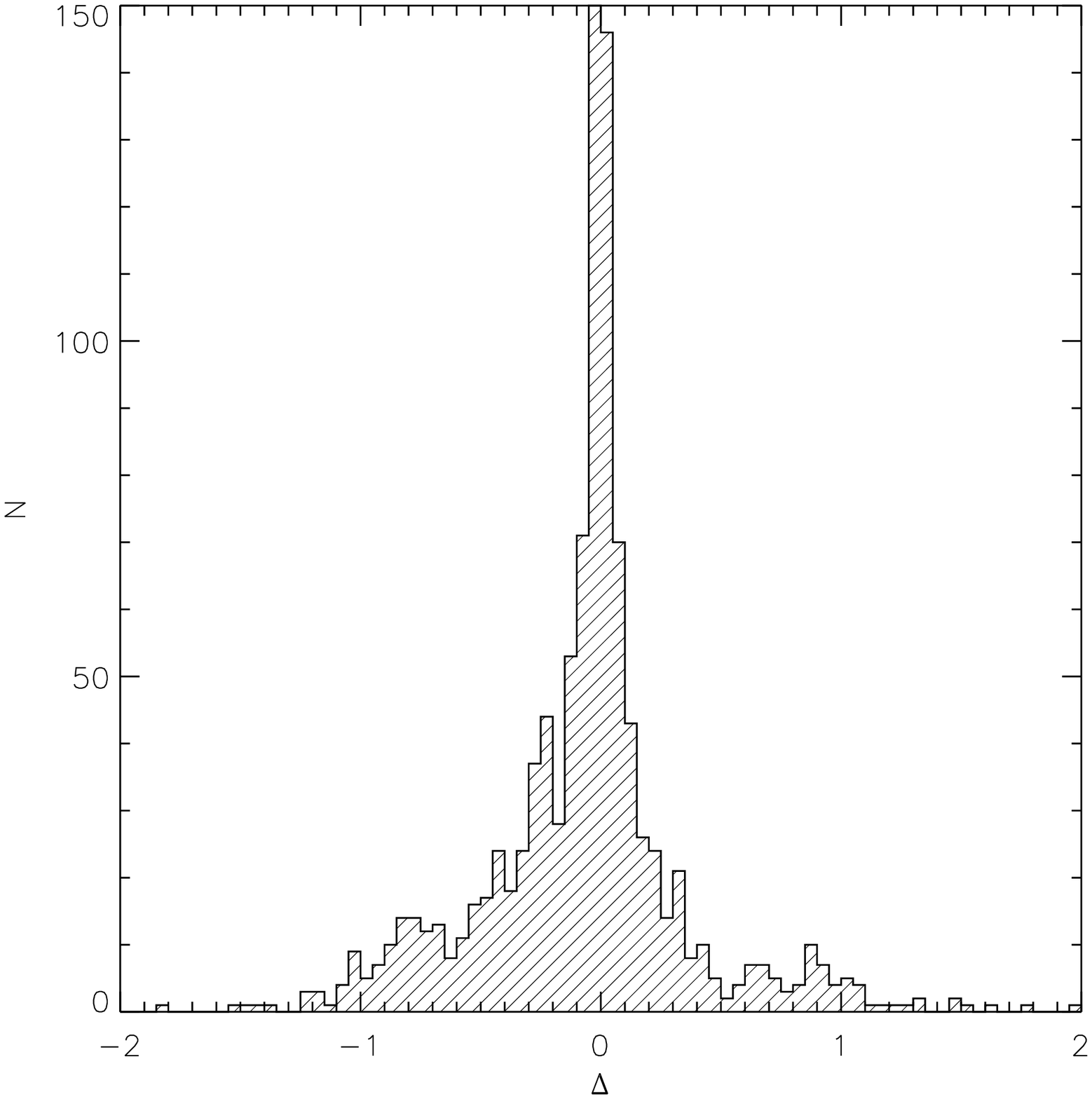}
  \caption{Deviation ($\Delta$) in magnitudes from the stellar locus in a three-color space. (Left)
    Distribution of $\Delta$ is shown for $\sim$67,000 sources with point-like rank of 4--10 in the top panel
    and against $V$ in the bottom panel. The solid lines correspond to a minimum $\Delta=\pm0.1$ mag and the
    $\pm3\sigma$ fluctuations expected from a combination of the $B$, $V$, \Rc, and $z\arcmin$ depth.
    (Right) The distribution of $\Delta$ for $K$-band selected stellar sources based on their
    $B-z\arcmin$ and $z\arcmin-K$ colors. This shows that the majority of stellar sources have
    $\Delta\approx0$.}
  \label{deltaplots}
\end{figure*}

For sources that are undetected in $K$ (or lack $K$ data), we use the technique (called the $\Delta$ method)
that is described in \citetalias{ly09}. First, objects were considered potential stellar candidates based on how
similar they are to the PSF. They were assigned a number between 0 and 10 (10 being most PSF-like).
However, unresolved galaxies might also be very compact. Therefore, to distinguish these galaxies from stars,
we then calculate their deviation from the stellar locus in the $B-V$, $V-\Rcf$, and $\Rcf-z\arcmin$ colors.
This deviation ($\Delta$) from the locus in a multi-color space is certainly affected by photometric scatter
for faint sources, so we limit stellar identification with the $\Delta$ method to $V=25.0$. $\Delta$ values
for $\sim$67,000 sources are illustrated in Figure~\ref{deltaplots}, and show a strong peak at $\Delta\approx0$
(the secondary peak is for low-$z$ galaxies).

One of the best tests of this method is to compare with the sample identified from $B-z\arcmin$ and
$z\arcmin-K$ colors. The distribution of $\Delta$ for the 1603 $K$-band selected stars is plotted in
Figure~\ref{deltaplots} and supports the $\Delta$ method. Among the $K$-band stellar sources, 82\% (1315/1603)
of them have point-like rank values of 4 and higher. Thus, we classify undetected $K$-band sources with rank
values of at least 4, $V<25.0$, and $\Delta$ values within $3\sigma$ of the expected fluctuations in
$\Delta$-magnitude relation as stellar sources. This yielded 1167 additional sources, for a total of 2770
stellar sources. We removed these stellar sources from our full catalog and report stellar contamination for
individual galaxy selection samples below.

\section{Photometric Redshifts and SED Modeling}\label{3}

\subsection{Photometric Redshifts}\label{3.1}
For this survey, the number of spectroscopic redshifts is limited ($N\sim 1000$), so well calibrated \zphot's
are used. These \zphot's are determined from 20 bands (\fuv, \nuv, $UBV\Rcf i\arcmin z\arcmin z_bz_rJHK$,
IA598, IA679, and five narrow-band filters) using the Easy and Accurate \zphot\ from Yale
\citep[EAZY;][]{brammer08} package. This photo-$z$ code has the following advantages: (1) it is easy to use
and fast, (2) it handles input measurements in fluxes rather than magnitudes \citep[e.g.,
{\it hyperz};][]{bolzonella00}, which is crucial for non-detections (see below), and (3) it has been
well calibrated against existing optical-to-NIR data with spectra for $\sigma_z/(1+z)\approx0.03$.

Throughout the paper, redshift distribution is described by N($z$), which uses the peak of the probability
redshift distribution. We have compared N($z$) for a subset of galaxies at high redshift to the distribution
made by adding the full redshift distribution for each galaxy. We find that the two distributions are in
agreement, justifying the use of the peak.

\begin{figure*} 
  \epsscale{0.45}
  \plotone{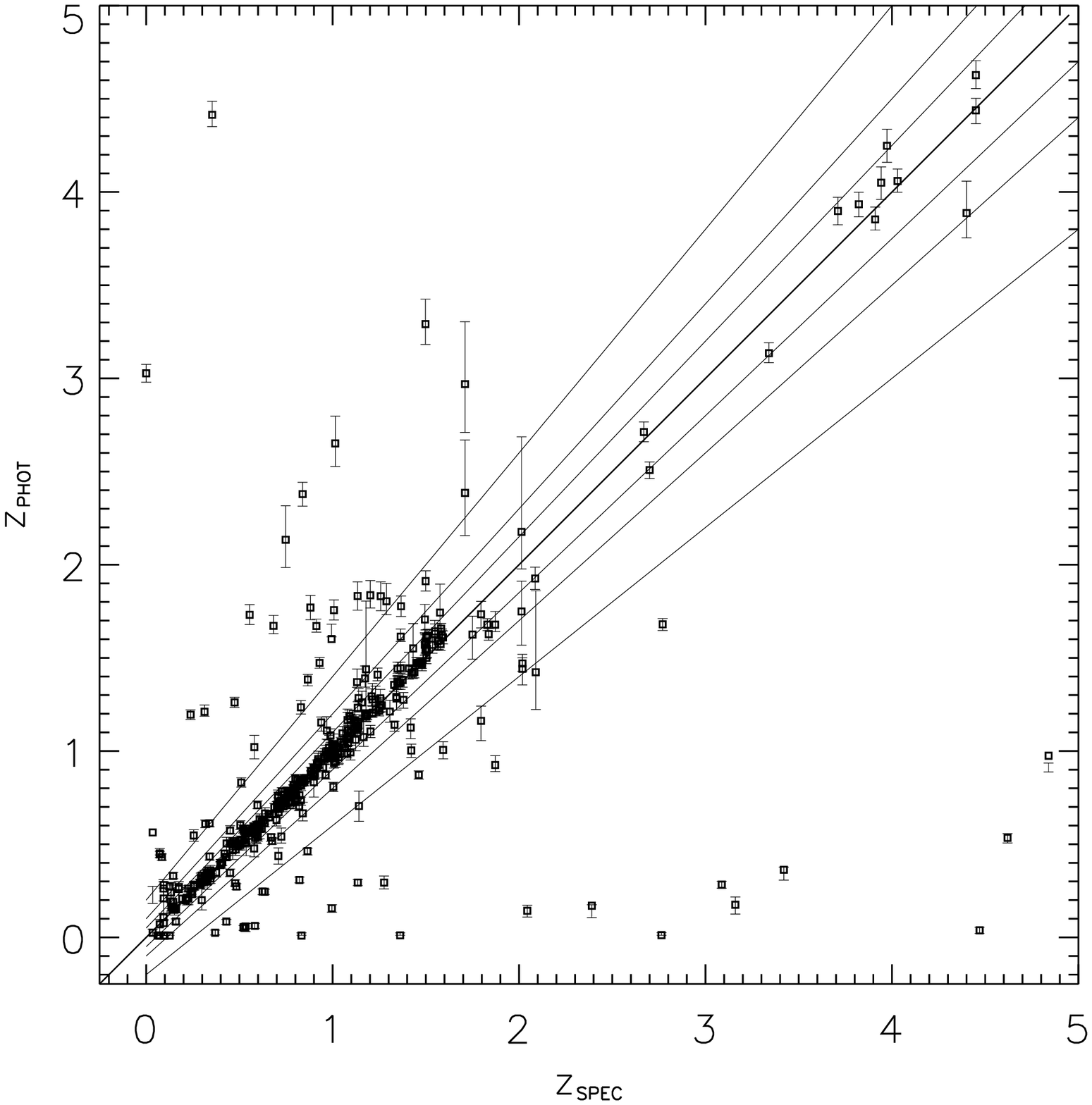} \plotone{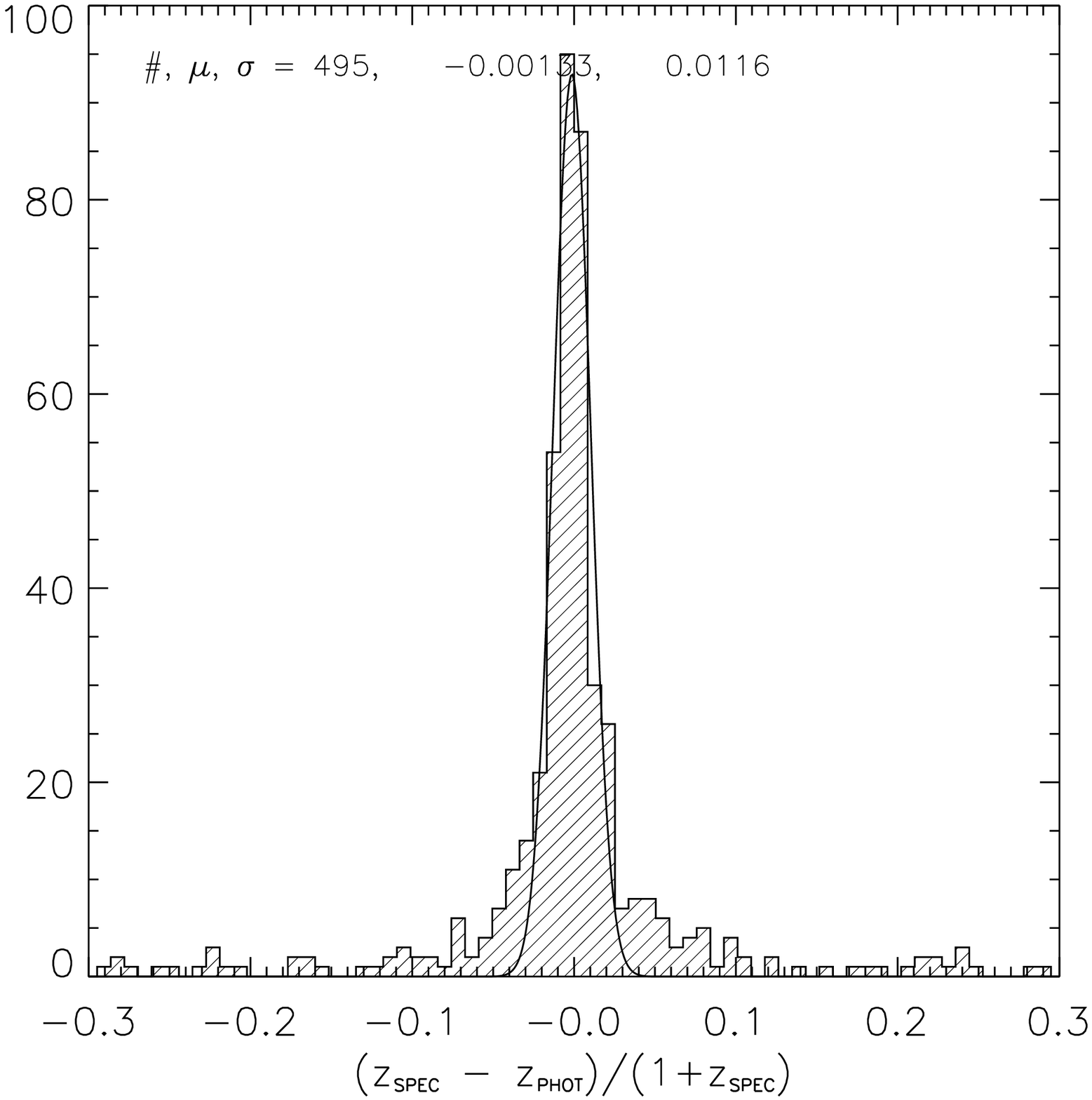}
  \caption{Comparison of photometric and spectroscopic redshifts. A total of \Nspec\ spectroscopic sources
    have \zphot\ estimates. The \zphot\ are derived from 20-band photometry and show accuracy of
    $(\zphotf-\zspecf)/(1+\zspecf) \approx \accurd$ with less than 1\% systematic offset. We find similar
    results using a smaller sample that has $>5\sigma$ detection in K. Solid lines are shown for
    $\pm$5\%, $\pm$10\%, and $\pm$20\% errors.}
  \label{zphotzspec}
\end{figure*}

Illustrated in Figure~\ref{zphotzspec} is a comparison of photometric and spectroscopic redshifts for
\Nspec\ sources using 20 bands. The spectroscopy was obtained from Subaru, MMT, and Keck.
The inclusion of the \nuv\ improved photo-$z$ accuracy by $\approx$17\% compared to
$UBV\Rcf i\arcmin z\arcmin z_bz_rJK$, IA598, and IA679 alone. $H$ and \fuv\ measurements added a further
$\sim$10\% improvement. Photo-$z$'s derived from 15 bands (excluding narrow-bands) are accurate to 2.5\%
out to $z\approx1.8$, and $\approx$\accur\ by including the narrow-band measurements. \spitzer/IRAC
measurements were also included, but produced little improvement. In particular, we found more problems
at lower redshift, and this is likely due to PAH features that have not been included in the photo-$z$
templates. We find that the catastrophic failure rate is $\sim$10\%. A preliminary examination indicates
that one-fourth of these failures are due to galaxies with very strong optical emission lines.

Our photometric redshifts of galaxies with emission lines identified with narrow-band filters
\citep[see][]{ly07} are also found to be reliable. Among 5029 narrow-band excess line emitters between
$z=0.24$ and $z=1.47$, 4568 have \zphot\ estimates. This is almost 10 times the number of spectra that we used
to compare \zphot\ with \zspec. We estimate the median \zphot\ and find good to reasonable agreement
prior to including narrow-band measurements and additional improvement with the narrow-band filters
included (see Table~\ref{tableNB}).
\begin{deluxetable}{lccc}
\tablewidth{0pt}
\tabletypesize{\scriptsize}
\tablecaption{Photometric Redshifts for Emission-line Galaxies}
\tablehead{
  \colhead{Sample} &
  \colhead{$z_{\rm NB}$} & 
  \colhead{\zphot (15-band)} &
  \colhead{\zphot (20-band)}
}
\startdata
  \Ha\ NB816   & 0.243 & 0.257$\pm$0.015 & 0.233$\pm$0.003\\
  \Ha\ NB921   & 0.401 & 0.410$\pm$0.008 & 0.396$\pm$0.007\\
  \OIII\ NB704 & 0.407 & 0.409$\pm$0.007 & 0.403$\pm$0.000\\
  \OIII\ NB711 & 0.423 & 0.428$\pm$0.004 & 0.431$\pm$0.000\\
  \OIII\ NB816 & 0.630 & 0.635$\pm$0.009 & 0.645$\pm$0.046\\
  \OIII\ NB921 & 0.837 & 0.788$\pm$0.010 & 0.835$\pm$0.000\\
  \OII\ NB704  & 0.891 & 0.797$\pm$0.088 & 0.890$\pm$0.000\\
  \OII\ NB711  & 0.912 & 0.890$\pm$0.013 & 0.909$\pm$0.000\\
  \OII\ NB816  & 1.189 & 1.093$\pm$0.272 & 1.193$\pm$0.002\\
  \OII\ NB921  & 1.467 & 1.537$\pm$0.033 & 1.473$\pm$0.000\\[-3mm]
\enddata
\label{tableNB}
\end{deluxetable}

These \zphot's derived without the inclusion of narrow-band data confirm (1) the ability of \cite{ly07} to
distinguish \Ha, \OIII, and \OII\ emitters accurately using simple two-color selection, and that (2)
the \zphot\ derived for SDF galaxies are reliable.

\begin{figure} 
  \epsscale{1.0}
  \plotone{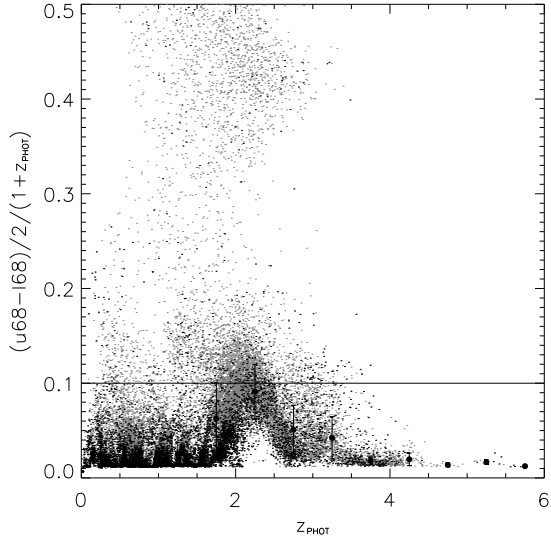}
  \caption{Illustration of the \zphot\ uncertainties for the combined \zphot\ sample. The $x$-axis shows
    the best-fitting \zphot\ while the $y$-axis shows \dz, the half-width of \zphot\ distribution at the 68\%
    confidence level normalized by $1 + \zphotf$. The horizontal lines correspond to the adopted 10\% accuracy
    per ($1+\zphotf$) used to exclude less accurate \zphot\ estimates. Black points denote sources in the
    \fBK\ catalog.}
  \label{sigplot}
\end{figure}

Another illustration of the accuracy of the \zphot's is shown in Figure~\ref{sigplot}, where half of the
upper 68\% minus the lower 68\% confidence ranges in redshift is shown versus photo-$z$. We refer to this as
\dz $\equiv \frac{\left(z_{+68} - z_{-68}\right)/2}{1+\zphotf}$, and use it to indicate \zphot\ accuracy.
The median and standard deviation of \dz\ are reported in Table~\ref{sigplot_table} and are shown in
Figure~\ref{sigplot}. As expected, $z\approx1.8$--2.8 suffers from larger \zphot\ errors due to the absence
of a Lyman or Balmer break within the CCD filters, lower sensitivity in the NIR, and in most cases,
non-detection in the rest-UV wave bands.

%
\begin{deluxetable}{crccrc}
\tablewidth{0pt}
\tabletypesize{\scriptsize}
\tablecaption{Summary of Photometric Redshift Uncertainties}
\tablehead{
  &
  \multicolumn{2}{c}{5BK3$\sigma$} & & 
  \multicolumn{2}{c}{5B5$\sigma$}\\
  \cline{2-3}
  \cline{5-6}
  \colhead{$z$} &
  \colhead{$N$} & 
  \colhead{Median $\pm$ $\sigma$} &
  \colhead{} & 
  \colhead{$N$} &
  \colhead{Median $\pm$ $\sigma$}
}
\startdata
0.0--0.5 &  3213 & 0.020$\pm$0.008 & &  6763 & 0.024$\pm$0.014\\
0.5--1.0 &  7029 & 0.019$\pm$0.009 & & 16172 & 0.020$\pm$0.008\\
1.0--1.5 &  4508 & 0.019$\pm$0.006 & & 10205 & 0.025$\pm$0.013\\
1.5--2.0 &  4242 & 0.033$\pm$0.016 & & 12827 & 0.062$\pm$0.038\\
2.0--2.5 &  1673 & 0.073$\pm$0.029 & &  9257 & 0.091$\pm$0.029\\
2.5--3.0 &  1253 & 0.044$\pm$0.023 & &  5313 & 0.050$\pm$0.025\\
3.0--3.5 &   522 & 0.050$\pm$0.032 & &  2388 & 0.041$\pm$0.021\\
3.5--4.0 &   353 & 0.019$\pm$0.005 & &  1333 & 0.018$\pm$0.004\\
4.0--4.5 &    45 & 0.020$\pm$0.005 & &   374 & 0.019$\pm$0.006\\
4.5--5.0 &     2 & 0.017$\pm$0.006 & &    30 & 0.014$\pm$0.002\\[-3mm]
\enddata                                  
\label{sigplot_table}
\tablecomments{Reported are the median values and standard deviations of the 68\% \zphot\ uncertainty:
  $\Delta z_{68} \equiv (z_{+68}-z_{-68})/2/(1+\zphotf)$.}
\end{deluxetable}

\begin{figure*}[htc] 
  \begin{center}
    \epsscale{0.4}
    \plotone{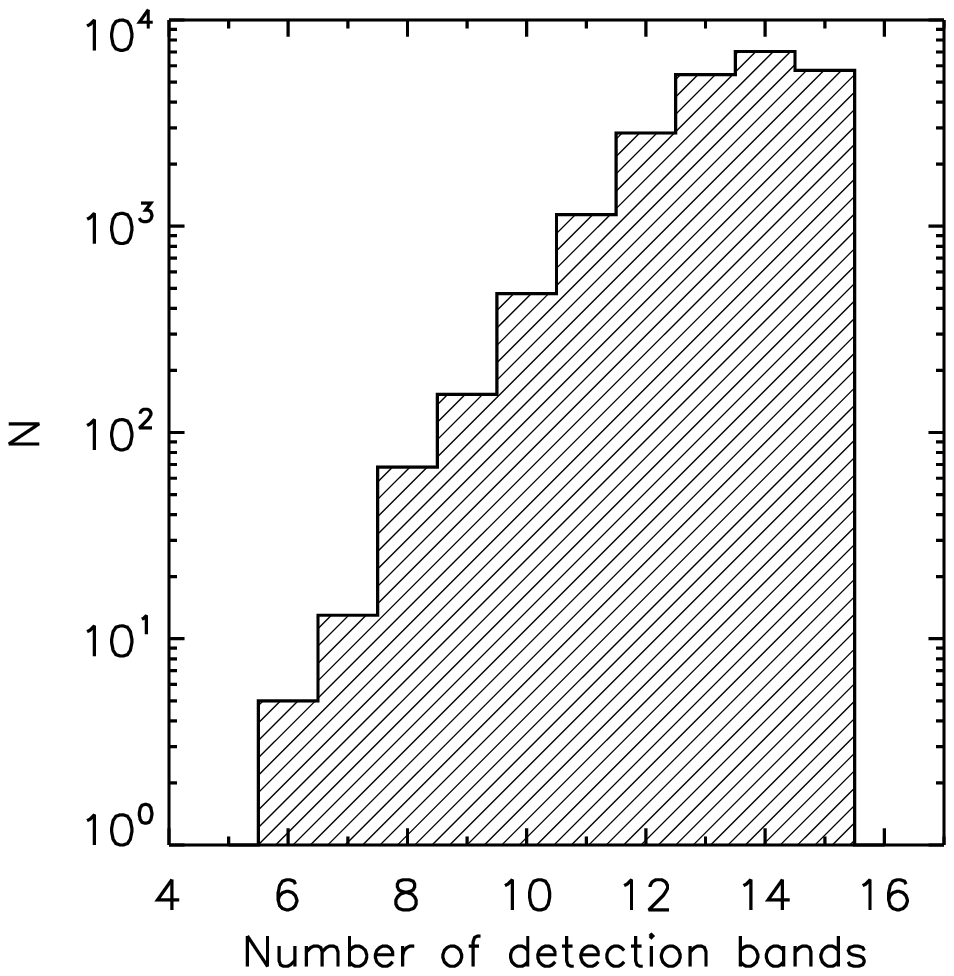} \plotone{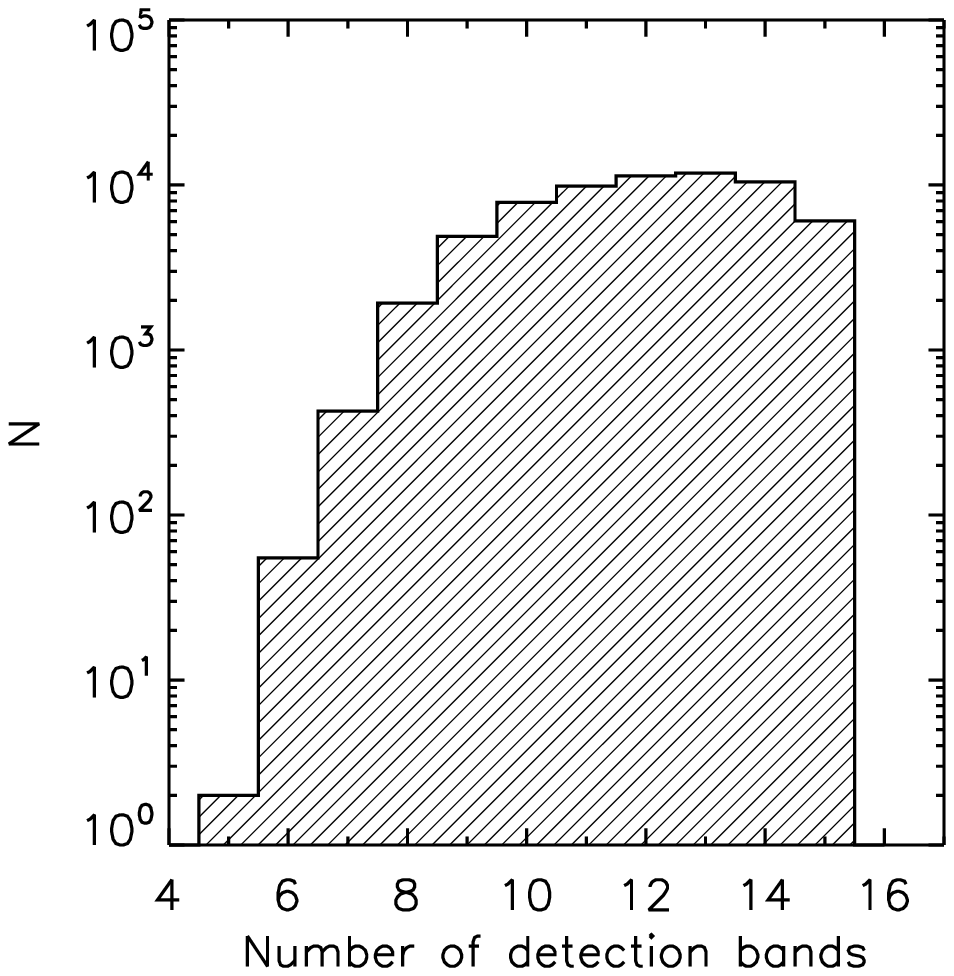}
    \caption{Summary of the number of band detections for sources in the \fBK\ (left) and \fBf\ (right)
      samples. Optical and NIR bands are included for a total of 15 bands. We have detections in at least
      10 wave bands for 99\% of 5BK$3\sigma$ sample and 88.7\% of \fBf\ sample.}
  \end{center}
  \label{detections}
\end{figure*}

We did a test excluding $K$-band measurements (our deepest NIR data) for a set of 24,000 $K$-band galaxies
detected at $3\sigma$ in at least five bands. We find that the \zphot\ errors increased by 20\% (mostly at
$z>1$), that the number of outliers (see below) doubled, and the greatest ($\Delta\zphotf\approx0.4$)
systematic offsets occurred at $\zphotf=1.3$--2.2. However, for the majority ($\approx$75\%) of sources,
the photo-$z$'s without $K$ are good to within $\Delta z=\pm0.05$, particularly at $\zphotf<1.2$.
Therefore, we generate two photo-$z$ samples consisting of bright sources with $K$-band data and those
that lack it. The first sample requires a minimum of five-band detection at the $3\sigma$ limit (hereafter \fBK)
with one of the bands being $K$. The second sample consists of at least five-band detection (without any $K$
restrictions) at the $5\sigma$ level (hereafter \fBf). Note that only the measurements from broad and
intermediate bands (i.e., narrow bands are excluded) are used to construct these photo-$z$ samples.
These samples contain 22853 and 64691 galaxies (after stellar removal).
Combining the non-stellar \fBf\ and \fBK\ samples yield 65117 galaxies.
Since 15-band detection is not available for all sources (e.g., some are undetected in the \nuv),
we summarize the mixture of detections that we have across 15 bands in Figure~\ref{detections}.
Detections in at least 10 wave bands represent 99\% of 5BK$3\sigma$ sample and 88.7\% of \fBf\ sample.

\begin{figure}[htc] 
  \epsscale{1.10}
  \plotone{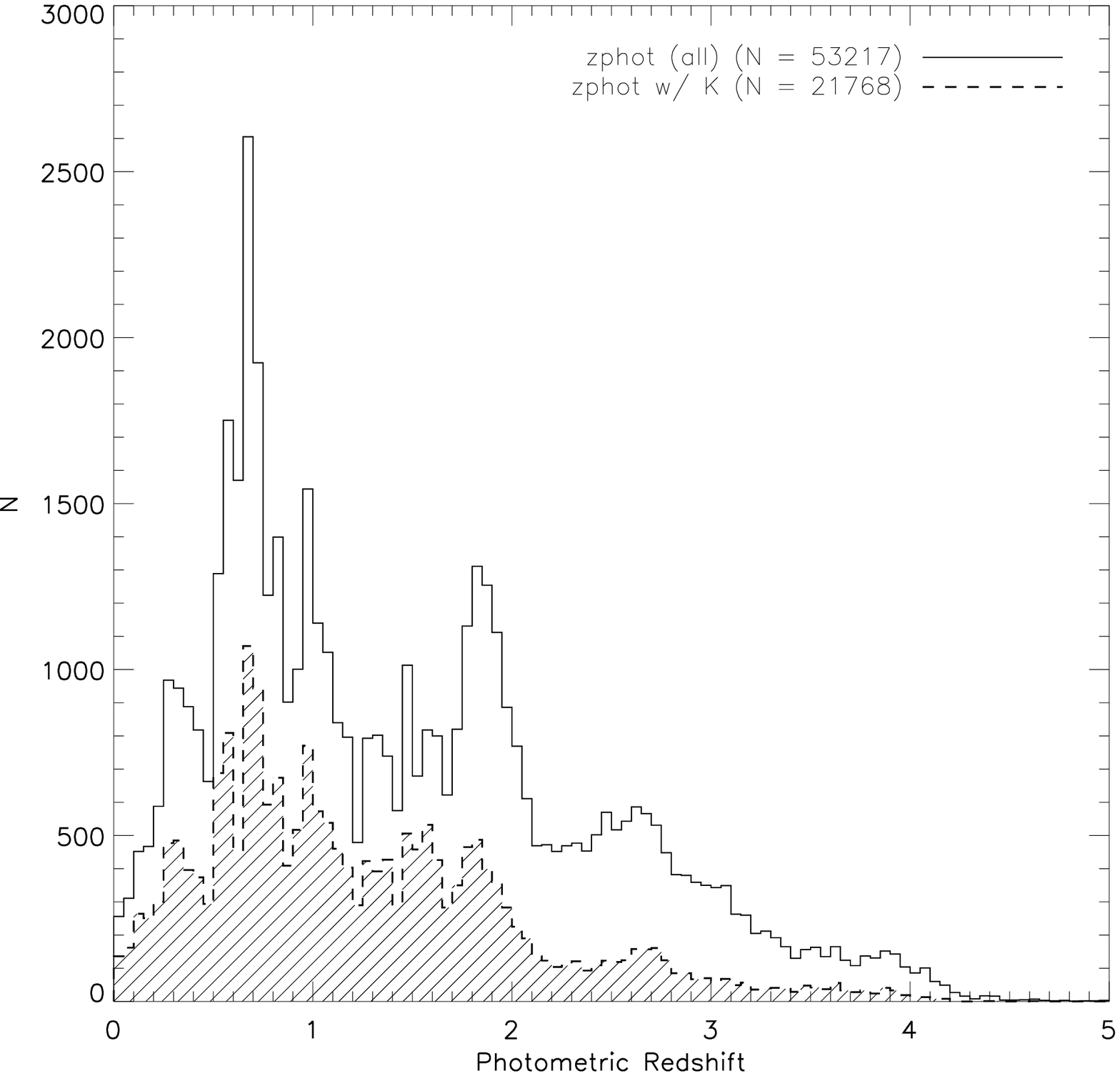}
  \caption{\zphot\ distribution for the combination (solid line) of the \fBK\ and \fBf\ samples
    and only the \fBK\ sample (shaded region). Sources classified as stars were removed
    and objects with \dz $>$ 0.1 were excluded.}
  \label{zphot}
\end{figure}

We exclude photo-$z$ measurements that are unreliable at the 10\% level based on \dz.
This leaves us with a total sample of \deltaall\ galaxies.
Among this accurate photo-$z$ sample, \deltaK\ galaxies are detected at $K$ above 3$\sigma$
while the rest were not.
The \zphot\ distribution for both catalogs is illustrated in Figure~\ref{zphot} where \ncensus\ sources are
identified to be at $z_{\rm phot}=1$--3 while \nlowz\ are located below $z=1$ and \nhighz\ are above $z=3$.
The spike in the distribution at $\zphotf\sim1.8$ is an artifact of poorer sensitivity between
1 \mm\ and 2 \mm\ that would capture the Balmer/4000 \AA\ break, and is less significant since the photo-$z$
accuracy is lower (see Figure~\ref{sigplot}). This feature does not effect census results, since the
spike is not located near either edge of the redshift window.

Because of the complicated selection for the photo-$z$ samples, one concern is whether derived photo-$z$'s
represent the entire population of galaxies. The photo-$z$ samples span the full range of galaxy colors
observed in the SDF. Of course, galaxies of lower mass and luminosity are systematically missed at higher
redshifts, due to our flux sensitivity limits.

\subsection{SED Modeling}\label{3.2}
To determine physical properties of our galaxies, the Fitting and Assessment of Synthetic Templates
\citep[FAST;][]{kriek09} code is used to model the SED. The spectral synthesis models are generated
from the \cite{bc03} code\footnote{\cite{ma05} models were also considered, but results do not
  differ significantly from those using the \cite{bc03} models.}, and consist of a star formation
history that follows an exponential decay (i.e., a $\tau$ model). For simplicity, we choose
$\log{(\tau/{\rm yr})} = $ 8.0, 9.0, and 10.0. These $\tau$ values were selected to include a bursty,
intermediate, and roughly constant star formation history. In addition to the star formation history,
the grid of models spans a range of $\log{({\rm age/yr})}$, between 7.0 and up to the age of the
universe at a given redshift (in increments of 0.1 dex), and dust extinction with $A_V = 0.0$--3.0
mag with 0.1 mag increments. FAST adopts the \cite{calzetti00} dust extinction law, and we only use
the solar metallicity models and assume a \cite{salpeter55} initial mass function (IMF).

The inputs to the SED fitting are the \zphot\ from EAZY, the photometric measurements, and the total
system throughput for each filter (the same as those used in EAZY). The outputs determined from
$\chi^2$ minimization are the stellar masses, stellar ages, dust reddening ($A_V$), star formation rates (SFRs), and
$\tau$ values. Narrow-band filter measurements were excluded in the SED modeling (thus up to 17
bands), since FAST has yet to include emission lines.

In the process of modeling the SEDs, we found a problem with the inclusion of IRAC data for 20\%
of our galaxies. First, the UV extinction-corrected\footnote{We use the derived $\EBV$ from SED
  modeling and assume a \cite{calzetti00} extinction law.} SFRs agree well with those derived
from SED modeling with 15 bands (excluding IRAC bands). However, when the IRAC bands were
included, the same comparison of SFRs yielded a factor of two higher SFRs from SED modeling.
We have not fully understood the cause(s) of this problem, so results using information from
the SED modeling are determined with 15 bands instead of 17 bands.

\section{Photometric Techniques to Identify $z=1$--3 Galaxies}\label{4}
The following two-color methods are used to select galaxies in the redshift desert: BX, BM, BzK
(star-forming and passive), and LBG. The sizes of these galaxy population samples are given in
Table~\ref{table3}. Since the $J$-band data are shallower than the $K$, the SDF DRG sample alone is $\sim$200
in size, so we do not include this in the census. We refer readers to \citet[hereafter L07]{lane07}
where a 0.5 deg$^2$ NIR survey was conducted and a large sample of bright ($K\lesssim21$ AB) DRGs was
obtained and compared against other NIR-selected galaxies.
\defcitealias{lane07}{L07}
\begin{deluxetable*}{llrrcrr}
\tablewidth{0pt}
\tabletypesize{\scriptsize}
\tablecaption{Summary of Photometric Selection Samples}
\tablehead{
  \colhead{Method} &
  \colhead{Section} & 
  \colhead{$N$} &
  \colhead{$N_{z_{\rm phot}}$} &
  \colhead{$\bar{z}_{\rm phot}$} & 
  \colhead{\zphot $ < 1.0$} & 
  \colhead{\zphot $ < 0.5$}
}
%
\startdata
  BX $\Rcf\leq25.5$                    & \ref{4.1}   &\nBX    &\nBXz    & 2.263$\pm$0.39 & 1282  (25.3\%)& \nBXi\ (\nBXif\%)\\
  BX $\Rcf\leq26.0$                    & \ref{4.1}   &\nBXb   &\nBXbz   & 2.240$\pm$0.40 & 1395  (20.3\%)& 1326 (\nBXbif\%)\\
  BM $\Rcf\leq25.5$                    & \ref{4.1}   &\nBM    &\nBMz    & 1.631$\pm$0.33 &  943  (15.5\%)& \nBMi\ (\nBMif\%)\\
  BM $\Rcf\leq26.0$                    & \ref{4.1}   &\nBMb   &\nBMbz   & 1.640$\pm$0.35 & 1112  (14.3\%)&  862 (\nBMbif\%)\\
  sBzK                                 & \ref{4.2}   &\nsBzK  &\nsBzKz  & 1.939$\pm$0.56 &  431  ( 5.5\%)&  179 ( 2.3\%)\\
  pBzK                                 & \ref{4.2}   &\npBzK  &\npBzKz  & 1.663$\pm$0.51 &    1  ( 0.5\%)&    0 ( 0.0\%)\\
  $z\sim3$ LBG $\Rcf\leq25.5$          & \ref{4.3.1} &\nzULBG &\nzULBGz & 2.913$\pm$0.49 &  307  (14.9\%)&  277 (\nUif\%)\\
  $z\sim3$ LBG $\Rcf\leq26.0$          & \ref{4.3.1} &\nzULBGb&\nzULBGbz& 2.862$\pm$0.46 &  348  (10.7\%)&  310 ( 9.5\%)\\
  $z\sim2$ LBG $V\leq25.4$             & \ref{4.3.2} &\nzNLBG &\nzNLBGz & 1.816$\pm$0.47 &  619  ( 8.1\%)&  238 ( 3.1\%)\\
  $z\sim2$ LBG $V\leq26.0$             & \ref{4.3.2} &\nzNLBGb&\nzNLBGbz& 1.823$\pm$0.45 &  690  ( 7.3\%)&  259 ( 2.8\%)\\
  $z\sim1$ LBG                         & \ref{4.3.3} &\nzFLBG &\nzFLBGz & 0.834$\pm$0.15 &\ldots         &\ldots\\
  Total (Shallow)\tablenotemark{a,b}   & \ldots      &\ntot   &\ntotz   &  \ldots        & 3583          & 2707\\ 
  Unique (Shallow)\tablenotemark{a,b,d}& \ldots      & \nUa   &\nUaz    &  \ldots        & 4120          & 2481\\
  Total (Faint)\tablenotemark{a,c}     & \ldots      &\ntotb  &\ntotbz  &  \ldots        & 3977          & 2936\\ 
  Unique (Faint)\tablenotemark{a,c,d}  & \ldots      & \nUb   &\nUbz    &  \ldots        & 4482          & 2688\\[-3mm]
\enddata
\label{table3}
\tablenotetext{1}{The combination of BX, BM, \Udrop\ ($z\sim3$), \sbzk, pBzK, and \nuv-dropout ($z\sim2$) samples.}
\tablenotetext{2}{BXs, BMs, and $U$-dropouts selected with $\Rcf \leq 25.5$, and \nuv-dropouts selected with $V\leq25.4$.}
\tablenotetext{3}{BXs, BMs, and $U$-dropouts selected with $\Rcf \leq 26.0$, and \nuv-dropouts selected with $V\leq26.0$.}
\tablenotetext{4}{``Unique'' refers to the sample which accounts for overlap between the photometric selection samples.}
\end{deluxetable*}


\subsection{BX/BM Selection}\label{4.1}
\begin{figure*} 
  \epsscale{0.4}
  \plotone{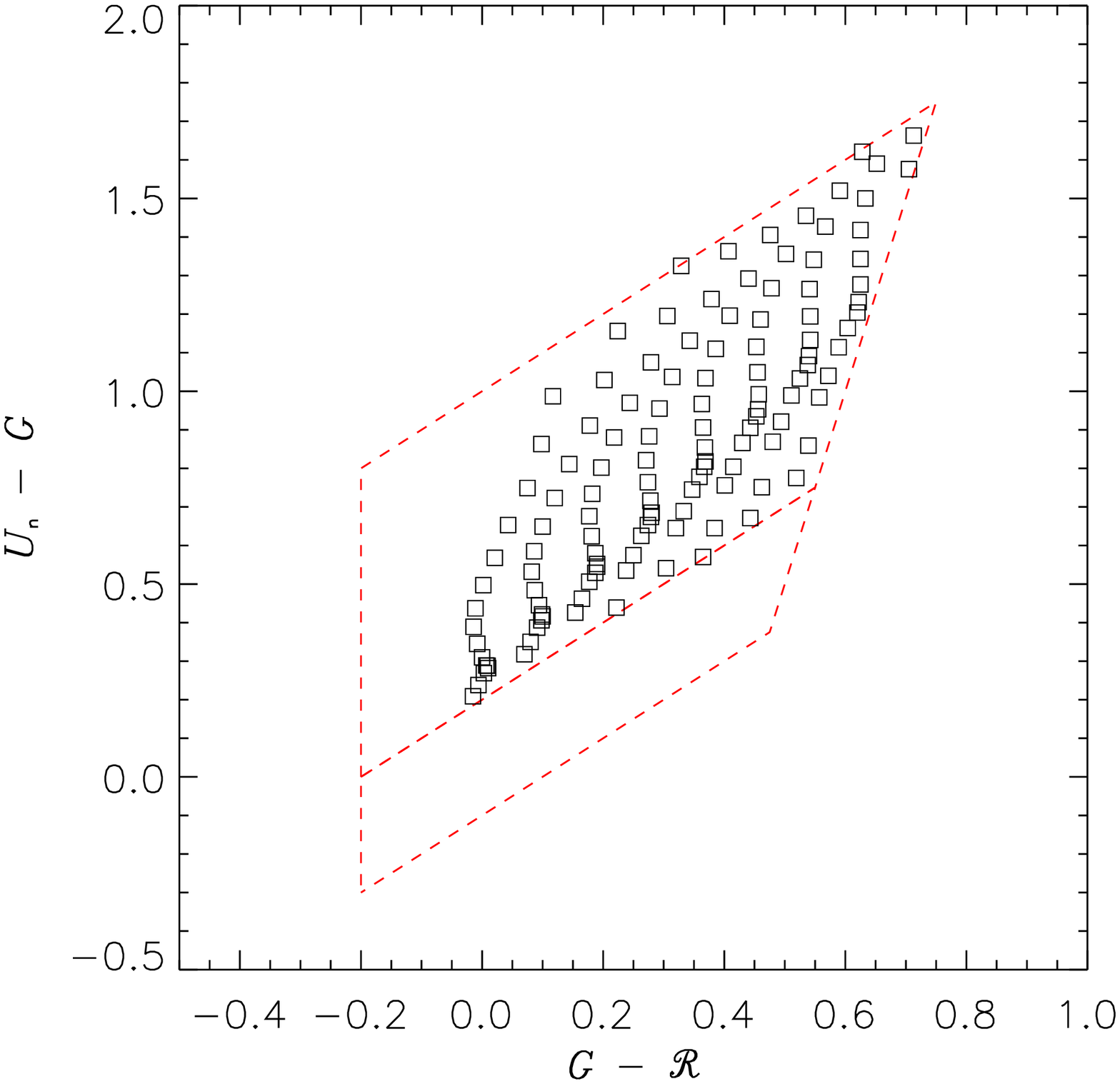} \plotone{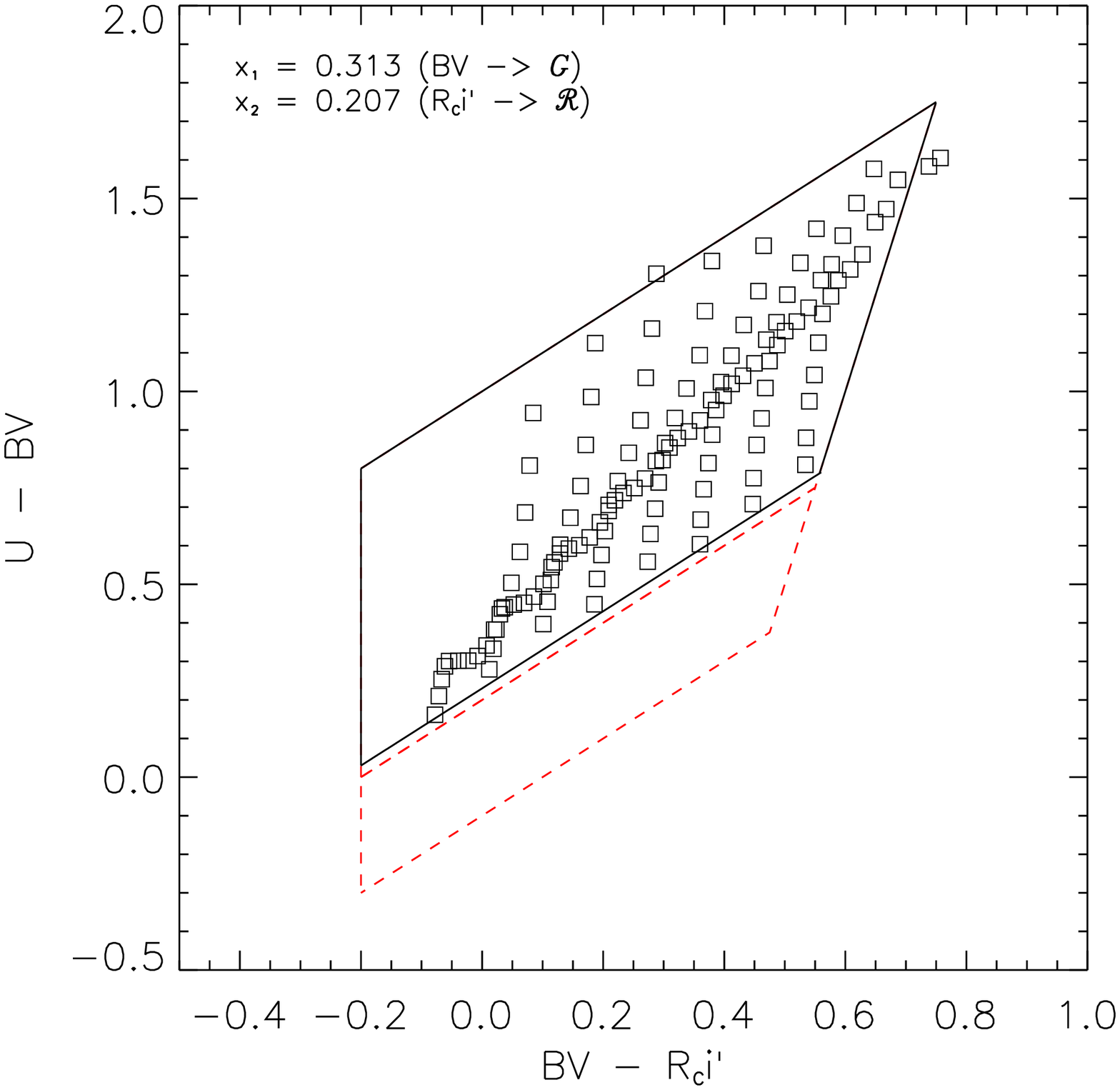}
  \caption{BX transformation for $\Un G\Rs$ to $UBV\Rcf i\arcmin$.
    (Left) The originally defined selection of BX galaxies in the $\Un G\Rs$ space.
    (Right) The $\UBV$ and $\BVRi$ colors for these BX galaxies, as determined by
    Equations~(\ref{eqnBV}) and (\ref{eqnRi}) with $x_1 = 0.314$ and $x_2 = 0.207$.
    The solid black lines refer to the final BX selection (see Equations~(\ref{eqn12})--(\ref{eqn15}))
    while the red dashed lines illustrate the \citetalias{steidel04} selection.
    These plots indicate a proper transformation has been made to select all BX galaxies.
    These transformation is further discussed in Section~\ref{4.1}. \tcolor}
  \label{BXtrans}
\end{figure*}

\citet[hereafter S04]{steidel04} defined BX and BM galaxies by two polygons in the $\Un-G$/$G-\Rs$ color space.
For BX, it is $G-\Rs \geq -0.2$, $\Un-G \geq G-\Rs+0.2$, $G-\Rs \leq 0.2(\Un-G)+0.4$, and $\Un-G \leq G-\Rs+1.0$.
The BM selection consists of $G-\Rs \geq -0.2$, $\Un-G \geq G-\Rs-0.1$, $G-\Rs \leq 0.2(\Un-G)+0.4$, and
$\Un-G \leq G-\Rs+0.2$.\defcitealias{steidel04}{S04}
However, the $\Un G\Rs$ system differs from existing filters of similar wavelengths for the SDF, so a
transformation is necessary between $UBV\Rcf i$\arcmin\ and $\Un G\Rs$. The transformation is derived by
using \cite{bc03} spectral synthesis models of star-forming galaxies.\footnote{This model is identical
  that used in \citetalias{ly09} and is similar to that used by \cite{steidel99}.}
Mock galaxies are produced from a grid of synthetic spectra between $z=0.85$ and $z=3.8$ and reddened
with the \cite{calzetti00} law with $E(B-V)=0.0$--0.4. We also include neutral hydrogen intergalactic medium (IGM) absorption
following \cite{madau95}. Then magnitudes in the $\Un$, $G$, $\Rs$, $U$, $B$, $V$, \Rc,
and $i$\arcmin\ filters are determined for these 540 mock galaxies.\footnote{The $\Un G\Rs$ filter profiles
  were provided by A. E. Shapley.} Among the artificial galaxies that meet the original BX or BM selection,
least-squares fitting is used to represent the $G$ band ($\Rs$ band) with a combination of $B$ and $V$ 
(\Rc\ and $i$\arcmin) such that $\Un-G = \UBV$ and $G-\Rs= \BVRi$, where
\begin{eqnarray}
\label{eqnBV}
BV            &=& -2.5\log{\left[\frac{x_1f_B + (1-x_1)f_V}{3630~\mu{\rm Jy}}\right]},\textrm{~and}\\
\label{eqnRi}
\Rcf i\arcmin &=& -2.5\log{\left[\frac{x_2f_R + (1-x_2)f_{i\arcmin}}{3630~\mu\textrm{Jy}}\right]}.
\end{eqnarray}
Here, $f_X$ is the flux density per unit frequency (erg s$^{-1}$ cm$^{-2}$ Hz$^{-1}$) in band ``\textit{X}.''
An illustration of this technique is shown in Figure~\ref{BXtrans} for BX galaxies.

\begin{figure*} 
  \epsscale{0.4}
  \plotone{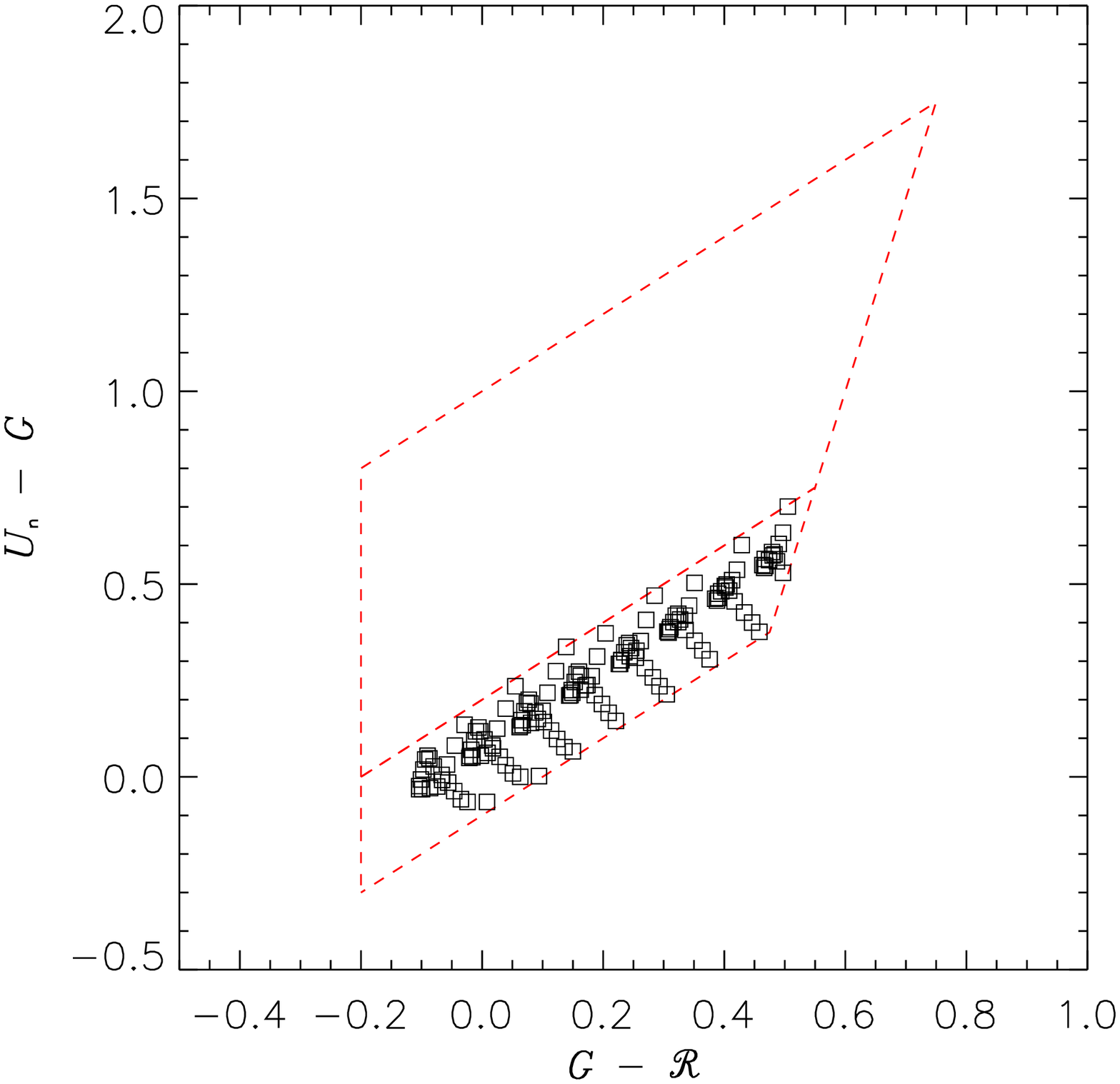} \plotone{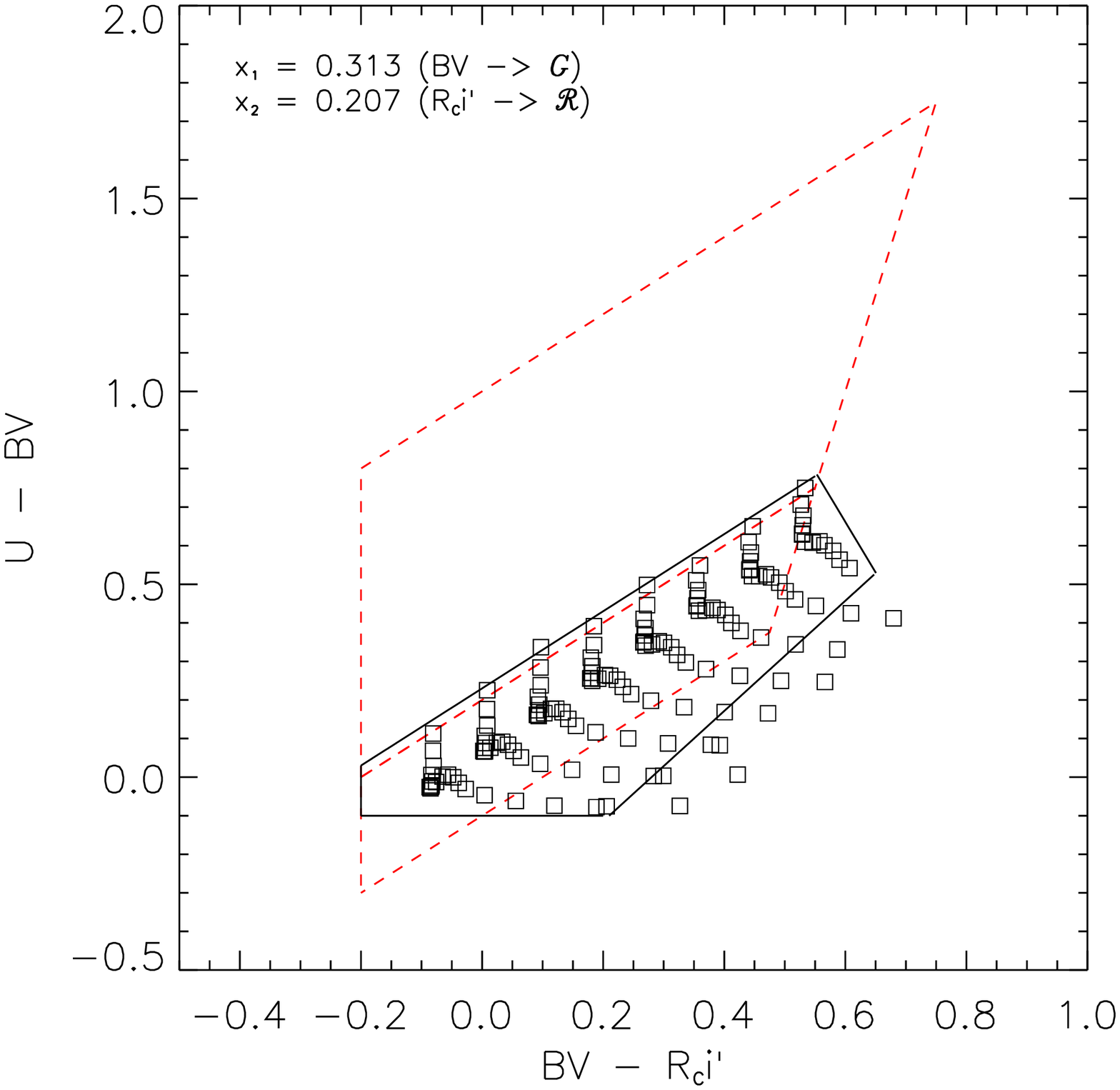}
  \caption{Similar to Figure~\ref{BXtrans}, but illustrated here for BM galaxies. These plots
    indicate a reasonable transformation for BM galaxies when using the BX transformation except for
    $z\sim1$ BM galaxies. We adopt a different selection for BM galaxies, as shown by the solid black
    lines (see Equations~(\ref{eqn16})--(\ref{eqn20})) compared to the \citetalias{steidel04} selection
    illustrated by the red dashed lines. Further discussion of the transformation is available in
    Section~\ref{4.1}. \tcolor}
  \label{BMresid}
\end{figure*}

While different values of $x_1$ and $x_2$ {\it can} be adopted for the BX ($x_1 = 0.314$ and $x_2 = 0.207$)
and BM ($x_1 = 0.491$ and $x_2 = 0.790$) selections,
the net result is that some BX galaxies would also meet the BM selection and vice versa.
However, the BX and BM galaxies occupy distinct regions in the default $\Un G\Rs$ plane, so only {\it one}
transformation set can be used to identify the two galaxy populations simultaneously.
As a compromise, we decided to adopt the BX $x_1$--$x_2$ transformation, since it occupies the region
between the BM and \Udrop\ techniques. With this approach, the BX selection parallelogram remains roughly the
same, while the BM and \Udrop\ selection regions are modified.
The only modification that we make to the BX selection is to move the lower bound up by 0.03 mag in $\UBV$:
\begin{eqnarray}
\label{eqn12}
  \BVRi &\geq& -0.2,\\
\label{eqn13}
  \UBV &\geq& \BVRi + 0.23,\\
\label{eqn14}
  \BVRi &\leq &0.2(\UBV) + 0.4,\textrm{\, and}\\
\label{eqn15}
  \UBV &\leq& \BVRi + 1.0.
\end{eqnarray}
It should be noted that the BM transformation did change significantly when adopting the BX photometry
transformation (see Figure~\ref{BMresid}). It now spans a wider range in $\BVRi$ color, so more $z<1$ sources
could potentially contaminate our BM sample at the cost of identifying $z>1$ BM galaxies. As a compromise, we
choose not to extend the selection region to encompass the lowest-redshift BM galaxies, in order to exclude
some low-$z$ interlopers. This may sound alarming, but it is expected, since the original BM selection spanned
a small range of $G-R$ color for a given $\Un-G$ color. Thus scatter from photometric uncertainties causes
missed faint $z>1$ galaxies and the inadvertent inclusion of low-$z$ interlopers. The final selection that we
use for BM galaxies is
\begin{eqnarray}
\label{eqn16}
  \BVRi &\geq& -0.2,\\
\label{eqn17}
  \UBV &\geq& -0.1,\\
\label{eqn18}
  \BVRi &\leq &0.382(\UBV) + 0.853,\\
\label{eqn19}
  \BVRi &\leq &0.70(\UBV) + 0.280,\textrm{\, and}\\
\label{eqn20}
  \UBV &\geq& \BVRi + 0.23.
\end{eqnarray}

\begin{figure} 
  \epsscale{1.1}
  \plotone{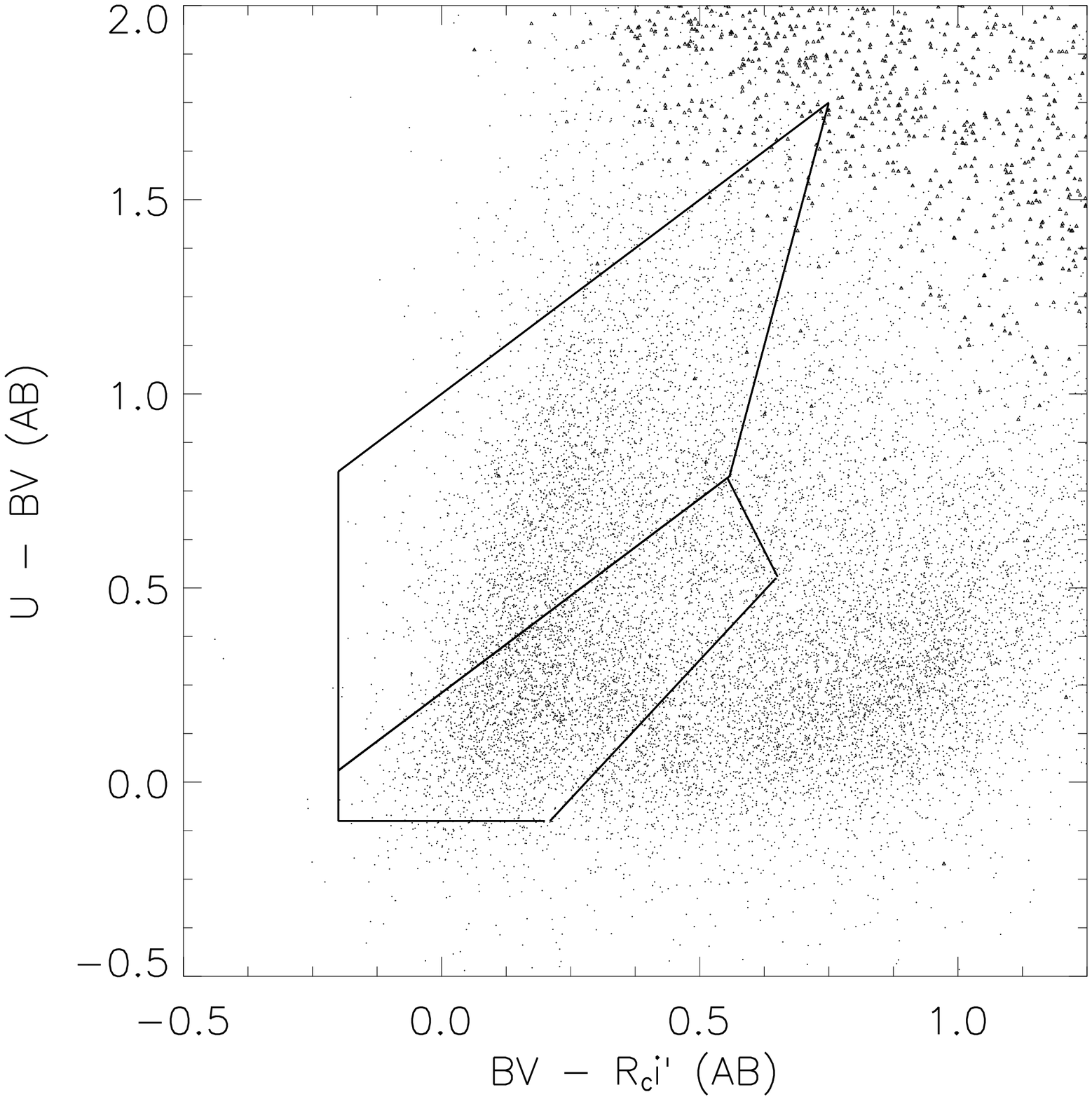}
  \caption{$\UBV$ and $\BVRi$ colors to select BX and BM galaxies. Triangle points are sources that are
    undetected in $U$ at $1\sigma$. Black lines indicate the selection boxes for BX (top) and BM (bottom).
    Selection is performed for \LyRcut\ mag.}
 \label{BXBM_select}
\end{figure}

The $\UBV$ and $\BVRi$ colors for SDF galaxies are shown in Figure~\ref{BXBM_select} for the selection
of BX and BM galaxies. Following \cite{steidel03}, we defined a non-detection in the $U$ band when its
flux falls below the 1$\sigma$ limit.

\begin{figure} 
  \epsscale{1.1}
  \plotone{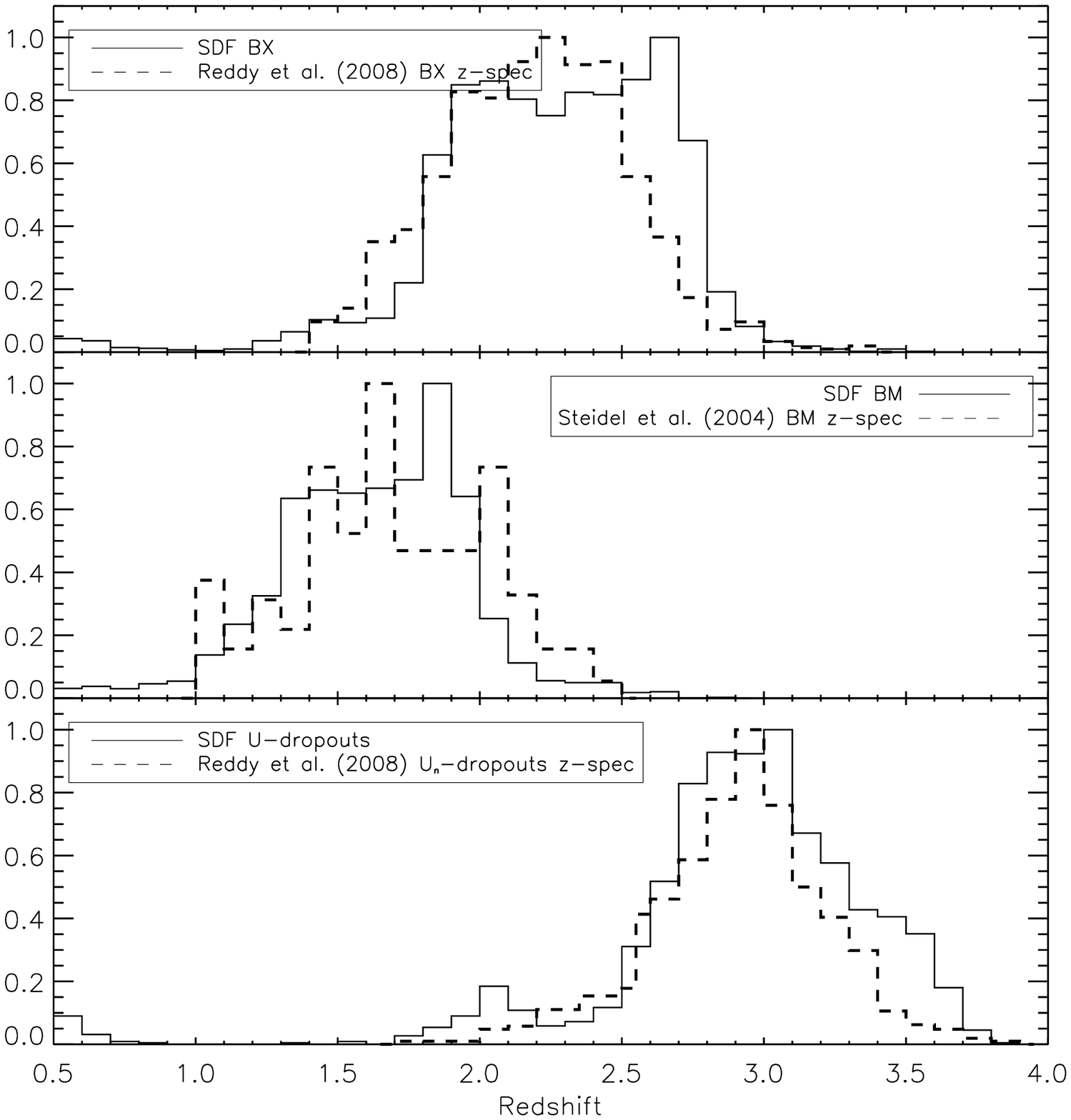}
  \caption{Distribution of \zphot's for the
    UV-selected BX, BM, and \Udrop\ (solid lines) compared to the spectroscopic samples of \citetalias{steidel04} or
    \cite{reddy08}. Galaxies were selected down to $\Rcf = \LyRcutn$ mag.}
  \label{BXBMLBG_zdist}
\end{figure}

\begin{figure*} 
  \epsscale{0.45}
  \plotone{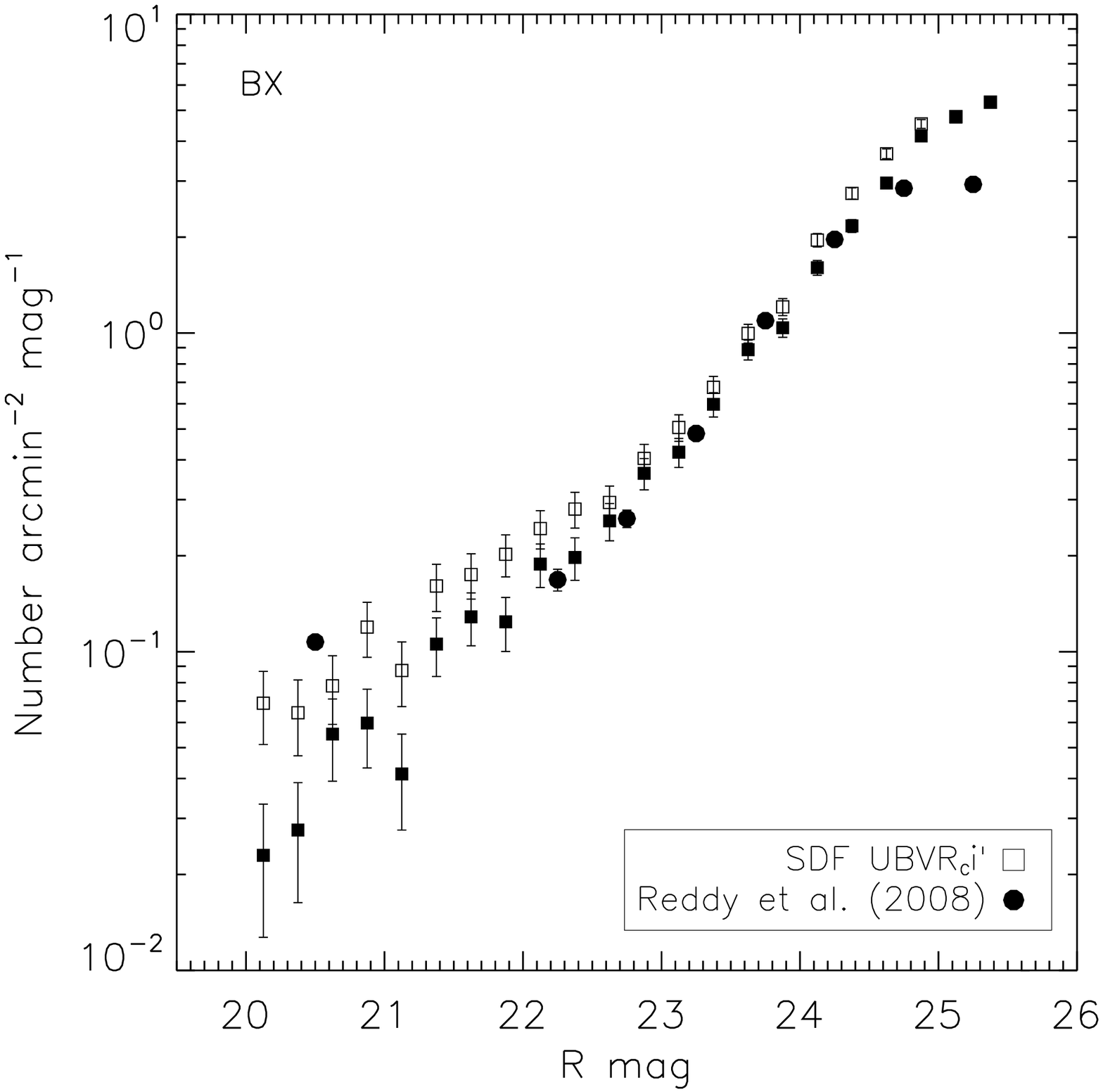} \plotone{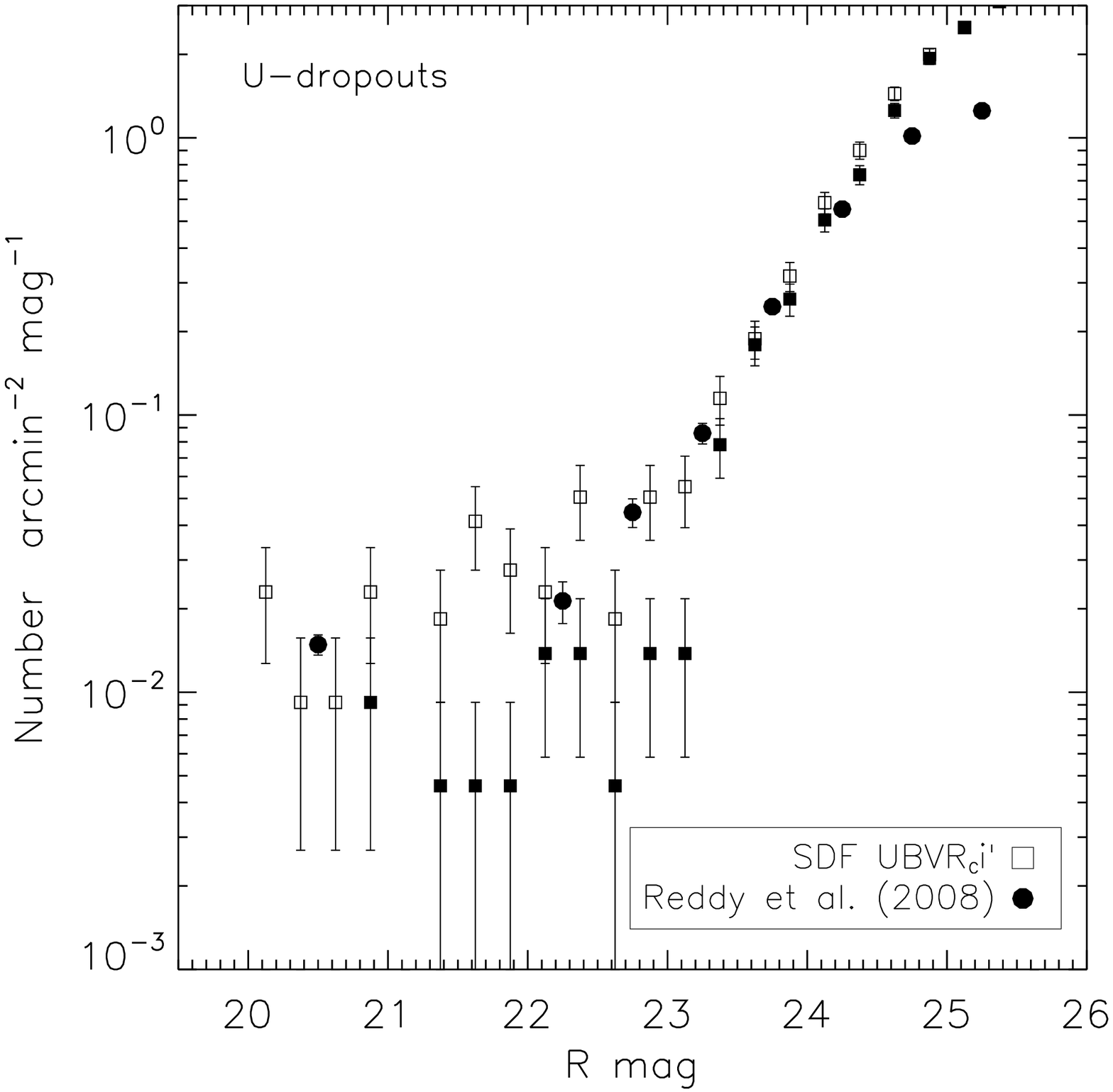}
  \caption{Observed surface density of BX galaxies (left) and \Udrops\ (right) shown as filled
    (exclude stars) and open (include stars) squares. There is good agreement with the filled circles from
    \cite{reddy08}.}
  \label{BXLBG_SD}
\end{figure*}

In total, \nBXs\ BX (\nBX\ after stellar removal) and \nBMs\ BM (\nBM\ after stellar removal) galaxies are
identified with \LyRcut. In Section~\ref{5.2}, we also discuss the BX and BM samples with $\Rcf \leq 26.0$,
which consist of \nBXsb\ BX (\nBXb\ after stellar removal) and \nBMsb\ BM galaxies
(\nBMb\ after stellar removal).
In Figure~\ref{BXBMLBG_zdist}, the photometric redshift distributions of the BX and BM samples are illustrated.
Also overlaid are the spectroscopic redshift distributions from \citetalias{steidel04} and \cite{reddy08}, which
demonstrates the accuracy of EAZY at $z\sim2$ and shows that the filter transformations did indeed approximate
the original BX and BM definitions.
Further evidence that the transformation was done correctly is the surface density of BXs, which is
shown in Figure~\ref{BXLBG_SD}. It is compared with measurements reported by \cite{reddy08} and shows good
agreement over a range of $\sim$3 mag. 

Although not illustrated in Figure~\ref{BXBMLBG_zdist}, a subset of the BX and BM samples have $\zphotf=0.1$--0.5.
The BX (BM) \LyRcut\ sample has \nBXi\ (\nBMi) with $\zphotf\leq0.5$ corresponding to \nBXif\% (\nBMif\%).
These numbers decrease to \nBXbif\% (\nBMbif\%) when considering the sample with $\Rcf\leq26.0$. These galaxies
have accurate \zphot's (\dz\ $<$ 4\%), so it cannot be argued that these galaxies are at $z>1$. Furthermore, we
inspected the $B-z\arcmin$ and $z\arcmin-K$ colors and find that these objects occupy a region below the
star-forming BzK selection (see Section~\ref{4.2}), which provides further evidence that they are at
$z\lesssim1.3$ (i.e., the Balmer/4000 \AA\ break is observed in the optical). The full \zspec\ distributions for
BX and BM galaxies were not provided in \citetalias{steidel04}, but the $\zspecf<1.0$ interloper statistics were
stated to be 17.1\% and 5.8\% for BX and BM ($\mathcal{R}\leq\LyRcutn$) galaxies, respectively.
\citetalias{grazian07} found similar results to ours for their BX sample.

\subsection{BzK Selection}\label{4.2}
\begin{figure} 
  \epsscale{1.1}
  \plotone{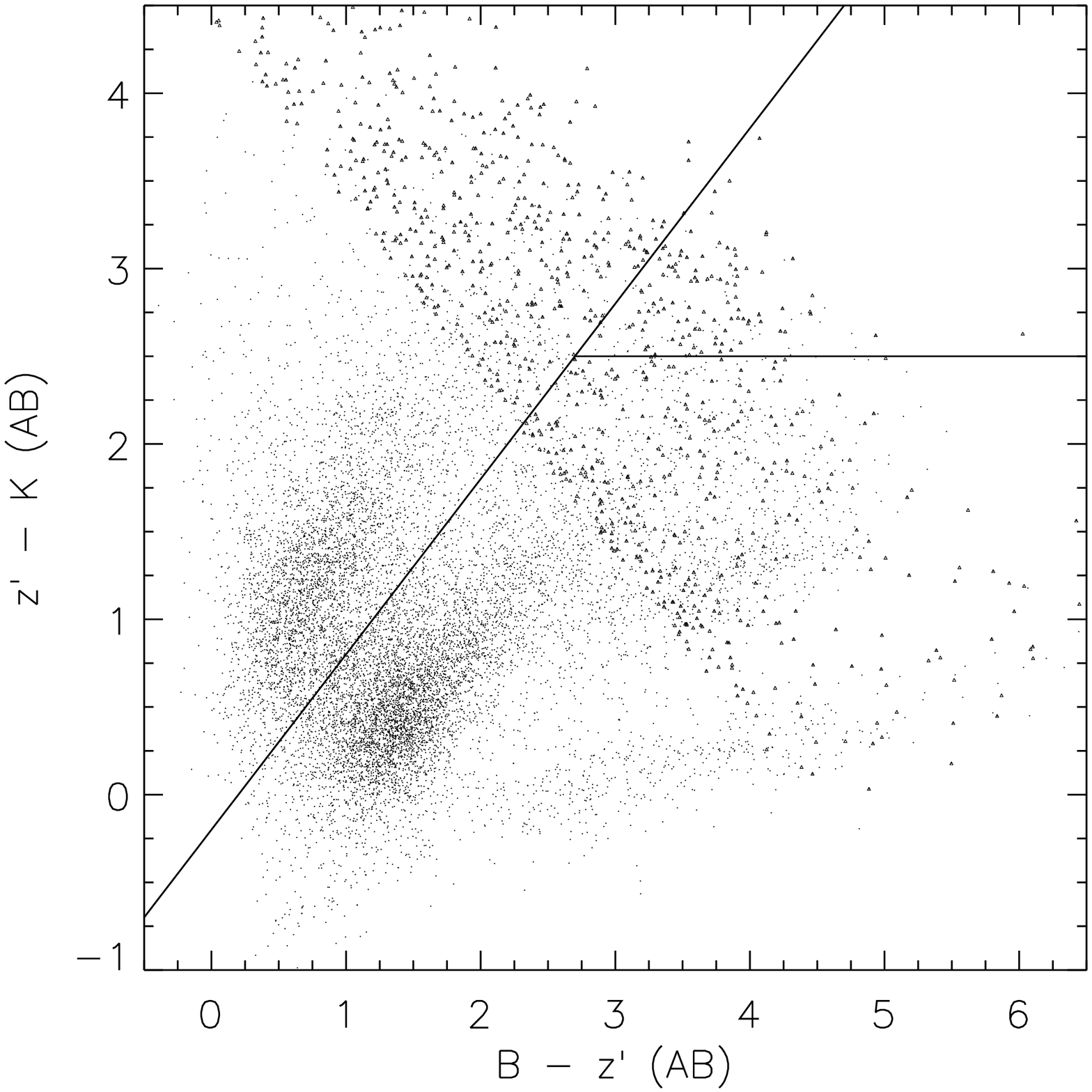}
  \caption{$B-z\arcmin$ and $z\arcmin-K$ colors for $\sim$25,000 sources detected in the
    $K$ band above $3\sigma$. Star-forming BzKs are identified as those above the slanted line (BzK = $-0.2$)
    while the selection of passive BzKs is through a combination of the horizontal and the slanted lines.
    Triangle points indicate $K$-band sources that are undetected in the $B$ band at $3\sigma$
    (lower limit on the $B-z\arcmin$ color).}
  \label{BzKselect}
\end{figure}
\begin{figure*} 
  \epsscale{0.4}
  \plotone{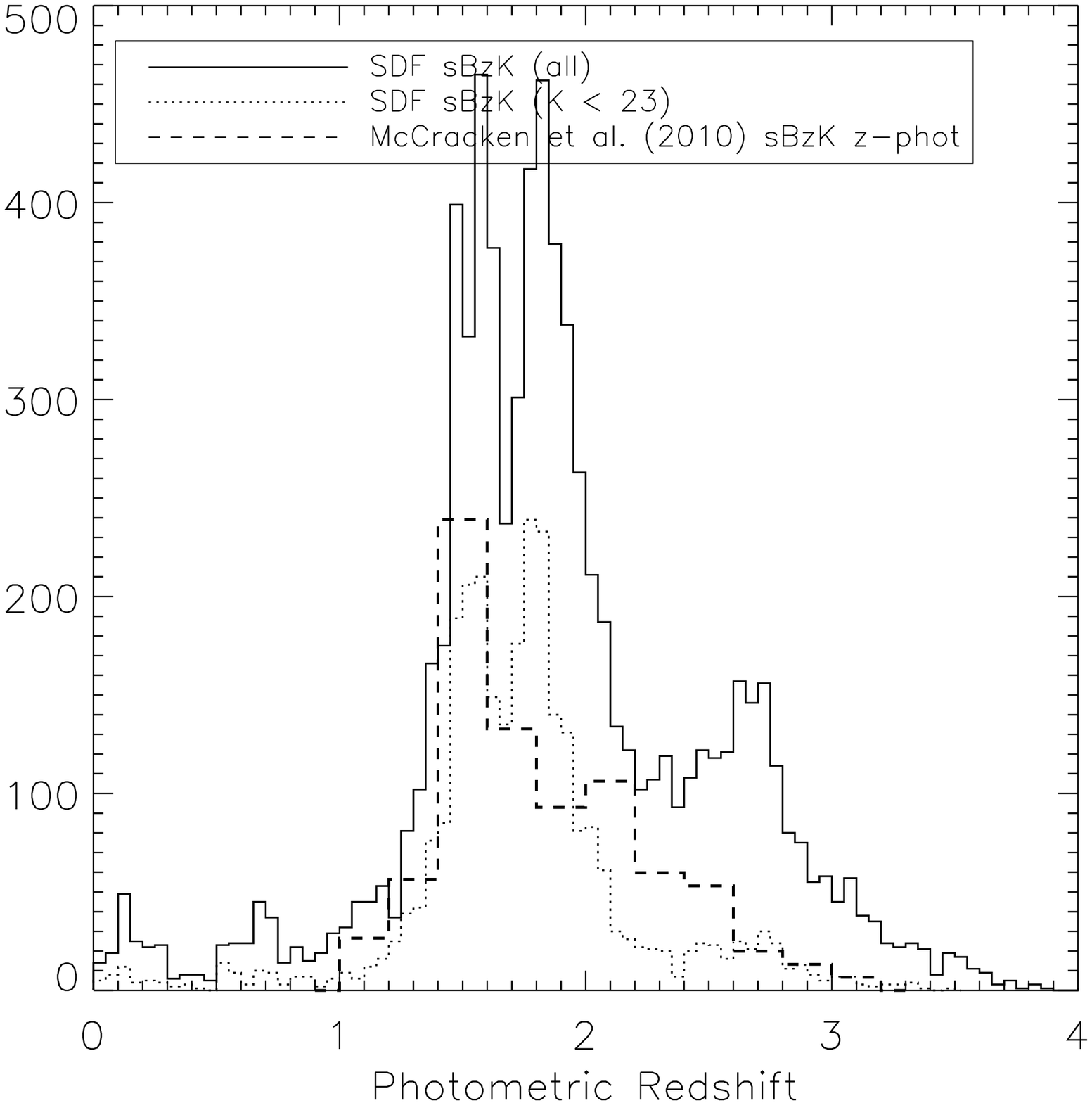} \plotone{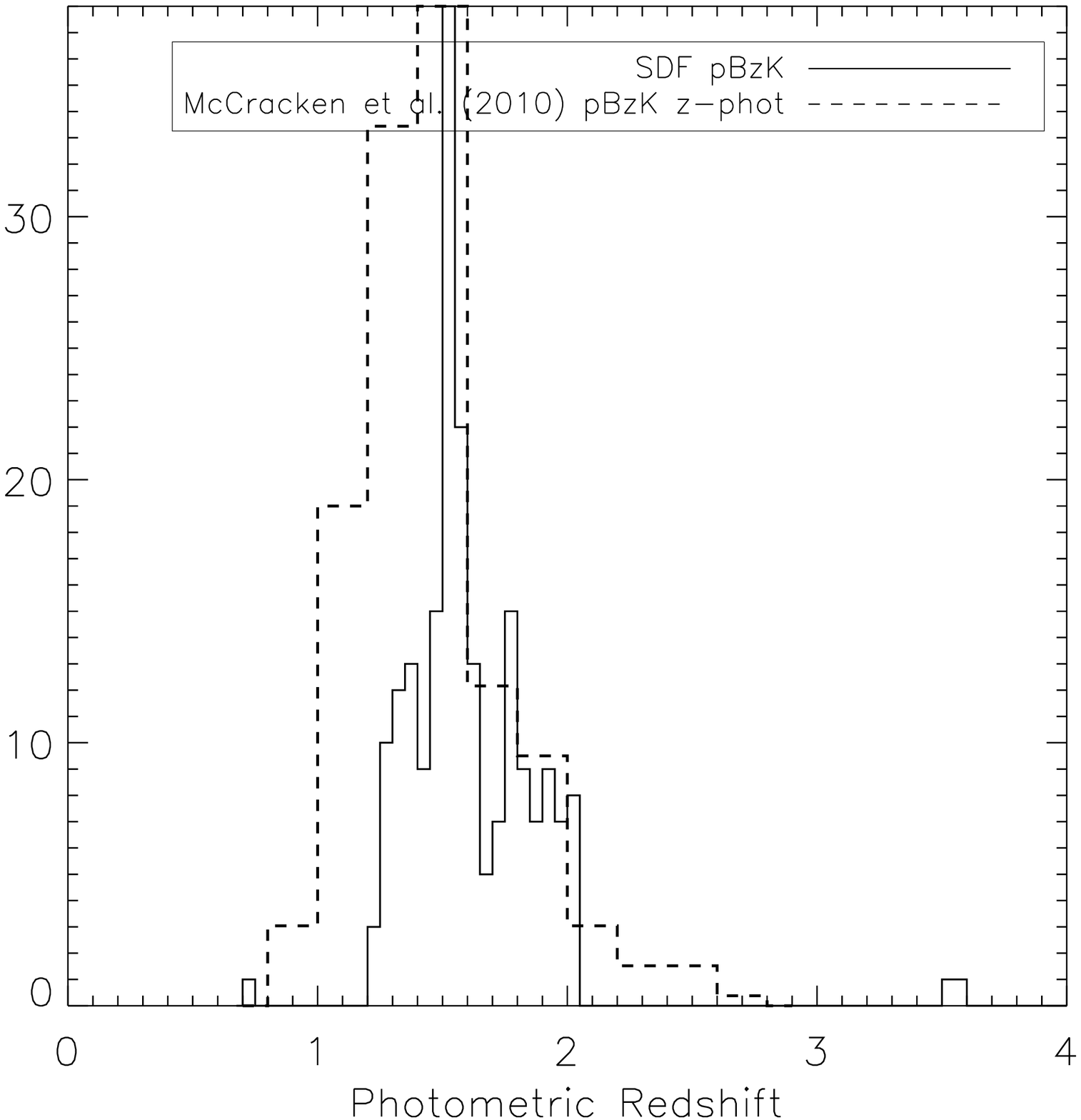}
  \caption{\zphot\ distribution for \nsBzKz\ of \nsBzKs\ \sbzk\ galaxies (left) and \npBzKz\ of
    \npBzKs\ pBzK galaxies (right) indicated by the solid lines. The dotted line represents sBzK galaxies
    with $K<23$ mag to compare with the photo-$z$ distribution of \sbzk\ galaxies from the COSMOS survey
    \citep{mccracken10}, which are overlaid as dashed lines. The COSMOS distributions have been normalized to
    the peak in the SDF $K<23$ mag \zphot\ distributions.}
  \label{BzKphotoz}
\end{figure*}

\cite{daddi04} defined star-forming BzK (\sbzk) galaxies as those with $BzK \equiv (z-K)-(B-z) \ge -0.2$ with
detection required in $B$ and $K$. This corresponds to the solid line in Figure~\ref{BzKselect}, which has
\nsBzKs\ star-forming BzK (hereafter \sbzk) galaxies falling above it. In addition, the passive BzK (pBzK)
selection, which requires $BzK < -0.2$ and $z-K \ge 2.5$, yields \npBzKs\ galaxies (\npBzK\ after stellar
removal) with $z=1$--1.5. The photo-$z$'s of the pBzK and \sbzk\ populations are shown in Figure~\ref{BzKphotoz}.
They show good agreement with photo-$z$'s derived for BzK galaxies in the COSMOS field \citep{mccracken10}.
The COSMOS BzK sample probed $K<23.0$ while our sample reaches $K\sim24.0$. We find better agreement with
COSMOS at the $\zphotf = 1.7$--3.0 range when limiting our sample to $K<23.0$. The COSMOS and SDF samples of
BzK galaxies independently confirm that \sbzk\ galaxies are at $z\approx1.5$--2.5 and pBzK galaxies are at
$z\approx1.0$--1.5.

\begin{figure} 
  \epsscale{1.1}
  \plotone{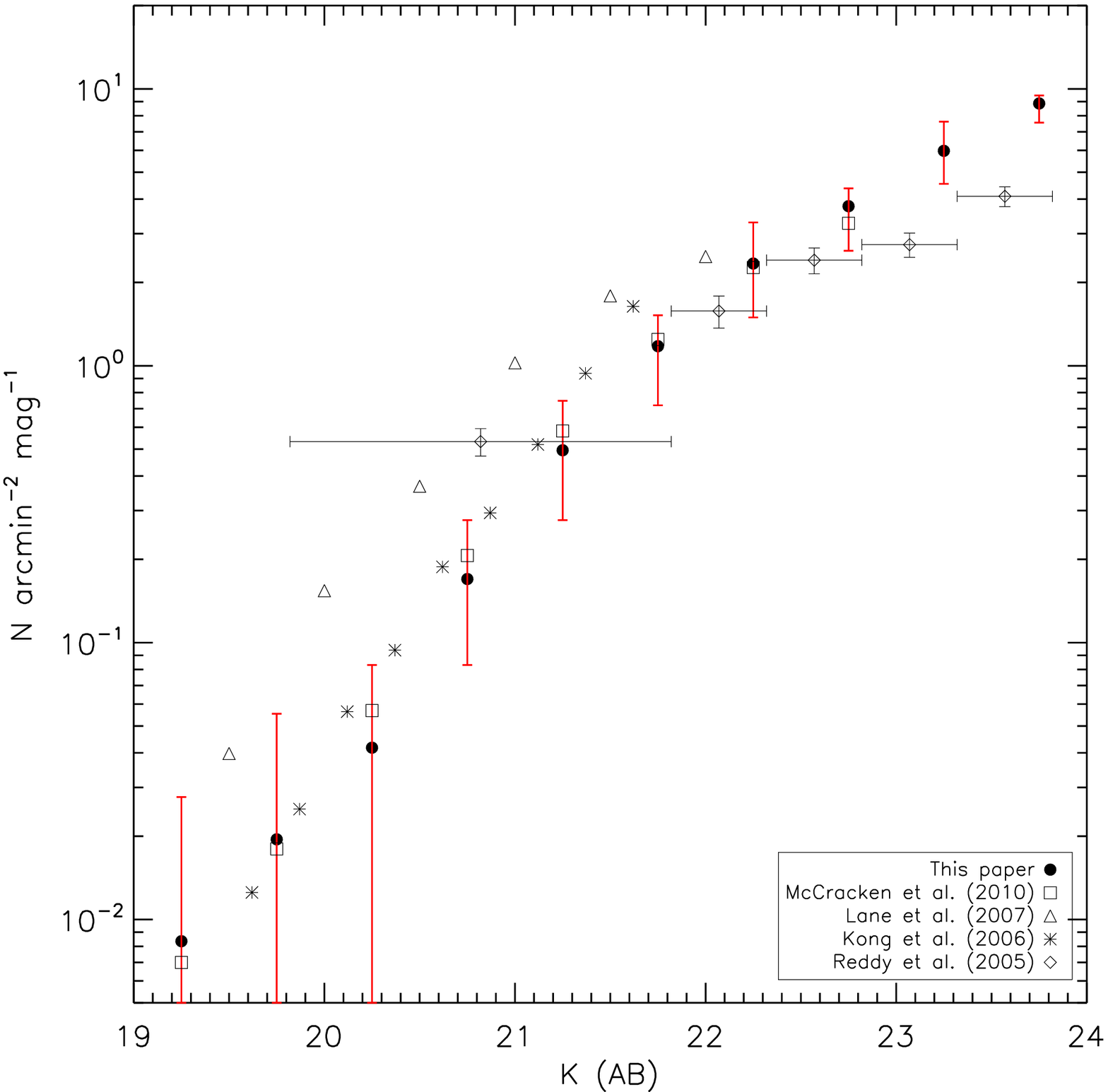}
  \caption{Surface density of \sbzk\ galaxies in the SDF (filled circles) compared to other surveys:
    \citet[squares]{mccracken10}, \citetalias{lane07} (triangles), \citet[asterisks]{kong06}, and
    \citetalias{reddy05} (diamonds). Red error bars illustrate the minimum and maximum surface density for eight
    independent 8\farcm5$\times$8\farcm5 fields in the SDF. \tcolor}
  \label{sbzk_sd}
\end{figure}

In Figure~\ref{sbzk_sd}, we plot the surface density of the SDF \sbzk\ galaxies and previous \sbzk\ surveys
\citep{reddy05,kong06,lane07,mccracken10}. We find that our measured surface density for \sbzk\ is
consistent with those reported in \cite{mccracken10} down to their $K$-band depth of $\sim$23.0 AB.
\cite{kong06} show a higher surface density but is consistent with the photometric scatter.
\citetalias{lane07} also find a higher surface density, although \cite{mccracken10} pointed out that this is
likely due to a photometric calibration issue. Finally, the surface density of \citetalias{reddy05}'s sBzKs\
is consistent with ours for $K<23$ AB, and then is lower than ours by a factor of at most two. We will
discuss this discrepancy at faint $K$ in Section~\ref{5.3}.

\subsection{LBG Selection}\label{4.3}
The above techniques were designed to select $z=1$--3 galaxies when the Lyman break technique is not
detectable with optical photometry. This has now changed since \citetalias{ly09} used ultraviolet imaging for 
the first survey of $z\sim2$ LBGs in the SDF. To fully span the $z=1$--3 era with the Lyman break technique,
we select $U$-, \nuv-, and \fuv-dropouts. We will interchangeably use $U$/\nuv/\fuv-dropouts and $z\sim3$/2/1
LBGs to refer to the Lyman-limit break selected sources.

\subsubsection{\Udrops}\label{4.3.1}
\begin{figure} 
  \epsscale{1.1}
  \plotone{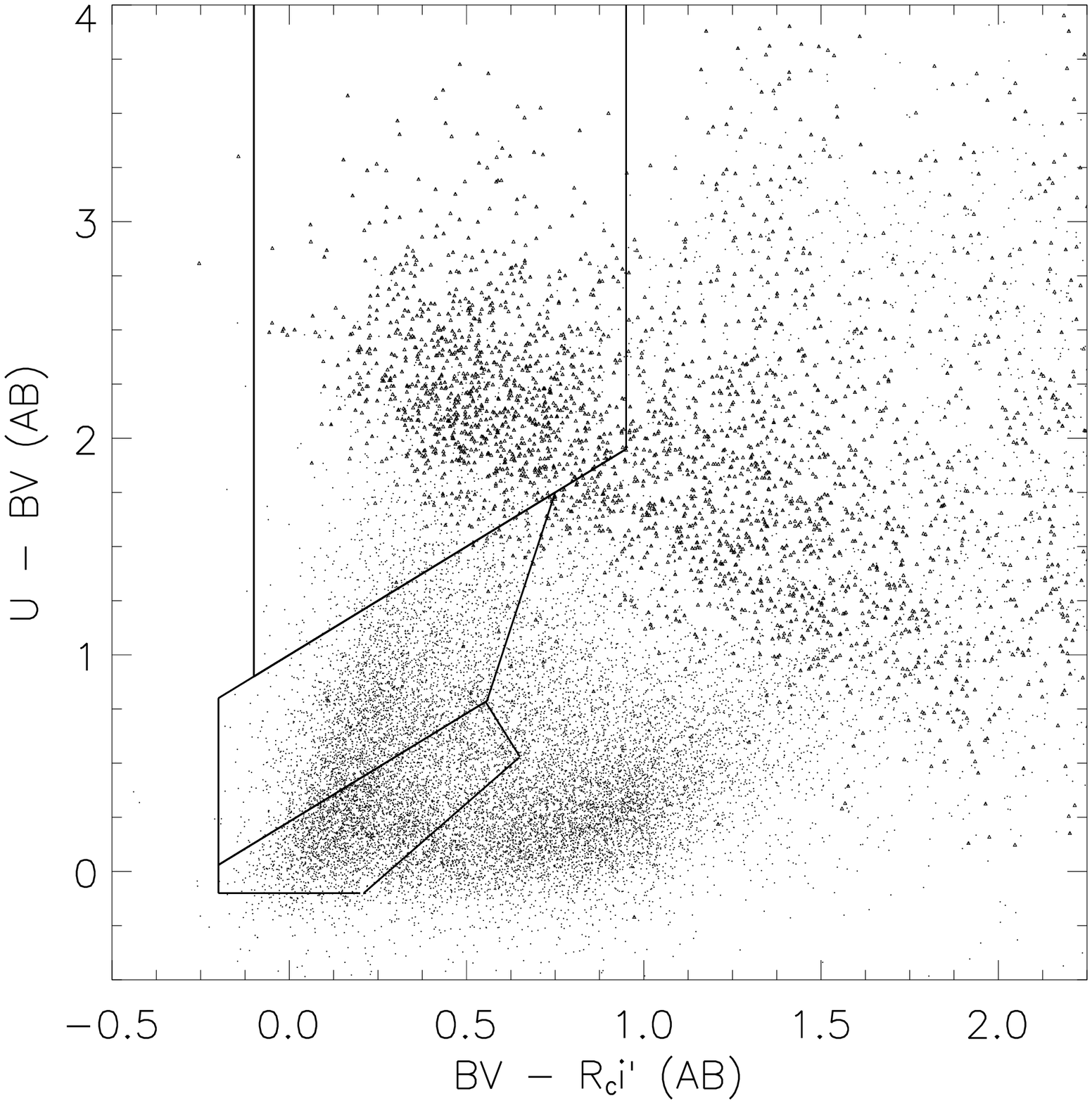}
  \caption{$\UBV$ and $\BVRi$ colors to select \Udrop\ galaxies. Triangle points are sources that are
    undetected in $U$ at $1\sigma$. The selection boxes for BXs, BMs, and \Udrops\ are shown.}
  \label{Udrop_select}
\end{figure}

Using the $UBV\Rcf i\arcmin$ to $\Un G\Rs$ transformation described in Section~\ref{4.1},
we identify \Udrops\ in a similar manner as \cite{steidel03}: $\UBV\geq-0.1$,
$\BVRi \leq 0.95$, and $\UBV\geq\BVRi+1.0$.
A modification is made to the right edge by shifting it 0.15 mag bluer in $\BVRi$.
This selection spans the region above the BX selection window, as illustrated in Figure~\ref{Udrop_select}.
We find \nzULBGs\ \Udrops\ with \LyRcut\ (\nzULBG\ after stellar removal). The surface
density for our \Udrops\ (see Figure~\ref{BXLBG_SD}) is similar to those reported by \cite{reddy08}.
Finally, we compare the \zphot\ for \Udrops\ against the \cite{reddy08} sample in Figure~\ref{BXBMLBG_zdist}
and find good agreement, which further supports the idea that our selection is probing the same galaxies
identified by Steidel et al.
  
\subsubsection{\nuv-dropouts}\label{4.3.2}
The selection of \nuv-dropouts is discussed in \citetalias{ly09}. The definition is: $NUV-B \geq 1.75$,
$B-V \leq 0.50$, and $NUV-B \geq 2.4(B-V) + 1.15$.
In defining the selection criteria, we used a spectral synthesis model that is similar
to that used for selecting \Udrops. The range of dust reddening probed by the above color
criteria, $\EBV=0.0$--0.4, is fairly similar to those selected by Steidel et al. should
galaxies that span this full range exists. Thus, the main difference between the two
populations of LBGs is the redshift sensitivity. This statement, of course, assumes that
there is no evolution in the properties (e.g., dust) of LBGs between $z\sim3$ and $z\sim2$.

There are two corrections that we make to the previous selection. First, a source is considered
undetected in the \nuv\ when it is below the 1$\sigma$ threshold. Previously, we adopted $3\sigma$
to be conservative; however, we decided to be consistent with \cite{steidel03} for the final
selection of all Lyman-limit galaxies. This will shift points in Figure 9 of \citetalias{ly09} by
1.2 mag redder in $NUV-B$. Second, we failed to apply an aperture correction for undetected
sources in \citetalias{ly09}. The aperture correction that we adopted previously was 1.81, and our
latest calculation, which we now use, indicates 1.85, which leaves a 0.6 mag offset in the
$NUV-B$ colors for faint galaxies.

By adopting a $1\sigma$ threshold in the NUV, the $z\sim2$ UV luminosity function discussed
in \citetalias{ly09} changes. To first order, the completeness corrections were underestimated at
$z=2$--2.5, since some of the simulated galaxies that fell below our $NUV-B=1.75$ mag will now
enter the $z\sim2$ LBG selection region. However, by adopting a lower threshold, more sources
will also enter our selection region that were bluer than $NUV-B=1.75$ mag. We note that the
comparison between BX/BM and $z\sim2$ LBG is more robust in this paper than in \citetalias{ly09}
since a direct comparison is made using the same data.

\begin{figure*} 
  \epsscale{0.45}
  \plotone{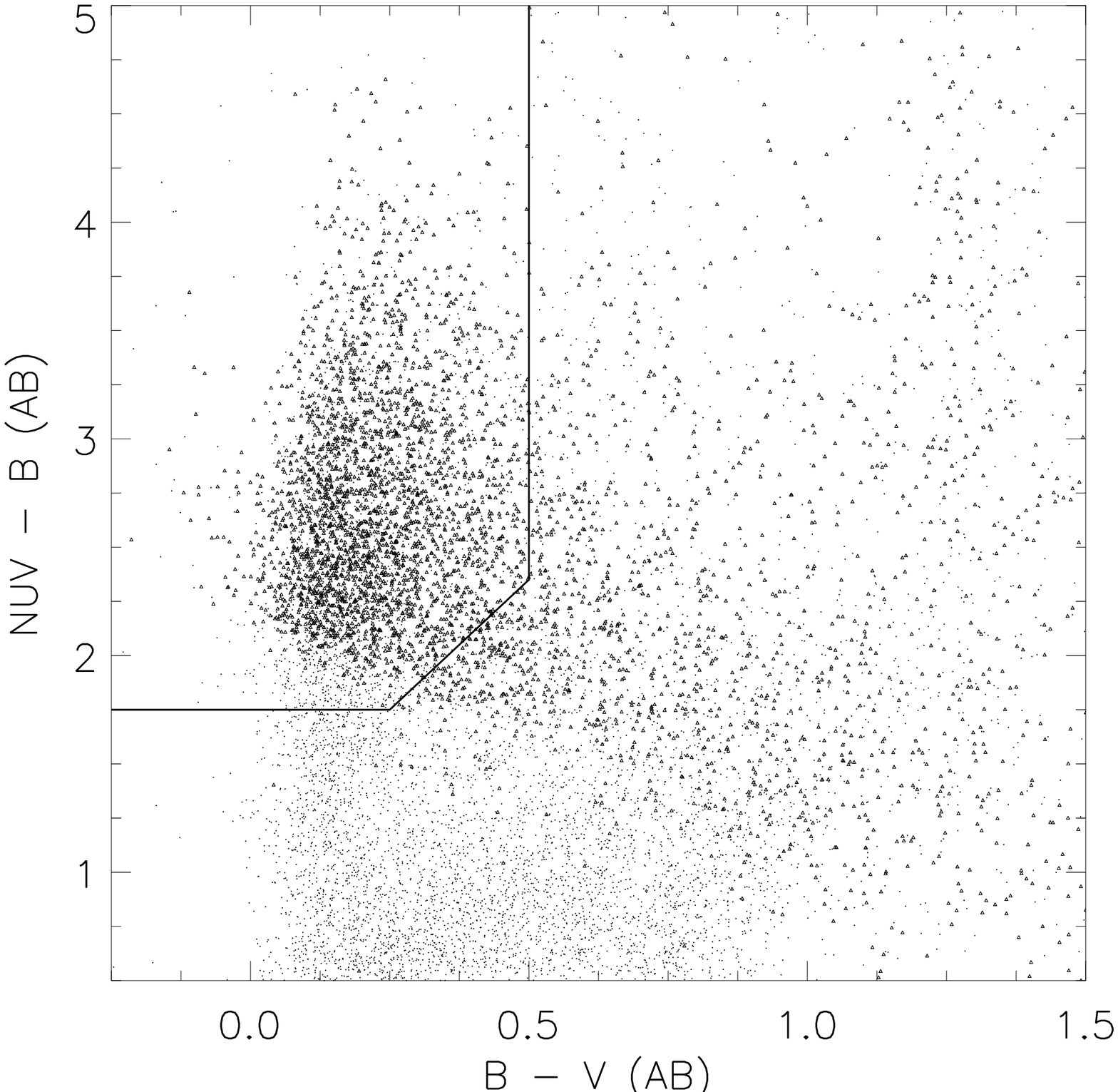} \plotone{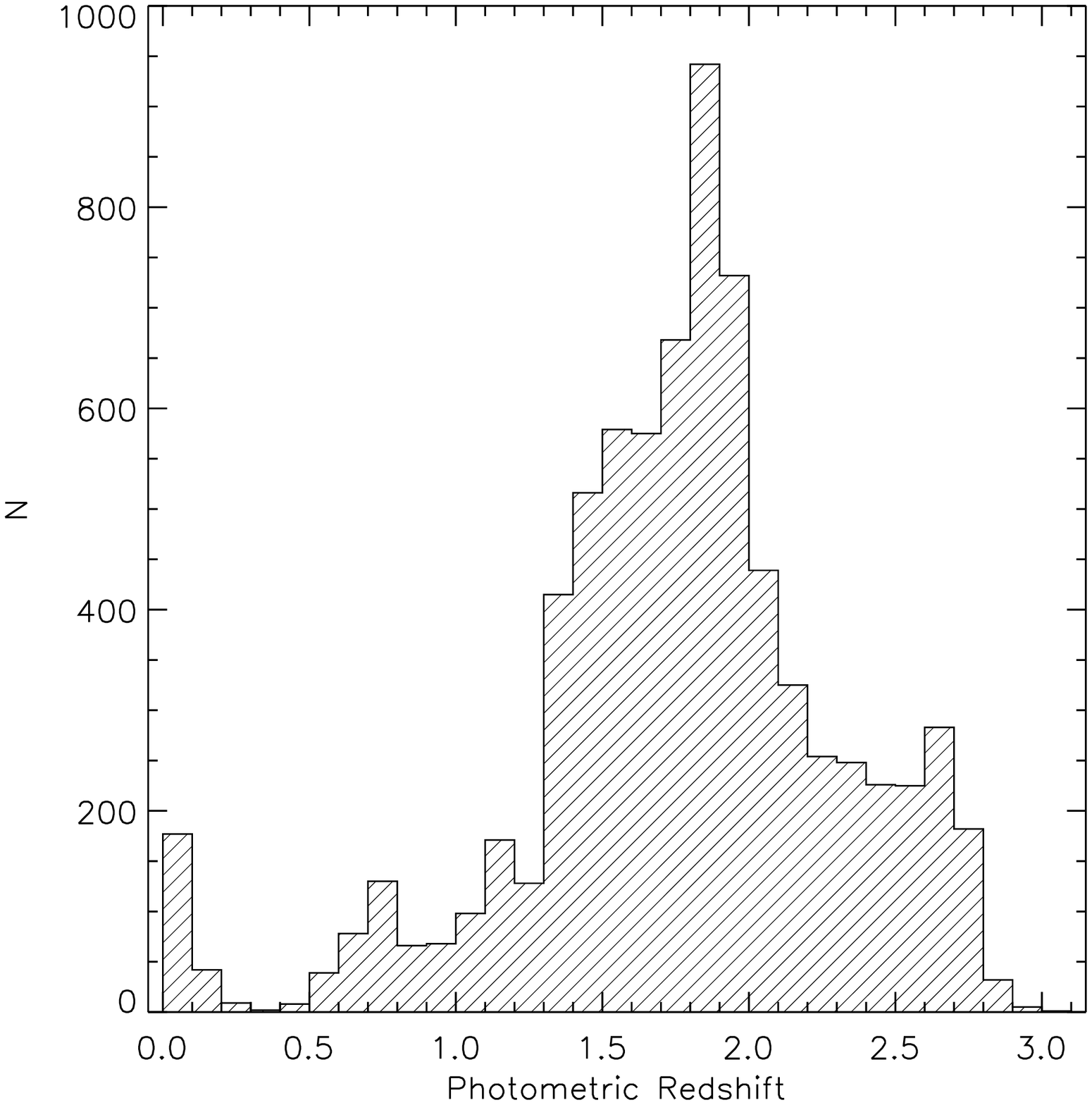}
  \caption{(Left) The selection of \nuv-dropouts with $NUV-B$ and $B-V$ colors with $V\leq25.4$.
    Triangle points indicate sources with $1\sigma$ limits in the \nuv\ band.
    (Right) The \zphot\ distribution for most \nuv-dropouts.}
  \label{NUVdrop}
\end{figure*}

The $NUV-B$ and $B-V$ colors for $V<25.4$ sources are shown in Figure~\ref{NUVdrop}, and
\nzNLBGs\ \nuv-dropouts (\nzNLBG\ after stellar removal) are identified.  Recall that the
\nuv\ fluxes were obtained by PSF-fitting bright sources in order to obtain reliable
fluxes for faint, confused sources. If this step was not performed, the \nuv-dropout sample
would be $\approx$30\% smaller. This is similar to the fraction of missed mock LBG galaxies
in the Monte Carlo simulation of \citetalias{ly09}. In Section~\ref{5.2}, we also discuss the
\nuv-dropout sample with $V\leq26.0$, which consists of \nzNLBGsb\ sources (\nzNLBGb\ after
stellar removal). The \zphot\ distribution is shown in Figure~\ref{NUVdrop}, and indicates
that the selection mostly identifies galaxies at $z=1.5$--2.5, with \nNLBGi\ contamination
from $z<1.5$ interlopers.

\subsubsection{\fuv-dropouts}\label{4.3.3}
\begin{figure*} 
  \epsscale{0.45}
  \plotone{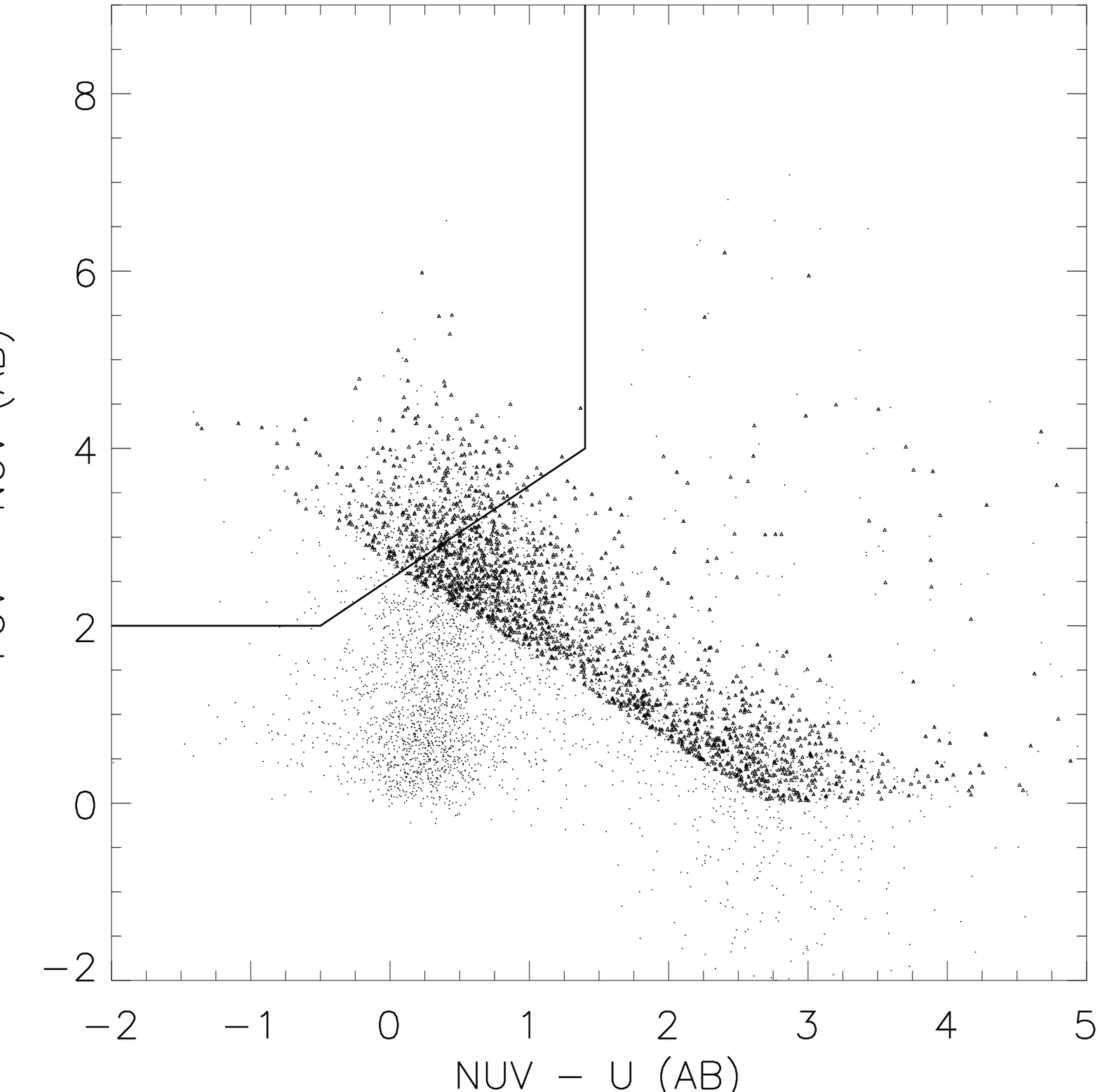} \plotone{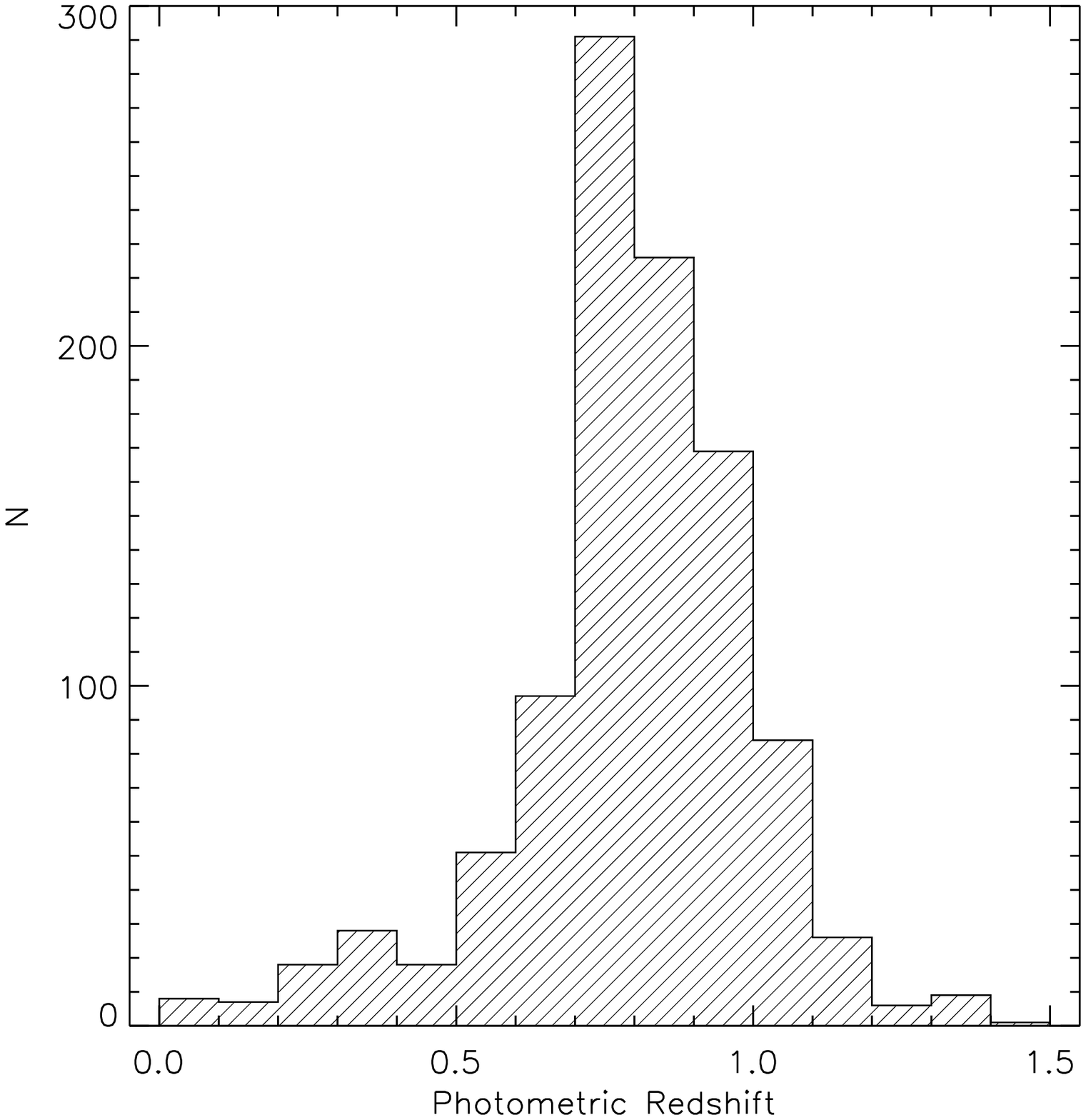}
  \caption{(Left) The selection of \fuv-dropouts with $FUV-NUV$ and $NUV-U$ colors. Triangle points
    indicate sources with $1\sigma$ limits in the \fuv\ band. (Right) The \zphot\ distribution
    for almost all \fuv-dropouts.}
  \label{FUVdrop}
\end{figure*}

Studies \citep[e.g.,][]{burgarella07} have used \galex\ \fuv\ imaging to select $z\sim1$ LBGs.
The selection of \fuv-dropouts is currently best described in conference proceedings. The
selection which has been adopted is $FUV-NUV\geq2.0$, $NUV-U\leq1.4$, and
$FUV-NUV\geq1.05(NUV-U)+2.52$. The $FUV-NUV$ and $NUV-U$ colors are shown in Figure~\ref{FUVdrop}.
We identified \nzFLBGs\ \fuv-dropouts (\nzFLBG\ after stellar removal) with $U < 25.0$ mag. We
find that the \zphot\ distribution is skewed towards $z\sim0.7$--0.8 (see Figure~\ref{FUVdrop}).
While \fuv-dropouts have been claimed to be at $z\sim1$, the Lyman continuum breaks occurs at
the \fuv\ filter center at $z=0.68$, reducing the flux by half. This peak is not surprising
given (1) the current UV sensitivity and (2) the majority of $z\sim1$ galaxies have $U>25$ mag.

\subsection{Summary}
We used color selection to identify \nBX\ BXs (\nBXb\ $\Rcf\leq26.0$), \nBM\ BMs
(\nBMb\ $\Rcf\leq26.0$), \nsBzKs\ sBzKs, \npBzK\ pBzKs, \nzULBG\ \Udrops\ (\nzULBGb\ $\Rcf\leq26.0$),
\nzNLBG\ \nuv-dropouts (\nzNLBGb\ $V\leq26.0$), and \nzFLBG\ \fuv-dropouts for a total of \ntot\
(adopting a shallower limit on \Rc\ or $V$) or \ntotb\ (adopting a fainter limit on \Rc\ or $V$) galaxies
potentially at $z\approx1$--3. However, since there is overlap between the different color-selected samples,
the non-redundant sample contains \nUa\ (\nUaz\ with \zphot) and \nUb\ (\nUbz\ with \zphot) galaxies in
the Shallow and Faint samples, respectively.

\section{Results}\label{5}
In this section we begin by discussing the properties of galaxies selected using a given photometric
technique. Then we compare the different color selections to discuss sample overlap and the selection
bias associated with each technique. We then compare the color-selected samples against photo-$z$'s
to estimate the completeness of using these techniques in acquiring a census of star-forming galaxies at
$z=1$--3.\footnote{We exclude the \fuv-dropouts since the sample overlap with others is small and the \zphot\
  distribution peaks at $z<1$.}

\subsection{The Physical Properties of $z=1$--3 Galaxies}\label{5.1}
\begin{figure} 
  \epsscale{1.1} 
  \plotone{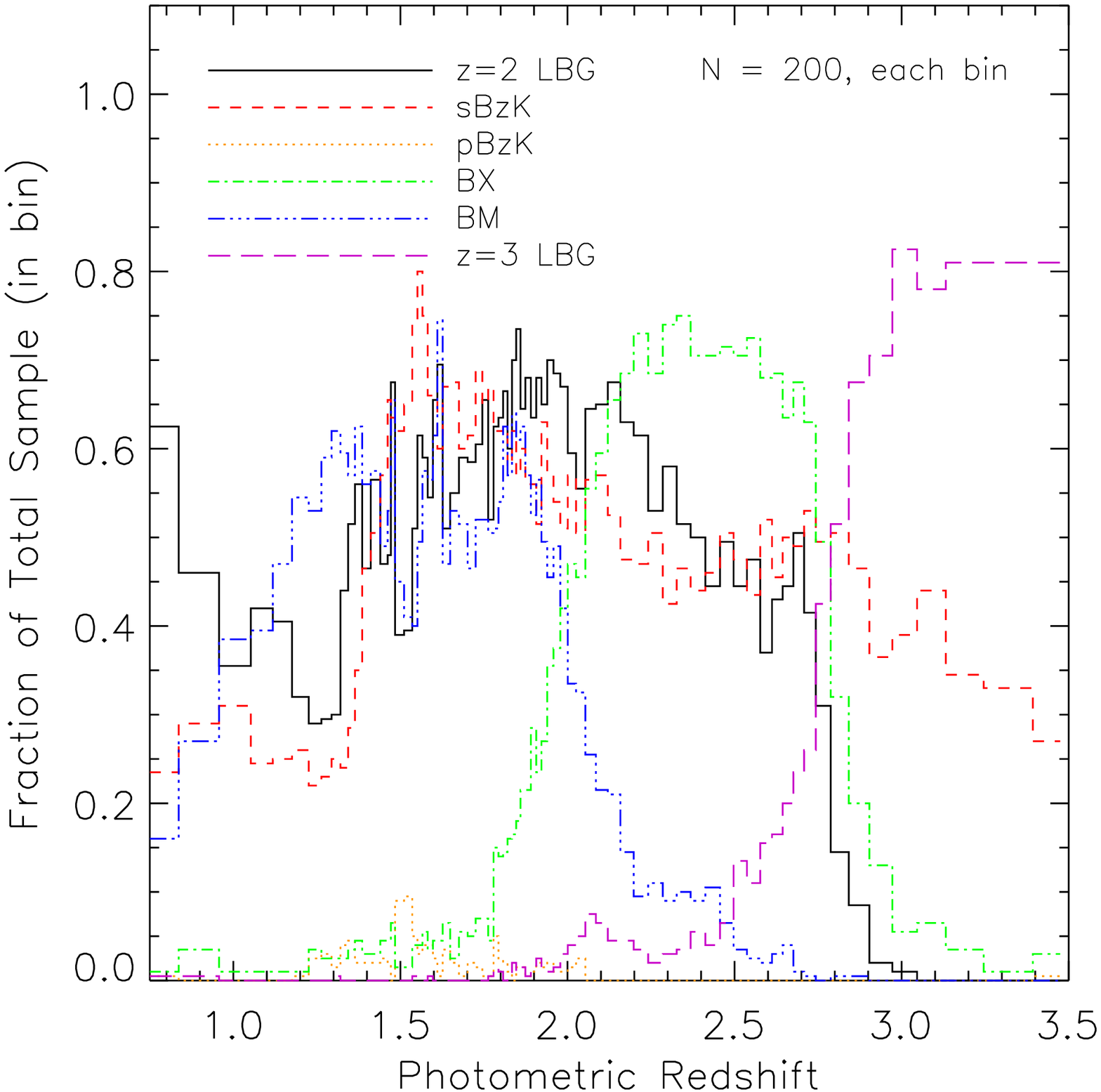}
  \caption{Fraction of the Shallow sample versus redshift for the different color selection
    techniques: \nuv-dropout (solid; black), \sbzk\ (dashed; red), pBzK (dotted; orange),
    BX (dot-dashed; green), BM (dot-dot-dashed; blue), and \Udrop\ (long-dashed; purple).
    We illustrate (in a given redshift bin) the fraction in terms of the total number of galaxies. \tcolor}
  \label{num_sfr}
\end{figure}

Using the color selection samples with \zphot\ and information from SED modeling, we illustrate in
Figure~\ref{num_sfr} the efficiency of these photometric techniques in terms of the relative fraction
of the total number. We find that the BX and BM methods identify 60\%--80\% of galaxies at the peak of
their \zphot\ distributions (i.e., where they are most sensitive). The \Udrop\ technique identifies
80\% of galaxies at $z\sim3$. However, this is optimistic since our ``Total'' sample is missing
low-mass $z\sim3$ red/dusty galaxies should they exist. Likewise, it appears that the \nuv-dropout
technique has similar efficiency as the BX/BM method with a redshift coverage that spans the redshift
of BX and BM galaxies. LBGs at $z\sim1$ would require UV imaging at $\approx$1800 \AA\ (between the
\fuv\ and \nuv\ band).
Finally, the star-forming BzK selection probes a wider range in redshift: capturing at least 40\% of
galaxies at $z>3$ and as much as $\sim$70\% at $z\sim1.5$.

Figure~\ref{num_sfr} also illustrates that the very different methods to identify galaxies at this epoch
are complementary, and that no single method identifies all galaxies. The union of the different color
selection techniques does yield a comprehensive galaxy census of $z=1$--3.

\begin{figure} 
  \epsscale{1.1}
  \plotone{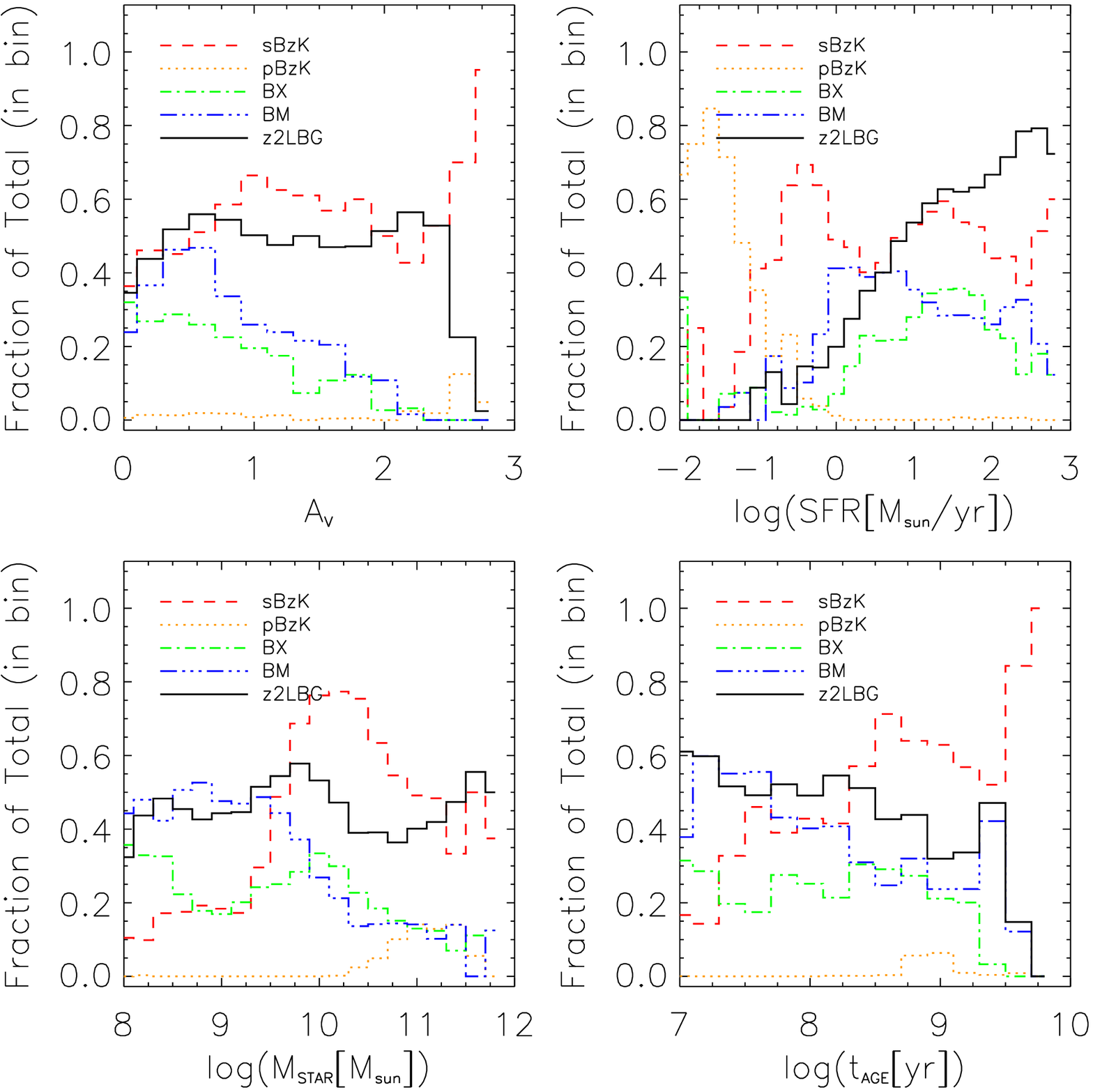}
  \caption{Physical properties of $z=1$--3 galaxies. The distribution of A$_{\rm V}$ (top left),
    $\log({\rm SFR}[M_{\sun}~{\rm yr}^{-1}])$ (top right), $\log(M_{\rm star}[M_{\sun}])$ (bottom left),
    and $\log({\rm age[yr]})$ (bottom right) for the different population. Color and line-style conventions
    follow those in Figure~\ref{num_sfr}.
    The \Udrops\ are not shown since most of them are at $\zphotf \gtrsim 3.0$. \tcolor}
  \label{hist_prop}
\end{figure}

To further understand the selection bias of these techniques, we illustrate in Figure~\ref{hist_prop} the
distribution of dust reddening, stellar mass, SFR, and stellar ages for the BX, BM, \sbzk, pBzK, and
\nuv-dropout samples. These distributions are normalized to the census of these techniques.
The majority of galaxies with stellar masses above $6\times10^9$ \Msun\ are acquired with the \sbzk\ technique.
UV selection methods are more efficient at selecting lower-mass galaxies. The BzK technique is also able to
identify the oldest galaxies and those with high dust extinction. UV techniques find young galaxies with little
dust extinction and typical SFRs of 1--10 \Msun\ \iyr. The $z\sim2$ LBG population appears to span a wider range
of dust reddening compared to the BX/BM method. We attribute this to the (1) larger range in rest-UV colors
($B-V<0.5$ mag spans a wider $G-\Rs$ color compared to the $G-\Rs$ color of BX/BM galaxies) and (2) an ability
to select massive galaxies. We discuss this further in Section~\ref{5.2.1}. Figure~\ref{hist_prop} also shows
that the most actively star-forming galaxies\footnote{The high SFRs are due to large dust extinction
  corrections.} will be identified using either the \sbzk\ or \nuv-dropout method. However, \lsfr\ of the
color-selected $\zphotf=1$--3 census consists of galaxies with SFRs between a few and 30 \Msun\ \iyr, and
\lsfrBXBM\ of this fraction is captured using the BX/BM method with \LyRcut\ mag.

\subsection{A Direct Comparison of Color Selection Techniques}\label{5.2}
\begin{deluxetable}{lccccc}
\tablewidth{0pt}
\tabletypesize{\scriptsize}
\tablecaption{Summary of Photometric Selection Census Surveys}
\tablehead{
 \colhead{Properties} & \multicolumn{5}{c}{Field}\\
  \cline{2-6}
  & 
  \colhead{GOODS} &
  \colhead{MUSYC} &
  \colhead{CDF-S} &
  \colhead{UDS\tablenotemark{a}} & 
  \colhead{SDF}
}
\startdata
Refer.        & \citetalias{reddy05} & \citetalias{quadri07} & \citetalias{grazian07} & \citetalias{lane07} & L11\\
Area          & 72.3                 &   413                 & 90.2                   & 2013                & 720\\
$N_{\rm total}$ & 931                  & 2959                  & 2630                   & 12503               & \nUa\\ 
$R$ depth     & 25.5                 & 25.5                  & 25.5                   & \ldots              & 25.5\\
$K$ depth     & 23.0                 & $\approx$23.45        & 23.5                   & 22.5                & 24.2\\
\zphot        & No                   & 9-band                & 13-band                & No                  & 20-band\\
\zspec        & Yes                  & No                    & $\sim$1000             & No                  & \ldots\BB\\
BX            & 620                  &  532                  & 1345                   & No                  & \nBX\\
BM            & \ldots\BC            &  741                  & \ldots\BC              & No                  & \nBM\\
$z3$LBG       & No                   &  401                  & \ldots\BC              & No                  & \nzULBG\\
sBzK          & 221                  & 2213                  & 747                    & 6736                & \nsBzK\\
pBzK          &  17                  &  326                  &  89                    &  816                & \npBzK\\
DRG           &  73                  &  480                  & 179                    &  330                & No  \\
ERO\E         & No                   & No                    & No                     & 4621                & No  \\
$z2$LBG       & No                   & No                    & No                     & No                  & \nzNLBG\\[-3mm] 
\enddata
\label{table7}
\tablenotetext{1}{UDS is short for the Ultra Deep Survey, which is part of the UKIRT Infrared Deep Sky Survey.}
\tablenotetext{2}{The SDF \zspec\ is mostly limited to low-$z$, so nothing is reported here.}
\tablenotetext{3}{The number here is merged with the above number for BX.}
\tablenotetext{4}{EROs (extremely red objects) are often distinguished in the literature from DRGs. This selection
  consists of using an $R-K > 5.3$ mag criterion.}
\tablecomments{``L11'' refers to this paper. Areas are in arcmin$^2$.}

\end{deluxetable}

Previous studies \citepalias{reddy05,quadri07,grazian07,lane07} have also investigated the completeness of
photometric selection and the selection bias in the GOODS-N, GOODS-S/CDF-S, MUSYC, and UKIRT fields. However,
as illustrated in Table~\ref{table7}, each study has its own advantages and disadvantages in terms of area
coverage, color selection used, and redshift accuracy.
Since these surveys could not identify $z\sim2$ LBGs, comparisons were made between
the BX/BM/\Udrop\ methods and the BzK/DRG methods given, and it was assumed that BX/BM's
are lower-redshift analogs of $z\sim3$ LBGs. We discuss the overlap between BX/BM and $z\sim2$ LBGs in
Section~\ref{5.2.1}. In previous studies, the term ``LBG'' is used to refer to the selection of \Udrops. We make a
clear distinction between the two.

\begin{figure*} 
  \epsscale{1.0}
  \plotone{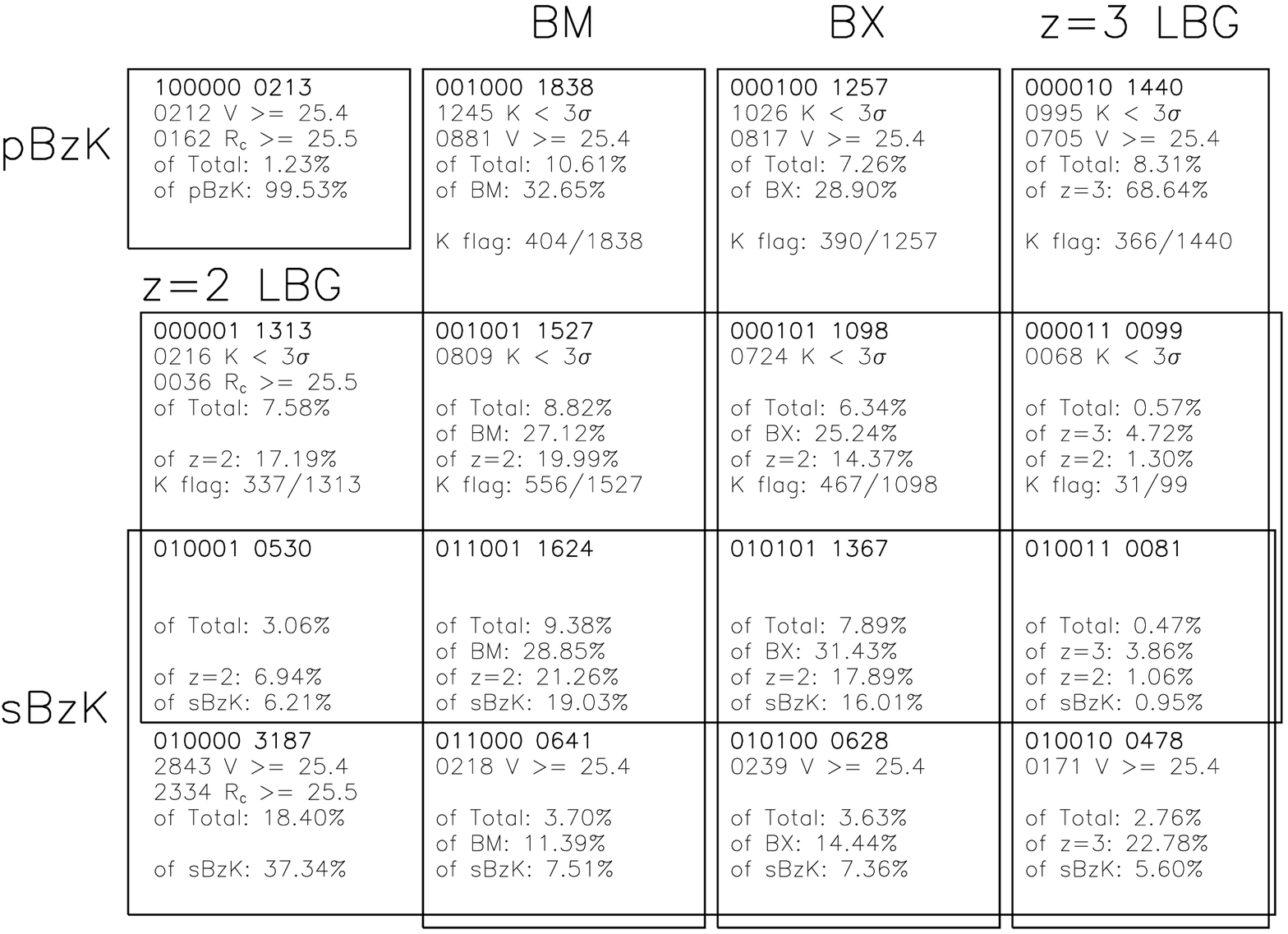}
  \caption{Diagram illustrating photometric galaxy sample overlap for the Shallow sample.
    Redshift increases to the right. Statistics are provided in each box including the size
    of a given sample, and the fraction relative to the full and each individual sample.
    ``K flag'' refers to sources that fall in lower sensitivity regions, so they cannot be
    classified in the BzK diagram. We exclude sources with $\zphotf < 0.5$.}
  \label{venn1}
\end{figure*}

A summary of the color selection sample overlap is provided in Table \ref{table4},
along with the average \zphot, $\EBV$, stellar age, stellar mass, and SFR.
We have created a 6-digit binary index to refer to the sample overlap. Each digit refers to a specific galaxy
population in the following order: pBzK, \sbzk, BM, BX, \Udrop, and \nuv-dropout. So for example, galaxies that
simultaneously satisfy the color criteria of \sbzk, BX, and \nuv-dropout are denoted as ``010101''. The diagram
in Figure~\ref{venn1} illustrates the sample overlap. The BX, BM, and \Udrops\ do not overlap, given the mutually
exclusive color selection. This is also the case for
the passive and star-forming BzK-selected galaxies.
Note that we have not placed any restrictions on the optical ($K$-band) magnitudes for NIR-selected (UV-selected)
galaxies for this table and this overlap diagram. We discuss the overlap with such restrictions in
Section~\ref{5.2.1}, to compare with previous estimates.

\newcommand{\med}{$_{\rm med}$}
\newcommand{\avg}{$$}
\newcommand\val[2]{#1$\pm$#2}
\begin{deluxetable*}{crrrrrr}
\tablewidth{0pt}
\tabletypesize{\scriptsize}
\tablecaption{Summary of Photometric Technique Overlap} 
\tablehead{
  \colhead{Type} &
  \colhead{$N$} & 
  \colhead{$z$\avg} & 
  \colhead{$E(B-V)$\avg} & 
  \colhead{$\log{(\rm{M})}$\avg} &
  \colhead{$\log{({\rm SFR})}$\avg} &
  \colhead{$\log{(t_{\rm age})}$\avg}\\
  &  &  &
  \colhead{(mag)} & 
  \colhead{(\Msun)} & 
  \colhead{(\Msun\ \iyr)} & 
  \colhead{(yr)}
}
\startdata
100000 &   207 & \val{1.66}{0.51} & \val{0.18}{0.14} & \val{10.93}{0.39} & \val{--1.66}{1.86} & \val{9.00}{0.23} \\      
010000 &  2525 & \val{1.85}{0.66} & \val{0.23}{0.17} & \val{10.14}{0.70} &  \val{0.78}{0.69} & \val{8.78}{0.56} \\      
011000 &   609 & \val{1.57}{0.36} & \val{0.15}{0.09} & \val{ 9.69}{0.50} &  \val{0.87}{0.42} & \val{8.56}{0.50} \\      
010100 &   559 & \val{2.29}{0.46} & \val{0.16}{0.12} & \val{10.15}{0.43} &  \val{1.13}{0.52} & \val{8.61}{0.44} \\      
010010 &   444 & \val{2.84}{0.41} & \val{0.13}{0.11} & \val{10.42}{0.34} &  \val{1.26}{0.44} & \val{8.80}{0.40} \\      
010001 &   530 & \val{1.57}{0.17} & \val{0.21}{0.14} & \val{10.42}{0.59} &  \val{1.20}{0.65} & \val{8.54}{0.47} \\      
011001 &  1618 & \val{1.73}{0.19} & \val{0.15}{0.09} & \val{ 9.89}{0.39} &  \val{1.13}{0.48} & \val{8.41}{0.45} \\      
010101 &  1325 & \val{2.21}{0.32} & \val{0.14}{0.10} & \val{10.07}{0.42} &  \val{1.33}{0.44} & \val{8.53}{0.46} \\      
010011 &    77 & \val{2.50}{0.35} & \val{0.13}{0.11} & \val{10.88}{1.28} &  \val{1.65}{0.75} & \val{8.74}{0.48} \\      
000001 &  1266 & \val{1.23}{0.40} & \val{0.27}{0.16} & \val{ 9.78}{0.82} &  \val{1.21}{0.82} & \val{8.19}{0.70} \\      
001001 &  1470 & \val{1.73}{0.26} & \val{0.15}{0.09} & \val{ 9.58}{0.80} &  \val{1.14}{0.54} & \val{8.23}{0.80} \\      
000101 &  1043 & \val{2.29}{0.29} & \val{0.10}{0.08} & \val{ 9.50}{0.81} &  \val{1.15}{0.45} & \val{8.15}{0.78} \\      
000011 &    95 & \val{2.59}{0.25} & \val{0.08}{0.11} & \val{ 9.82}{1.46} &  \val{1.38}{1.12} & \val{8.14}{0.63} \\      
001000 &  1622 & \val{1.47}{0.40} & \val{0.12}{0.09} & \val{ 9.17}{0.62} &  \val{0.71}{0.48} & \val{8.26}{0.66} \\      
000100 &   908 & \val{2.30}{0.53} & \val{0.10}{0.10} & \val{ 9.43}{0.77} &  \val{0.86}{0.50} & \val{8.22}{0.70} \\      
000010 &  1167 & \val{2.99}{0.51} & \val{0.07}{0.09} & \val{ 9.75}{0.69} &  \val{1.05}{0.59} & \val{8.33}{0.54} \\[-3mm]
\enddata
\label{table4}
\tablecomments{Average properties are reported. The first column is a 6-digit binary flag with each digit
  referring to a specific galaxy population in the order of pBzK, \sbzk, BM, BX, \Udrop, and \nuv-dropout.
  See Section~\ref{5.2} for further description. Sources with $\zphotf<0.5$ are excluded. }
\end{deluxetable*}

\subsubsection{Sample Overlap: Color--Color Plots}\label{5.2.1}
\begin{deluxetable}{cccc}
\tablewidth{0pt}
\tabletypesize{\scriptsize}
\tablecaption{Summary of Sample Overlap: NIR Selection}
\tablehead{
  \colhead{Type} &
  \colhead{FoV Limited} & 
  \colhead{$K >3\sigma$} & 
  \colhead{Total}
}
\startdata
  BX             & 3493 & 2065 & 1995 (\stataa)\\
  BM             & 4670 & 2955 & 2265 (\statab)\\
  $z$$\sim$3 LBG & 1701 &  824 &  559 (\statac)\\
  $z$$\sim$2 LBG & 6249 & 4784 & 3602 (\statad)\\[-3mm]
\enddata
\label{stat1}
\tablecomments{Sources with \zphot $< 0.5$ have already been excluded.}
\end{deluxetable}

\begin{deluxetable}{cccccc}
\tablewidth{0pt}
\tabletypesize{\scriptsize}
\tablecaption{Summary of Sample Overlap: $\Un G\Rs$ Method}
\tablehead{
  \colhead{Type} &
  \colhead{$R < 25.5$\tablenotemark{a}} & 
  \colhead{BX} &
  \colhead{BM} &
  \colhead{$z$$\sim$3 LBG} &
  \colhead{Total}
}
\startdata
  \sbzk\         & 6201 & 1995 & 2265 &  559 & 4819 (\statsd)\\
  $z$$\sim$2 LBG & 7603 & 2465 & 3151 &  180 & 5796 (\statNd)\\[-3mm]
\enddata
\label{stat2}
\tablenotetext{1}{Sources with \zphot $< 0.5$ are excluded.}
\end{deluxetable}

\begin{deluxetable}{cccc}
\tablewidth{0pt}
\tabletypesize{\scriptsize}
\tablecaption{Summary of Sample Overlap: \nuv-dropout Method}
\tablehead{
  \colhead{Type} &
  \colhead{$z_{\rm phot} \geq 1.5$} &
  \colhead{$V < 25.4$} & 
  \colhead{Total}
}
\startdata
  BX     & 3697 & 2934 & 2337 (\statca)\\
  BM     & 3499 & 2886 & 2512 (\statcb)\\
  \sbzk\ & 6300 & 4095 & 3160 (\statcc)\\[-3mm]
\enddata
\label{stat3}
\end{deluxetable}

To further understand and illustrate the selection effects of these techniques, we plot where each galaxy
population/technique (e.g., BX) falls in the color--color selection of another (e.g., BzK).
We first discuss the BzK selection, followed by the $\Un G\Rs$ methods, and finally the selection that
we have developed to identify $z\sim2$ LBGs. Note that the numbers below have excluded interlopers
with $\zphotf < 0.5$ (unless otherwise indicated).
When computing the overlap fraction and providing the two-color distribution of sources, we place
limits on either the $V$, \Rc, or $K$, as we did when we generated the high-$z$ photometric samples.

\begin{figure*} 
  \epsscale{0.45}
  \plotone{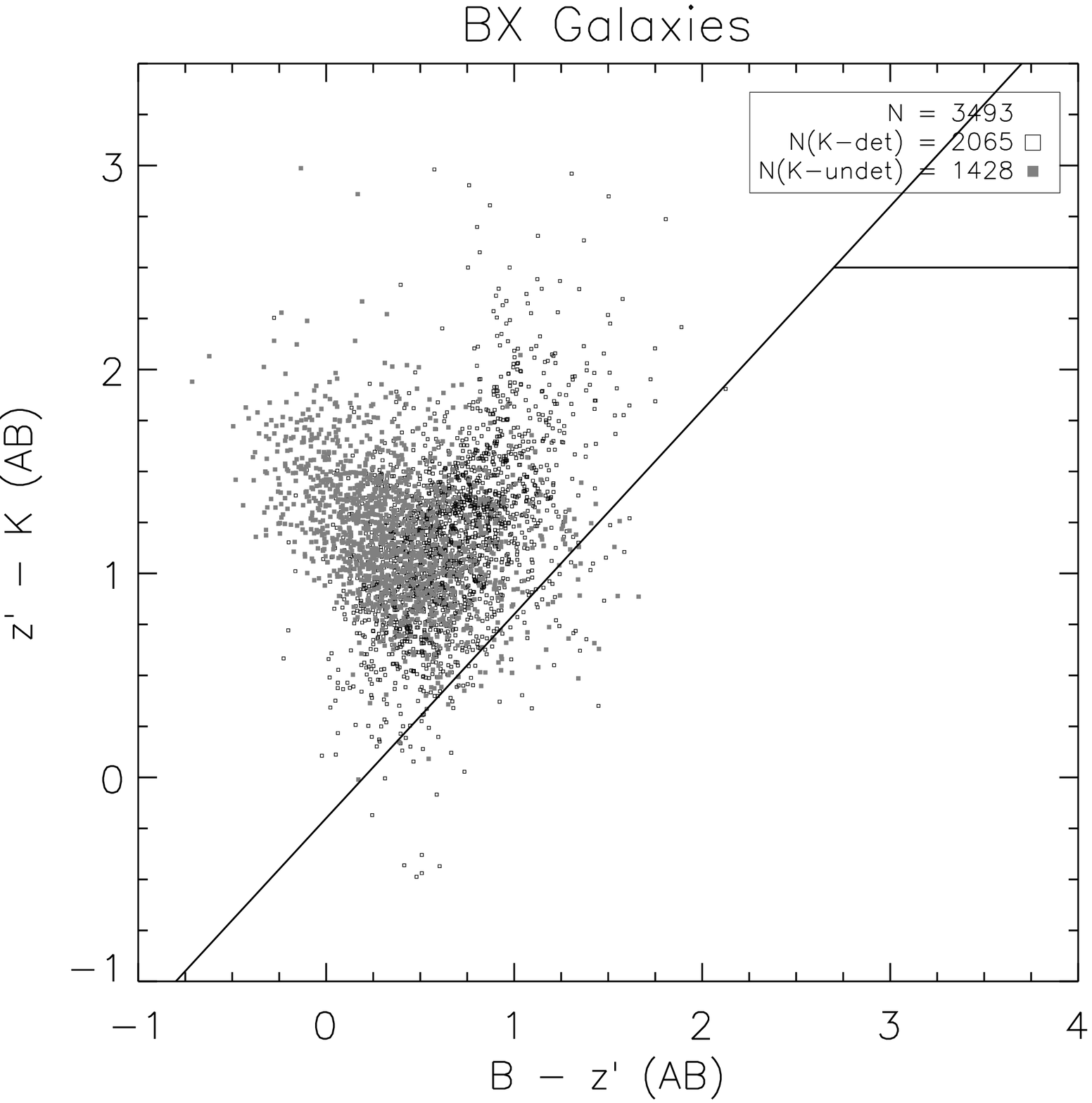} \plotone{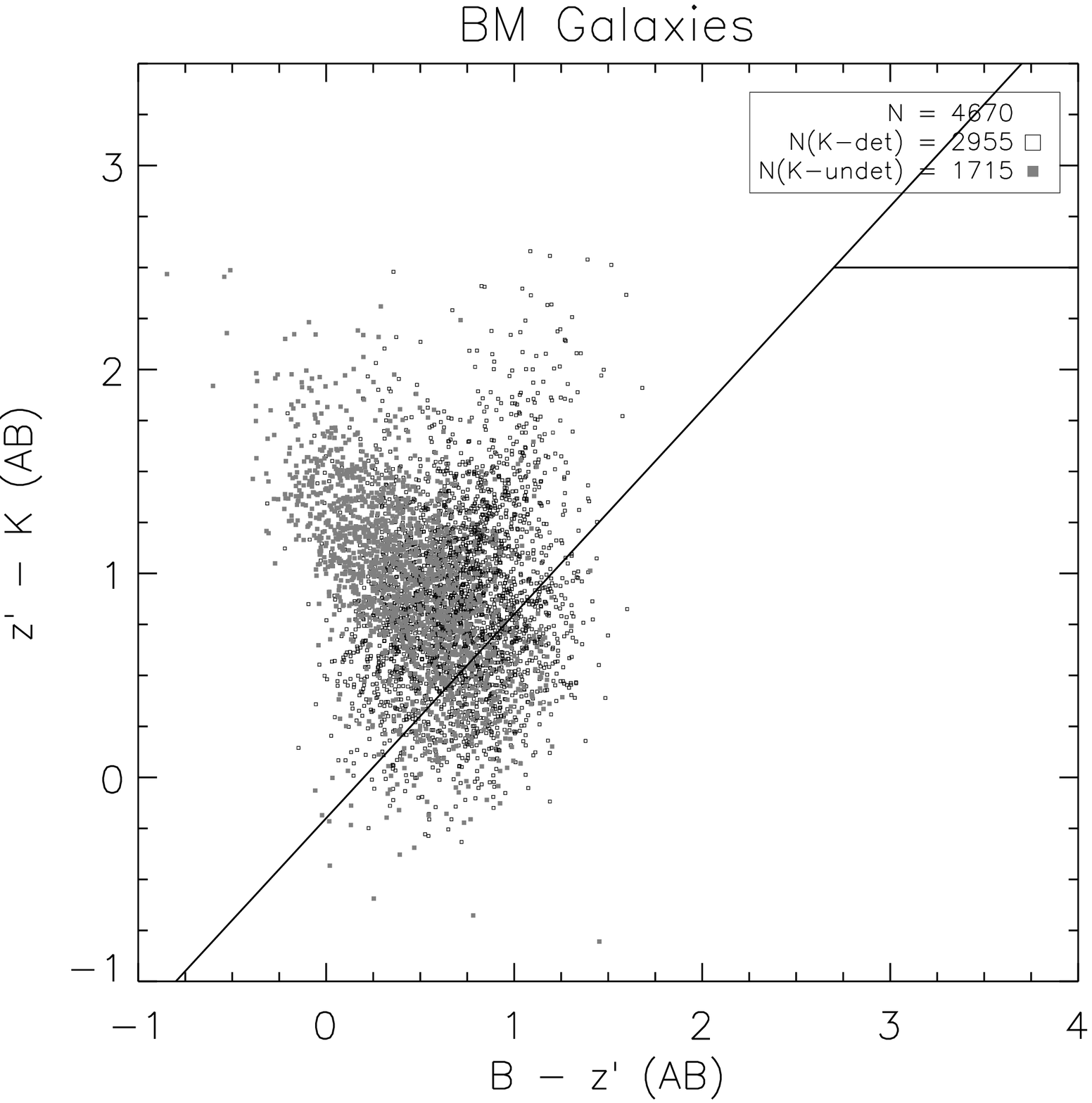}\\
  \plotone{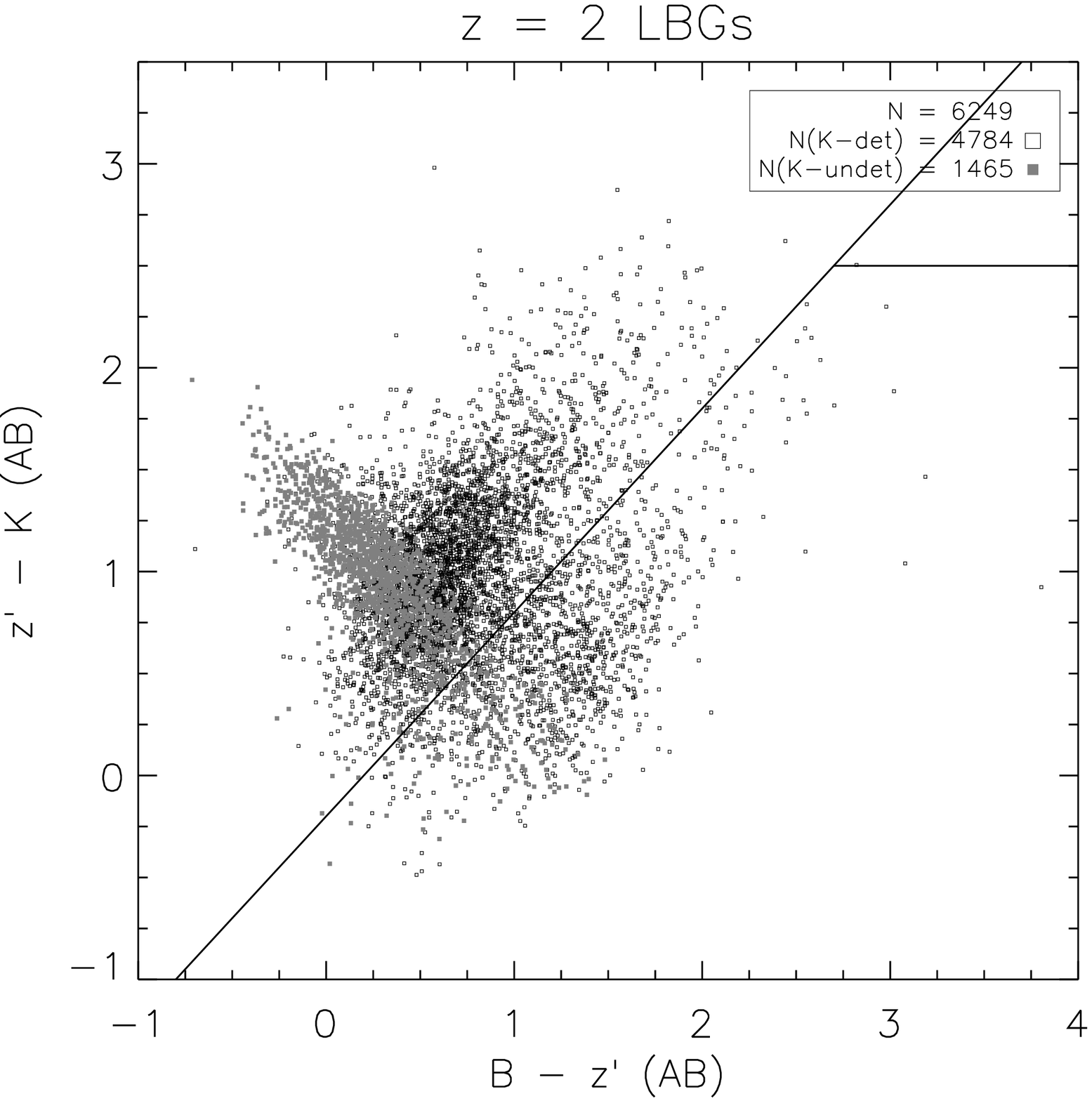} \plotone{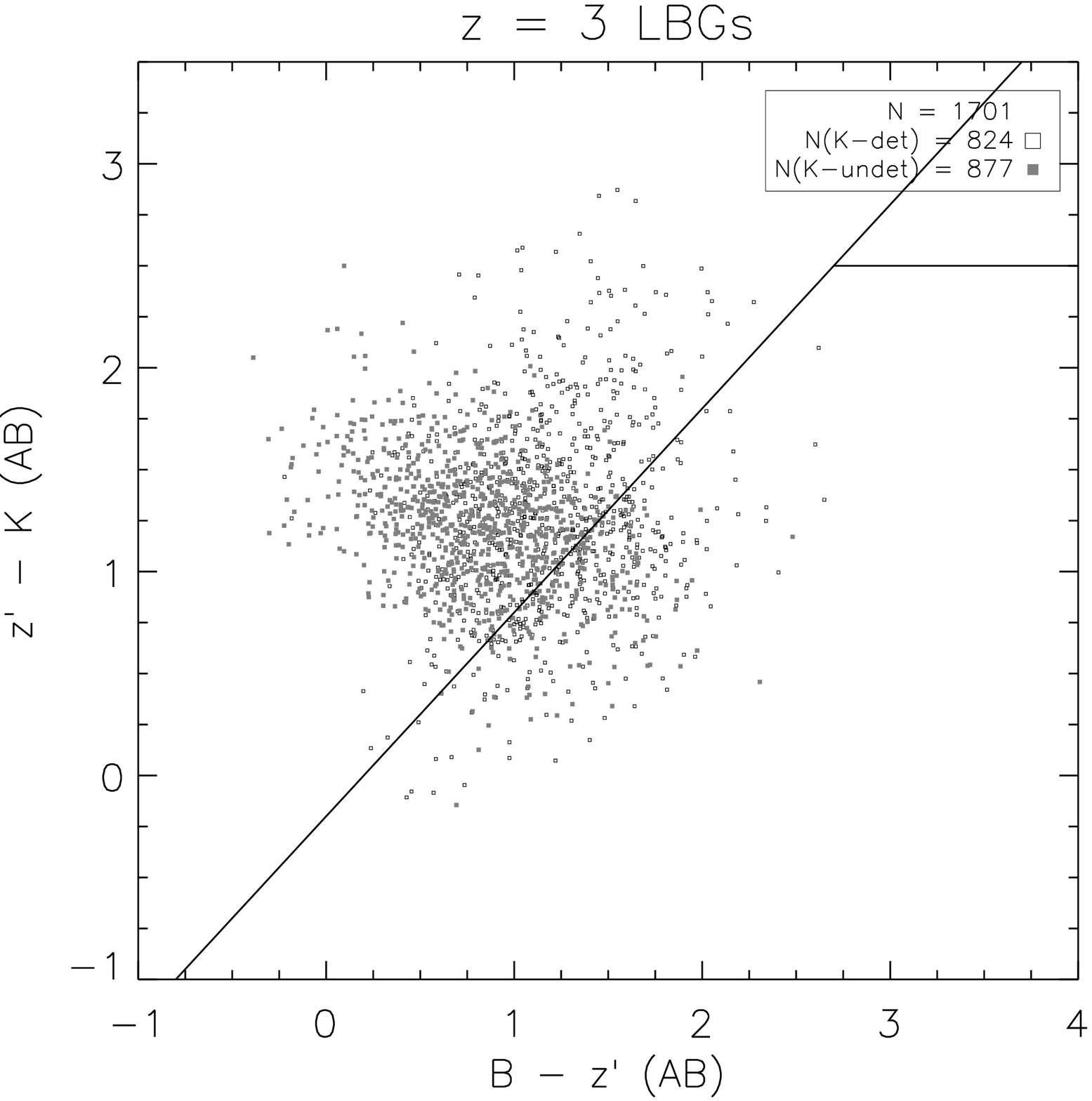}
  \caption{Sample overlap: $B-z\arcmin$ and $z\arcmin-K$ colors
    for BX galaxies (upper left), BM galaxies (upper right), \nuv-dropouts (lower left), and \Udrop\
    galaxies (lower right). Sources shown as black are those with $>3\sigma$ detection in $K$, while those
    in gray are sources undetected in $K$ (upper limits on their $z\arcmin-K$ color). Note that the axis
    scales differ from those in Figure~\ref{BzKselect}.}
  \label{compare_BzK}
\end{figure*}

A detailed summary of the statistics reported in this section is provided in Tables~\ref{stat1}--\ref{stat3}.\\
\indent{\it Near-IR selection.}
The $B-z\arcmin$ and $z\arcmin-K$ colors for BXs, BMs, \nuv-dropouts, and
\Udrops\ are shown in Figure~\ref{compare_BzK}.
When the BX, BM, \Udrop, and \nuv-dropout samples (1) are restricted to the region of 
sensitive $K$ photometry, (2) exclude sources with $\zphotf < 0.5$, and (3) only consider
sources detected above $3\sigma$ in $K$, we find that \stataa\ of BXs, \statab\ of BMs,
\statac\ of \Udrops, and \statad\ of \nuv-dropouts meet the \sbzk\ selection,
which indicates that \sbzk\ method is capable of selecting almost all high-$z$ UV-selected
galaxies having $K\lesssim24$ mag.
Table~\ref{stat1} provides more information regarding these statistics.
We note that if we did not exclude objects with $\zphotf<0.5$, the overlap fraction
would be \stataza\ (BX), \statazb\ (BM), \statazc\ (\Udrop), and \statazd\ (\nuv-dropout).
More than \statact\ of \Udrops\ with $K$-band detections are also identified as \sbzk. The
redshift where the Balmer/4000 \AA\ break enters the $K$-band window is $z\approx4$, so extremely
deep $K$-band imaging ($K=25$--26) can yield mass-limited samples of $z \lesssim 4$
galaxies with stellar masses above $\sim10^{10}$ \Msun. These high-$z$ sBzKs were also seen by
\citetalias{grazian07}, but in smaller numbers.
Unfortunately, they did not report a \sbzk--\Udrop\ overlap fraction to compare to ours. \citetalias{grazian07}
did find that $>86\%$ of $z>1.4$ BX/BM galaxies are \sbzk\ galaxies, which is consistent with the
above number. \citetalias{quadri07} determined that 80\% of BX/BM/\Udrop\ galaxies (regardless of \zphot) are
\sbzk-selected, which agrees with our number without any \zphot\ constraints. Likewise, similar numbers were
reported by \citetalias{reddy05}.

The majority of sources that are missed by the \sbzk\ method are BM and \nuv-dropout galaxies. We find that
these non-\sbzk\ galaxies have characteristic $\zphotf\approx1.9$, $\EBV\approx0.1$, stellar masses of
$\sim10^{10}$ \Msun, SFRs of $\sim$10 \Msun\ \iyr, and stellar ages of $3\times10^8$ yr.
It is not surprising that the BzK technique misses blue galaxies,
since it is more sensitive to redder galaxies.

None of the BX, BM, \Udrop, and \nuv-dropout sources are classified as a passive BzK, indicating that pBzK
galaxies represent a completely separate population.
This was also seen by \citetalias{quadri07}. We further discuss this in the Appendix.

\begin{figure} 
  \epsscale{1.1}
  \plotone{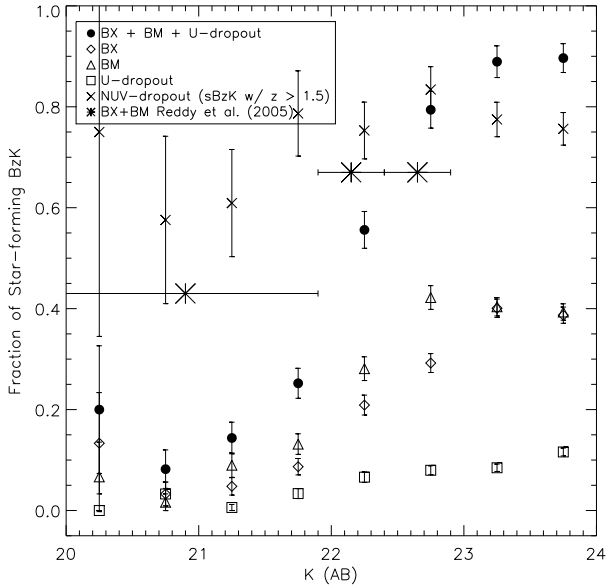}
  \caption{Fraction of $\zphotf\geq1.5$ \sbzk\ galaxies that meets the BX (diamonds), BM (triangles),
    \Udrop\ (squares) or \nuv-dropout (crosses) selection. The combination of the first three UV selection
    techniques is shown as filled circles. The sample
    overlap increases towards lower $K$-band luminosities, which is expected given the selection bias that UV
    techniques target lower mass galaxies. Overlaid in asterisks are estimates from \citetalias{reddy05}.
    This figure also illustrates that the Lyman break technique is able to identify many massive
    $z\sim2$ galaxies, which the BX/BM method misses.}
  \label{BzKoverlap}
\end{figure}

The fraction of \sbzk\ galaxies that are BXs, BMs, or \Udrops\ varies with $K$-band magnitude, with larger
overlap at fainter $K$ (see Figure~\ref{BzKoverlap}). This is to be expected, as the UV selection techniques
probe less reddened and young galaxies that are typically of lower masses. Our measurements agree with those
of \citetalias{reddy05} at $K\approx22$--23 mag and extend toward $K\approx24$ to show a slightly larger fraction
of overlap. However, our \sbzk--BX/BM/\Udrop\ overlap fraction is noticeably lower for $K<22.0$ mag. We find
that the overlap fraction for a magnitude bin of $20<K<22$ is typically 23\% with 1$\sigma$ variation of
10\% when dividing our SDF survey into cells of 8\farcm5 $\times$ 8\farcm5, which are
similar to the surveyed area of \citetalias{reddy05}. Thus, the discrepancy could be explained by clustering of
such galaxies and their smaller area coverage.

\begin{figure*} 
  \epsscale{0.45}
  \plotone{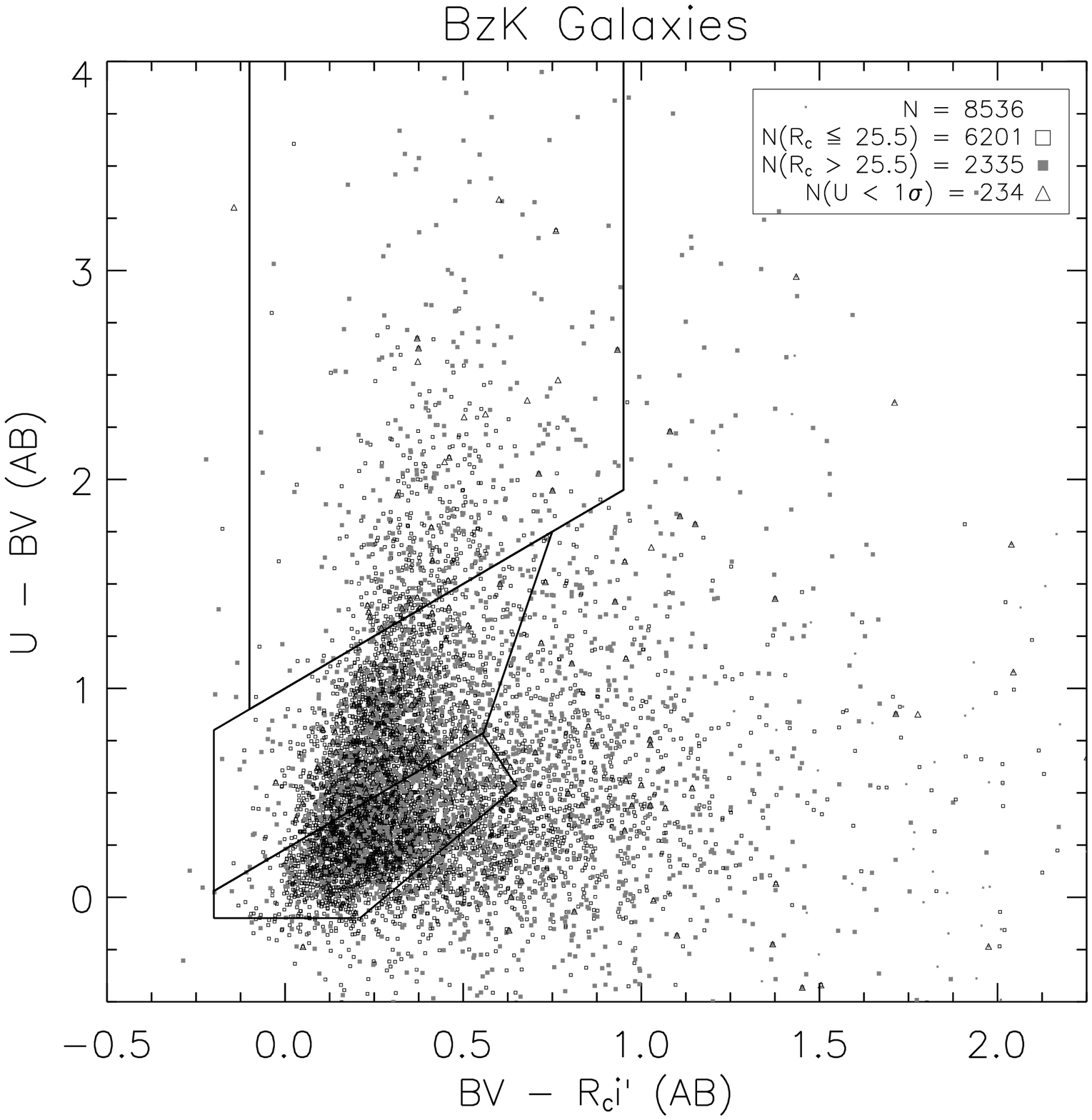} \plotone{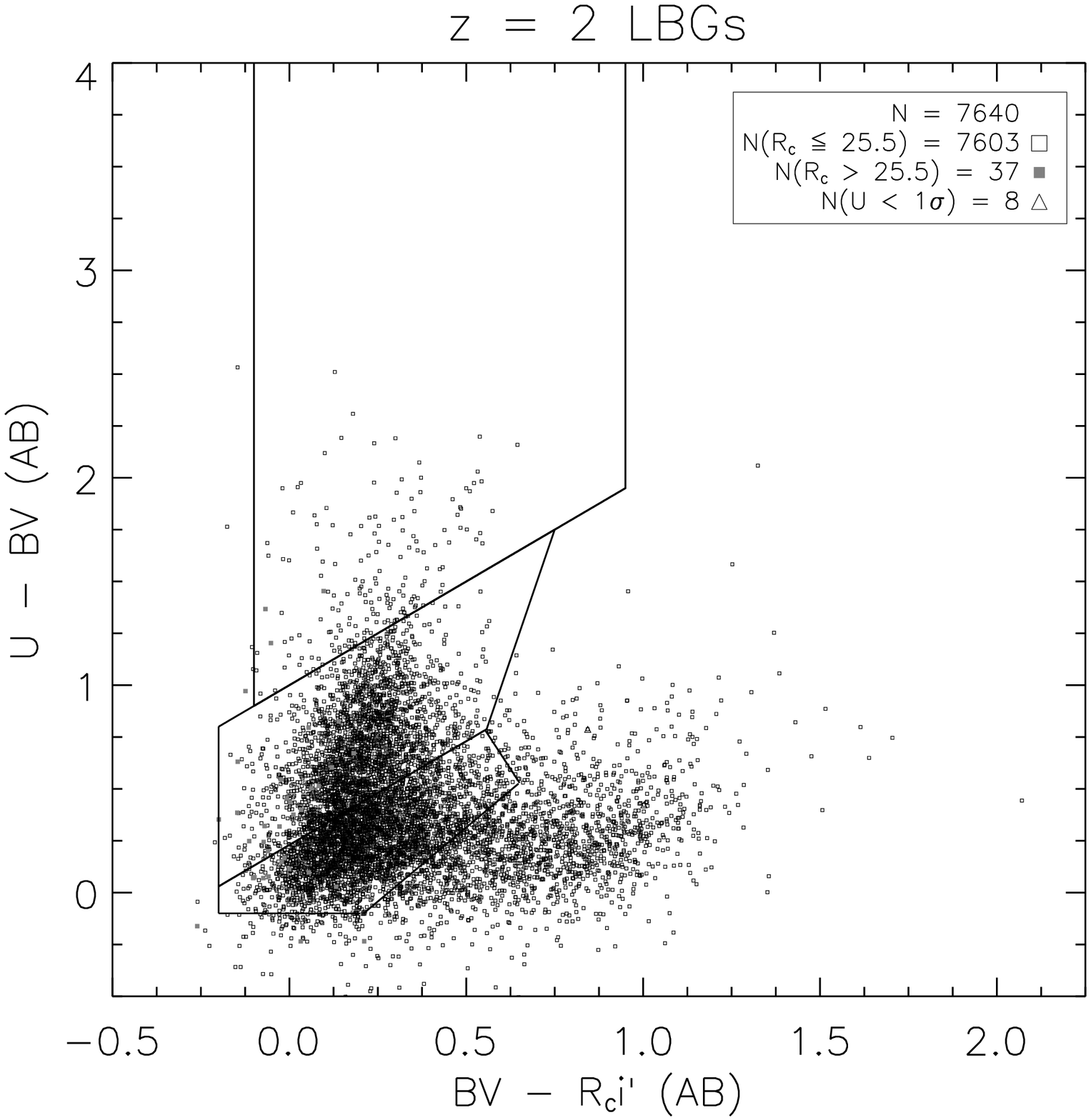}
  \caption{Sample overlap: $\UBV$ and $\BVRi$ colors for
    BzK galaxies (left) and $z\sim2$ LBGs (right). Sources shown as black are those with \LyRcut\ while
    those in gray are fainter than $\Rcf=\LyRcutn$. Triangle symbols indicate non-detections in $U$.}
  \label{compare_UGR}
\end{figure*}

{\it Non-ionizing UV continuum selection: the $\Un G\Rs$ method.}
For the selection of BX, BM, and \Udrop\ galaxies, the BzK galaxies and $z\sim2$ LBGs are plotted on the
$\UBV$ and $\BVRi$ color space in Figure~\ref{compare_UGR}.
We find that among the \sbzk\ galaxies with \LyRcut\ and $\zphotf>0.5$, \statsa\ are BXs, \statsb\ are BMs, and
\statsc\ are \Udrops, for a total overlap of \statsd. Detailed information of the overlap is provided in
Table~\ref{stat2}.
\citetalias{quadri07} determined $\sim$65\% for this fraction. To understand this small yet noticeable
difference, we point out that our sample probes much fainter $K$ magnitudes than they did, by $\approx$1 mag.
Since the majority of BX/BM galaxies have
low stellar masses, many of them are faint in $K$. As a result, our survey is more complete
in mass, and is weighted more toward the lower luminosity galaxies. We see this from the slightly higher
overlap fraction in Figure~\ref{BzKoverlap} for $K_{\rm AB}>23.0$ mag.
Likewise, among \nuv-dropouts with \LyRcut\ and $\zphotf>0.5$, \statNa\ are BXs, \statNb\ are BMs,
and \statNc\ are \Udrops, for a total overlap of \statNd. The small overlap between \nuv-dropouts and
\Udrops\ is to be expected, since we intended to exclude $z\gtrsim2.7$ galaxies with the $B-V\leq0.5$
mag criterion \citepalias[see][]{ly09}.

Recall that \citetalias{steidel04} defined the BX/BM selection to identify $z\sim2$ galaxies that are analogous to
$z\sim3$ LBGs. The presence of this large overlap shows that this statement is {\it mostly} true. We find,
however, that the LBG selection identifies a significant population of redder and more massive galaxies: at
least half of the $K<21.5$ \sbzk\ galaxies are selected using the \nuv-dropout method. This was previously
illustrated in Figure~\ref{compare_BzK} with the \nuv-dropouts spanning redder $B-z\arcmin$ and $z\arcmin-K$
colors compared to the BX galaxies, and again in Figure~\ref{BzKoverlap}, where the \nuv-dropout method
identified more $K$ bright galaxies compared to BX/BM. It also explains the cluster of points at
$\BVRi=0.5$--1.3 mag, which is also seen for \sbzk\ galaxies.\\

\begin{figure*}
  \epsscale{0.45} 
  \plotone{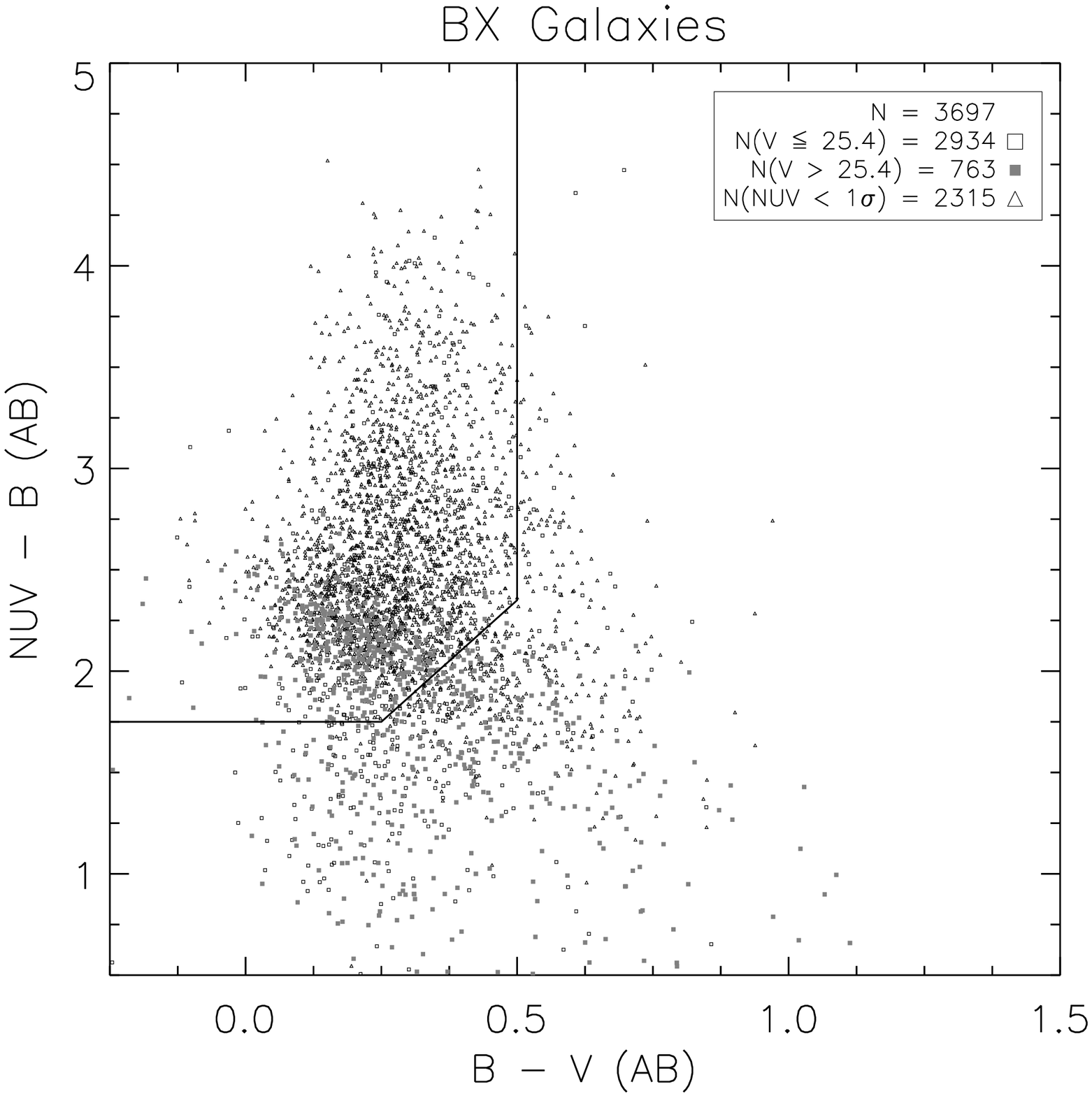} \plotone{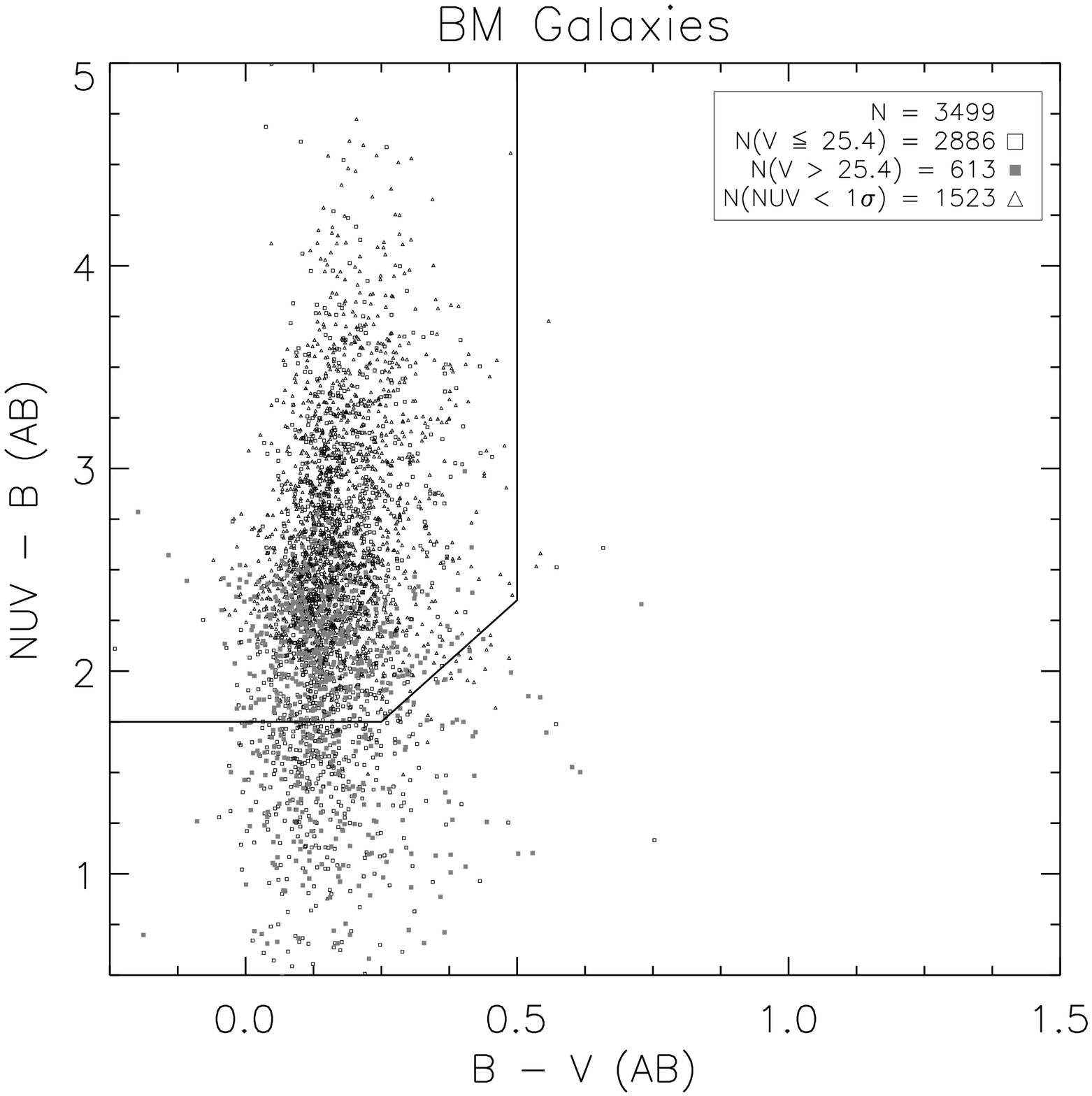}
  \epsscale{0.45}
  \plotone{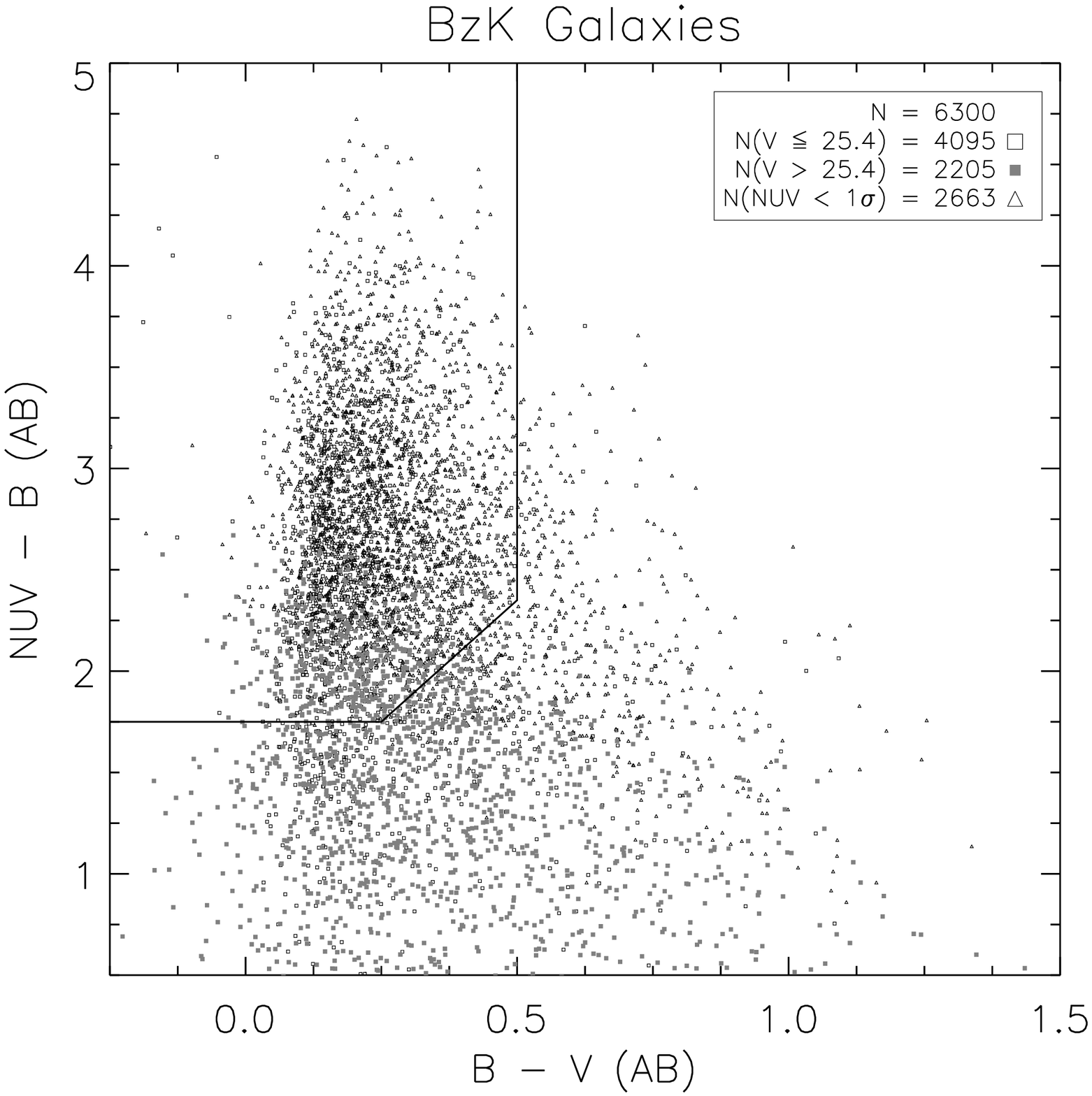}
  \caption{Sample overlap: $NUV-B$ and $B-V$ colors for BX-, BM-, and BzK-selected galaxies. We have
    limited these sources to those with $\zphotf\ge1.5$ since the \nuv\ data are only able to see the
    Lyman continuum break at $z\sim1.5$. Sources below $V=25.4$ are shown as gray squares. Triangles
    are sources that are undetected in the \nuv.}
  \label{compare_NUV}
\end{figure*}

{\it Ionizing UV continuum selection: the \nuv-dropout method.}
We show the $NUV-B$ and $B-V$ colors for BX, BM, and BzK galaxies in Figure~\ref{compare_NUV}.
We limit the BX, BM, and \sbzk\ samples to sources brighter than $V=25.4$ and require that
$\zphotf\geq1.5$, since the \nuv\ will only be sensitive to a Lyman break above this redshift.
We find the fractions of BX, BM, and \sbzk\ samples that meet the \nuv-dropout criteria are
\statca, \statcb, and \statcc, respectively. More information regarding these statistics are
given in Table~\ref{stat3}. The high overlap strongly suggests that using a Lyman-limit break
to select high-$z$ galaxies of low and high stellar masses is a successful alternative to
those methods.

{\it Summary.}
We find that the overlap fraction between UV- and NIR-selected samples
(with typical restrictions on optical and NIR depths) of $z=1$--3 galaxies is high:
74\%--98\%. The sample overlap was greater when examining the UV-selected samples in the BzK color space
rather than vice versa. This illustrates that UV selection techniques, which are good at identifying
blue high-$z$ galaxies, miss the redder galaxies, particularly the more massive ones. This has been seen
in previous studies \citepalias[e.g.,][]{quadri07}. Several comparisons between the
BX/BM galaxies and the $z\sim2$ LBGs confirm that most (80\%--87\%) BX/BM galaxies do show a
strong Lyman-limit break.
This implies that these UV-selected galaxies are nearly analogous to Lyman break selected galaxies.
However, the \nuv-dropout population does span a wider range in dust extinction, and shows greater
overlap with bright \sbzk\ galaxies.

\subsection{The Star Formation Rate Density from the Census}\label{5.3}
\begin{figure} 
  \epsscale{1.1}
  \plotone{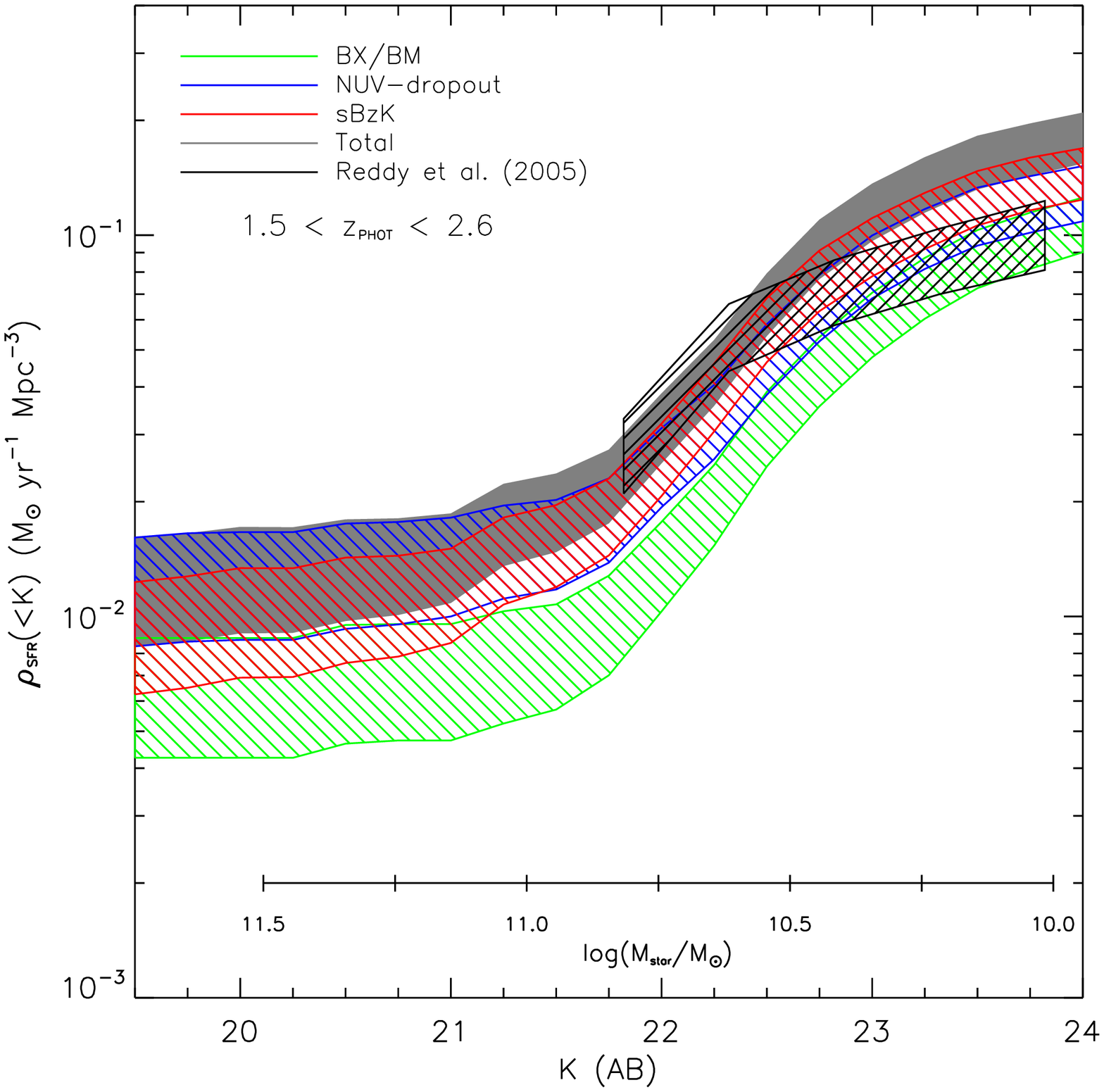}
  \plotone{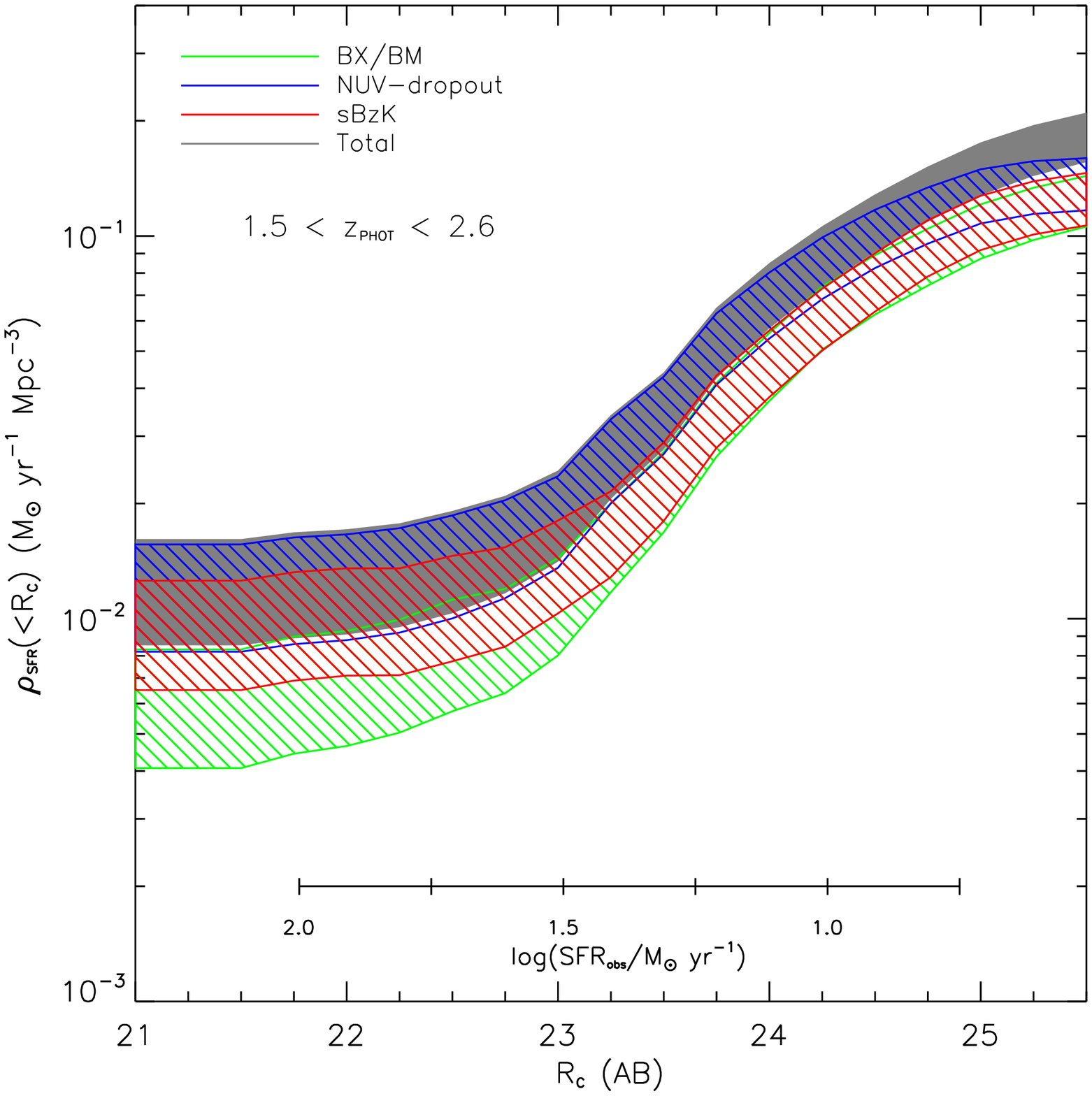}
  \caption{Accumulated SFR density against $K$- (top) and
    \Rc-band (bottom) magnitudes from a census of star-forming galaxies at $1.5 < \zphotf < 2.6$.
    Shaded regions include uncertainties in the SFR due to the \zphot\ accuracy and cosmic variance estimates
    from \cite{somerville04}. The census total is shown by the shaded gray region while those determined using
    the \sbzk, BX/BM, and \nuv-dropout methods are shown by the hatched red, green, and blue regions,
    respectively. A stellar mass or observed 1700 \AA\ SFR scale is provided for guidance. The stellar mass
    is from \cite{daddi04} SED-fit relation. For the UV SFR, we adopt \cite{kennicutt98} relation. \tcolor}
  \label{SFRd}
\end{figure}
With a census of star-forming galaxies from the combined color selection techniques, we can determine the SFR
density from the results of our SED modeling with FAST. \citetalias{reddy05} first estimated this with a census
of DRGs, BX and BM galaxies, and \sbzk\ galaxies at $1.4<\zspecf<2.6$. We consider \sbzk, BX and BM, and
\nuv-dropout galaxies, and limit our sample to $1.5 < \zphotf < 2.6$, since this is the overlap range where
all three techniques work. While our census does not include DRGs, \citetalias{reddy05} reported that
these galaxies contribute $\lesssim20\%$ towards the SFR density via their census.

We determine the integrated SFR density as a function of $K$- and \Rc-band magnitudes
for the three populations and the full census accounting for sample overlap.
To normalize our SFR measurements by the effective survey volume, we assume that the completeness $C(z)$ of the
census at any given redshift is proportional to the \zphot\ distribution normalized to unity at the \zphot\
peak. We note that this is a lower limit on incompleteness, so SFR density measurements reported here can be
underestimated. The following equation,
\begin{equation}
  V_{\rm eff} = \Omega \int_{z=1.5}^{z=2.6}dzC(z)\frac{dV}{dzd\Omega},
\end{equation}
yields volumes of \VeffK\ and \VeffR\ when we consider sources 
with $K\lesssim24$ and \LyRcut\ mag, respectively. We determined a total SFR density of \sfrKn\
(0.18$\pm$0.03) \Msun\ \iyr\ \vMpc\ for sources brighter than $K=24.0$ mag ($\Rcf=\LyRcutn$ mag).
The SFR density is illustrated in Figure~\ref{SFRd}, where the range of allowed values is $1\sigma$
from including uncertainties in the SFRs due to the \zphot\ accuracy and cosmic variance
\citep[estimated from][]{somerville04}.

It appears that the SFR density down to a stellar mass limit is mostly contributed by \sbzk\ galaxies. They
are not only numerous in the sky, but also have significant star formation that is obscured by dust.
For example, while \NEBV\ of the census is composed of star-forming galaxies with
$\EBV > \EBVlimit $, they account for \pEBV\ of the integrated SFR density. If a cut of $\EBV=0.4$ mag
(recall that the BX/BM and \nuv-dropout techniques are sensitive up to $\EBV \sim 0.4$ mag) is adopted,
then \pEBVb\ of the total SFR density is from ``less'' dusty galaxies.

Compared to \citetalias{reddy05}, we find a higher integrated SFR density by \highR.
However, \citetalias{reddy05} was unable to select \nuv-dropouts. Therefore, if we removed
galaxies that uniquely meet the \nuv-dropout selection, then the SFR density for $K <24.0$ mag is
reduced by \sfrKNUVp\ to \sfrKNUV, which is still \highRb\ higher than the reported value of
\citetalias{reddy05}.

One explanation for the comparatively higher SFR density is the effect of field-to-field fluctuations
on the \sbzk\ galaxy population. In Figure~\ref{sbzk_sd}, we compared different \sbzk\ surveys and found
that the SDF \sbzk\ counts agree well with those from COSMOS \citep{mccracken10} down to the COSMOS
survey limit of $K\sim23$. However, \citetalias{reddy05} begin to see a relative decline at $K\sim23$ and is
approximately a factor of two lower than our measurements at $K\sim24$. When limited to $z=1.4$--2.6, the
surface density reported by \citetalias{reddy05} is 4.93 arcmin$^{-2}$ for $K\lesssim24$, which is 27\% lower
when compared to the SDF down to the same limits and redshift range (6.78 arcmin$^{-2}$).

To understand the general discrepancy in surface density, we reasoned that field-to-field fluctuations is a
likely culprit since \citetalias{reddy05} covers 72.3 arcmin$^{-2}$. As an exercise, we subdivided the SDF into
eight independent fields each with the same surveyed area of \citetalias{reddy05}. We determined the surface
density as a function of $K$ for each field, and found that the variations observed can fully explain the
discrepancy for $K\lesssim23$ and about half of the differences for $K\sim23$--24. If this explanation is correct,
then such variations can easily affect both the cumulative SFR density from the census of color selection
techniques and the contribution to the census solely from \sbzk\ galaxies. For example, among the BX/BM and
\sbzk\ census, we find that 66\% and 89\% are from BX/BM and \sbzk\ galaxies, while \citetalias{reddy05}
report 68\% and 67\%. In addition, the smaller area coverage of \citetalias{reddy05} yields a census sample
of $\sim$500 galaxies in size, which is statistically less robust when compared to the SDF sample of $\sim$5200
for $z=1.5$--2.6.

\subsection{The Completeness of Color Selection Techniques}\label{5.4}
\begin{figure*}[thc]
  \epsscale{1.1} 
  \plottwo{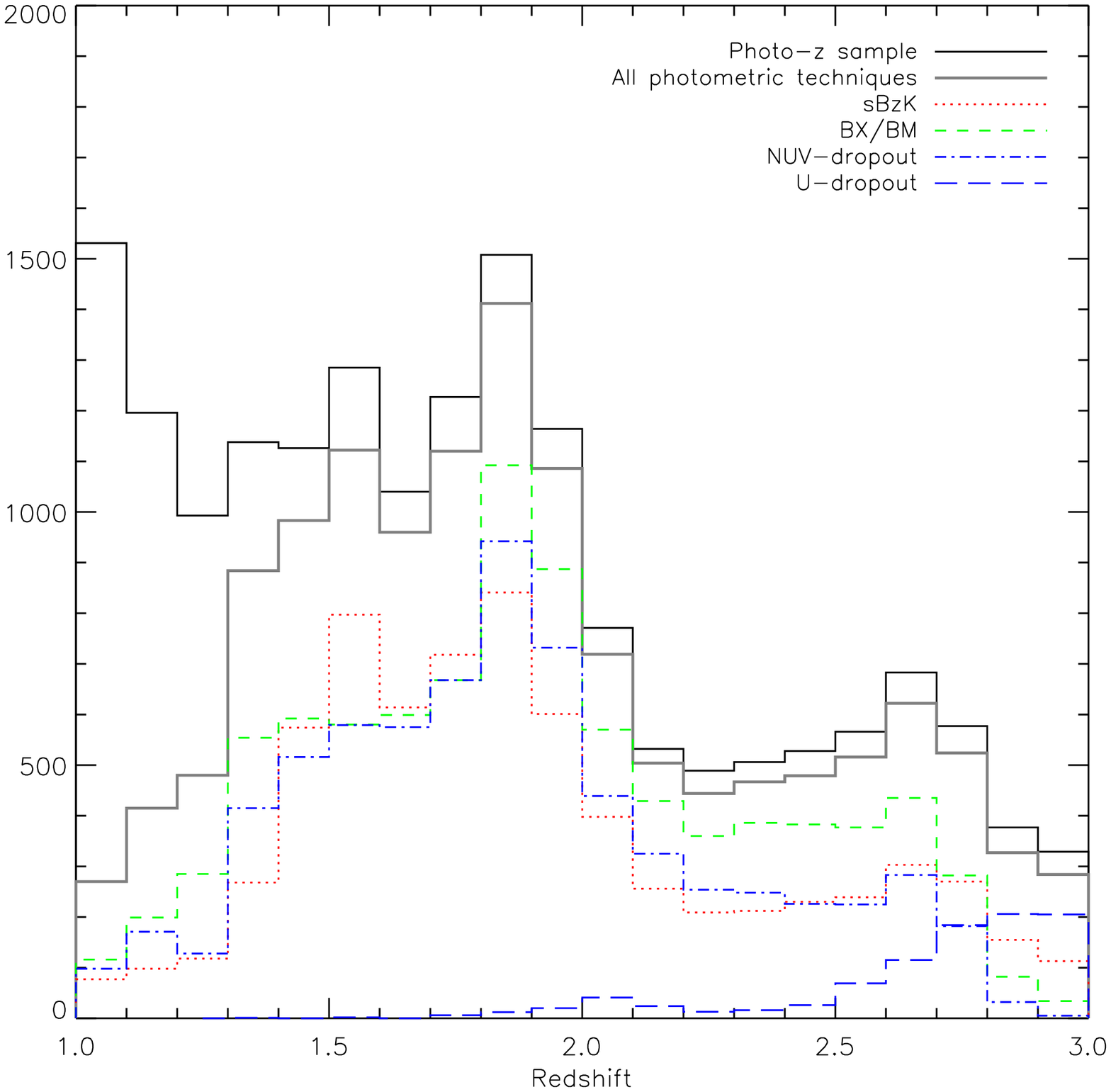}{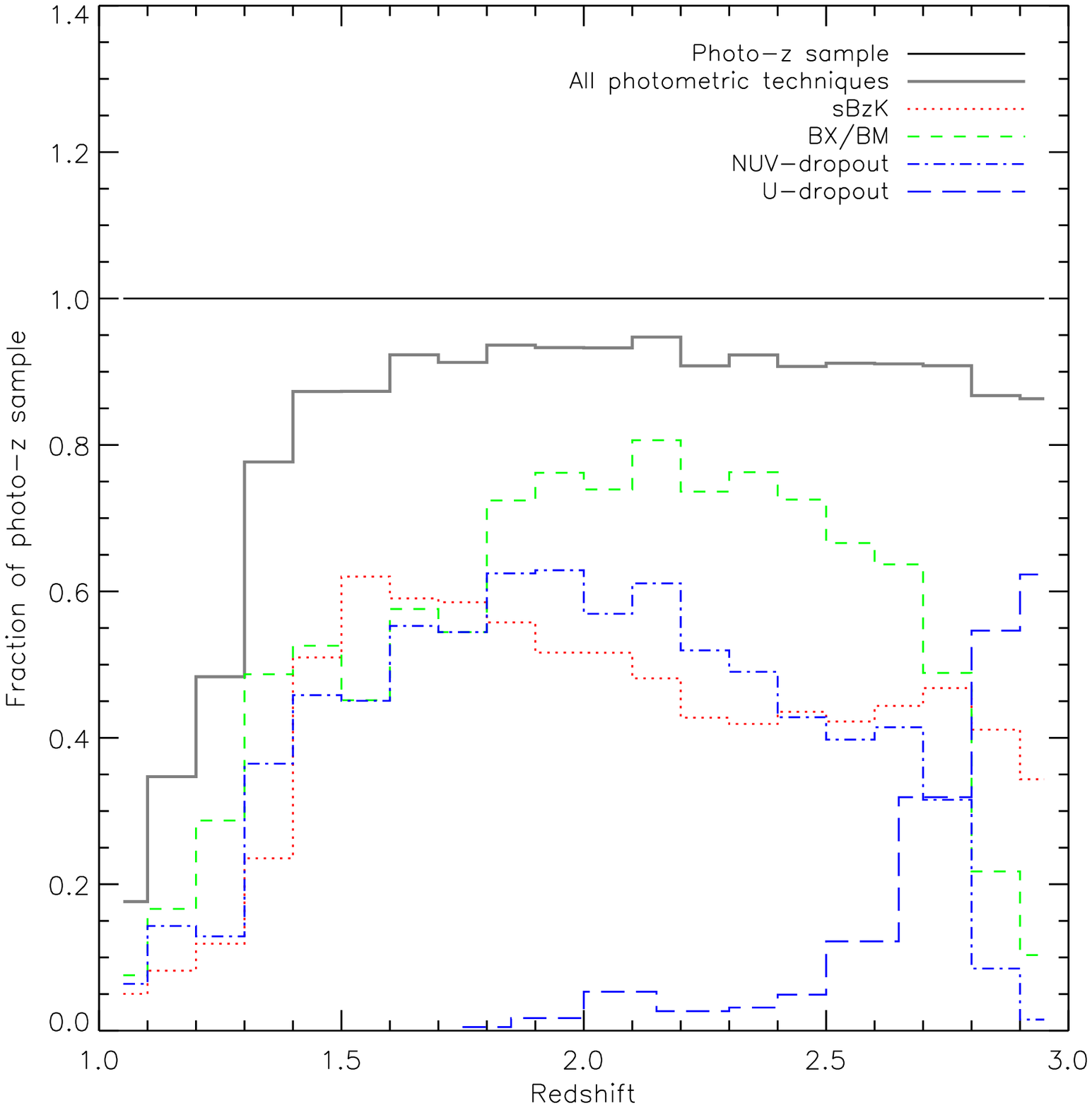}
  \caption{Comparison between galaxies from color selections and the photo-$z$ sample. The total
    numbers are shown on the left and the distributions are normalized to the photo-$z$ sample (thin
    black solid lines) on the right. The \sbzk, BX/BM, \nuv-dropout, and \Udrop\ samples are indicated by
    the dotted (red), short-dashed (green), dot-dashed (blue) and long-dashed lines (blue). The combination of
    these techniques, accounting for overlap between techniques, is shown by the thick gray solid lines. \tcolor}
  \label{photoz_pop}
\end{figure*}

In Figure~\ref{photoz_pop}, we show the redshift distribution for the different color selection
techniques and the photo-$z$ sample. For this comparison, we consider photo-$z$ sources that meet the
selection depth used for color selections (i.e., 3$\sigma$ detection in $K$, \LyRcut\ mag, or
$V<25.4$ mag). This yields a total of \ncensusC\ galaxies with $\zphotf=1.0$--3.0.
It is apparent from Figure~\ref{photoz_pop} that the color selection techniques each identify some
subset of galaxies with $\zphotf=1$--3. We find that the combination of different color selection
techniques is able to identify \censuscomp\ of galaxies (down to the above magnitude limits) with
$\zphotf=1.5$--2.5.
Since the photo-$z$ sample is limited to the depths of the color selection techniques,
the missed rate of 10\% is likely a result of photometric scatter that drives these galaxies
out of the color selection regions.

As expected, a noticeable ($\approx$50\%) fraction of galaxies with $\zphotf=1.0$--1.4 are missed. This is
because (1) the BzK selection was designed to search at $z\gtrsim1.5$, (2) two-thirds of the BM galaxy
population lie at $\zphotf>1.5$, (3) almost all BX galaxies are above $\zphotf>1.5$, and (4) the
\galex/NUV filter only detects a Lyman continuum break at $\zphotf>1.5$. However, as we discussed previously,
recently available {\it Hubble}/WFC3 filters are beginning to be used to select galaxies via the Lyman
break at $z\sim1.3$, and such samples may be able to address this loss.

\section{Modeled Predictions for a Mock Census}\label{6}
\indent{\it Motivation.}
To further understand the selection effects of the census derived from color selection techniques
(hereafter the ``Census''), we generate Monte Carlo realizations of our data and of the Census using
mock galaxies that represent the $z \approx 1$--3 population. This simulation intends to (1) examine
the stellar population incompleteness effects on the multi-band $\chi^2$-weighting scheme (discussed in
Section~\ref{2.1}), (2) create a census of mock LBGs, BX/BM, and BzK galaxies at $z=1.0$--3.5, and (3)
compare modeled predictions with observations. In summary, these analyses find both agreements and
disagreements with observational results. The former firmly supports many of our
main results while the discrepancies appear to have a minor effect.

The ability for a galaxy to be included in the Census is dependent on a few factors. First,
galaxies must be identified in the multi-band $\chi^2$ image. We required $\chi\geq 3.5$
for at least five connecting pixels. Because the multi-band image spans a wide range in observed
wavelengths (4500 \AA--2 \mm), these $\chi$ values are influenced by the shape of the SED,
which can be parameterized by redshift, stellar mass, dust reddening, galaxy ages, and
star formation histories. It is thus important to simulate a wide range of these galaxy properties
(see below). Second, we place restrictions on the sample such that they are bright enough to be
included in either the \fBK\ or \fBf\ photo-$z$ catalogs. Finally, we also apply the different
color selection techniques to cull LBGs, BX/BM, and BzK galaxies to generate a ``Mock Census.''
These steps that we follow are identical to those conducted in Sections~\ref{2.1}, \ref{3.1}, and 
\ref{4}.

\indent{\it Technique/approach.}
We begin with a grid of spectral synthesis models that adopt a range of magnitudes, redshifts,
$\EBV$ values, galaxy ages, and $\tau$ values for an exponentially declining star formation history.
These models allow us to generate the full SED for artificial galaxies, which are then
convolved with the bandpasses for all 15 bands to obtain apparent magnitudes. We then add noise based
on known sensitivity at each bandpass to determine the probability of (1) being included into
the multi-band $\chi^2$ image, (2) satisfying the \fBK\ and/or \fBf\ photo-$z$ criteria, and
(3) meeting the color selection criteria for the Census.

\indent{\it Assumptions.}
In estimating the $\chi$ values for mock galaxies, we assume that sources are unresolved
with an FWHM of 1\farcs1. This process involves normalizing a Gaussian PSF by the S/N of the
source in each band, and then taking the square-root of the sum of the squares for eight bands
($BV\Rcf i\arcmin z\arcmin K$, IA598, and IA679). We then determine if the number of connecting
0\farcs2 pixels that meets the minimum $\chi=3.5$ criteria is at least five to be included.

The grid of models spans redshifts from 1.0 to 3.2 (0.2 increments), $\EBV=0.0$--0.5 (0.1 increments),
and ages between $\log({\rm age/yr}) = 7.0$ dex and the age of the universe at a given
redshift\footnote{For $z=1$ and 3, the limits are 5.75 and 2.11 Gyr, respectively.}
(0.25 dex increments). We adopt exponentially declining star formation history with $\tau$ = 0.1, 1.0,
and 10.0 Gyr. The limits on these properties are identical to those assumed in modeling the SEDs
(see Section~\ref{3.2}). We also assume a \cite{chabrier03} IMF and solar metallicity for the
spectral synthesis models. 
We also include redshift-dependent IGM absorption from neutral
hydrogen for rest wavelengths blueward of 1216 \AA\ following \cite{madau95}. With these parameters,
we have a total of 2,268 SEDs. These SEDs are normalized in the $z$\arcmin-band between
20.0 and 26.0 mag (0.5 mag increments), which yields 27,216 possible models.
We chose to normalize at $z$\arcmin\ since it is the most sensitive band closest to $K$.
The use of the $z$\arcmin-band also provides a good compromise between the rest-frame UV
(sensitive to recent star formation) and optical (sensitive to stellar mass).

\begin{figure*}
  \epsscale{0.5} 
  \plotone{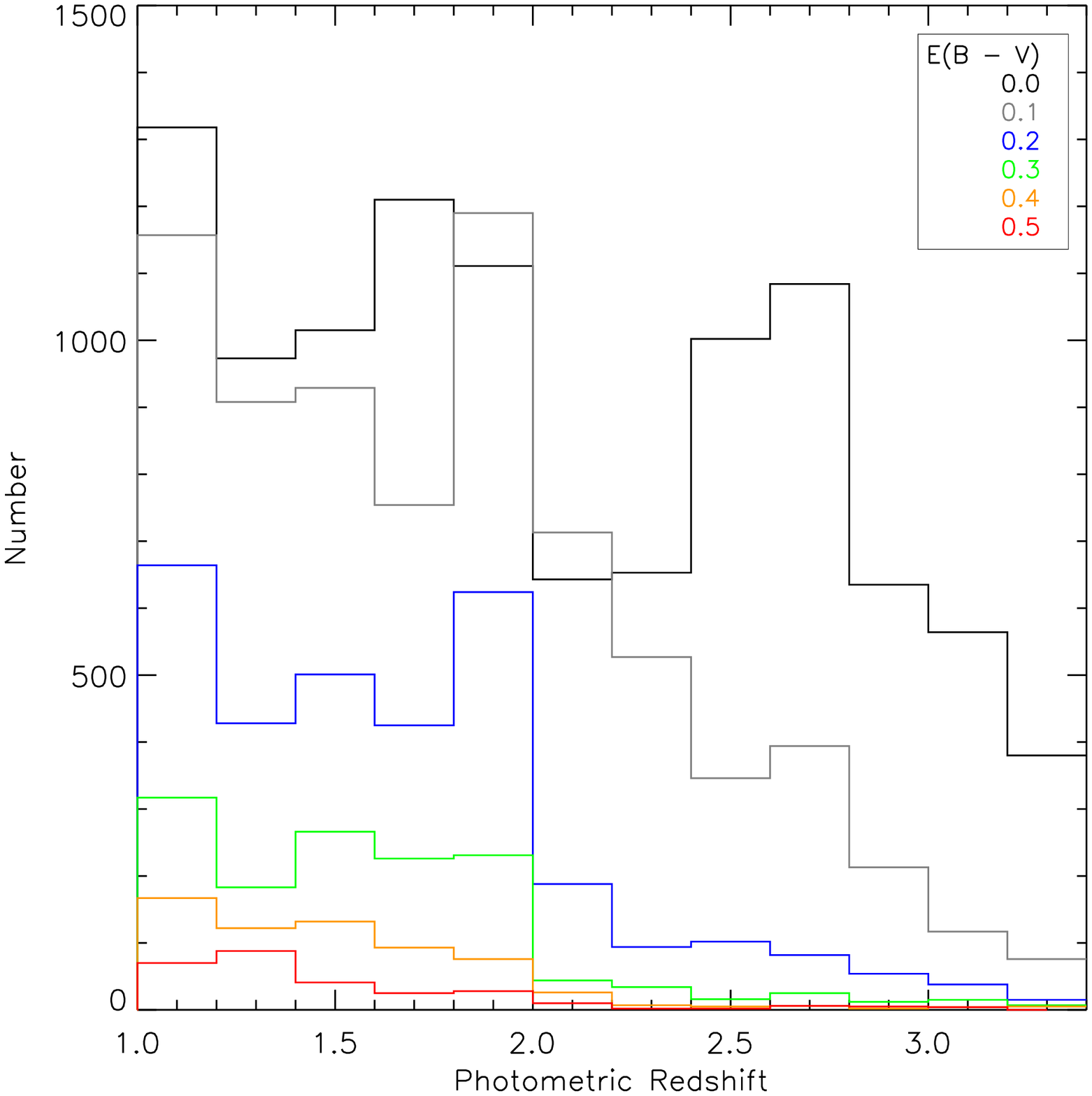} \plotone{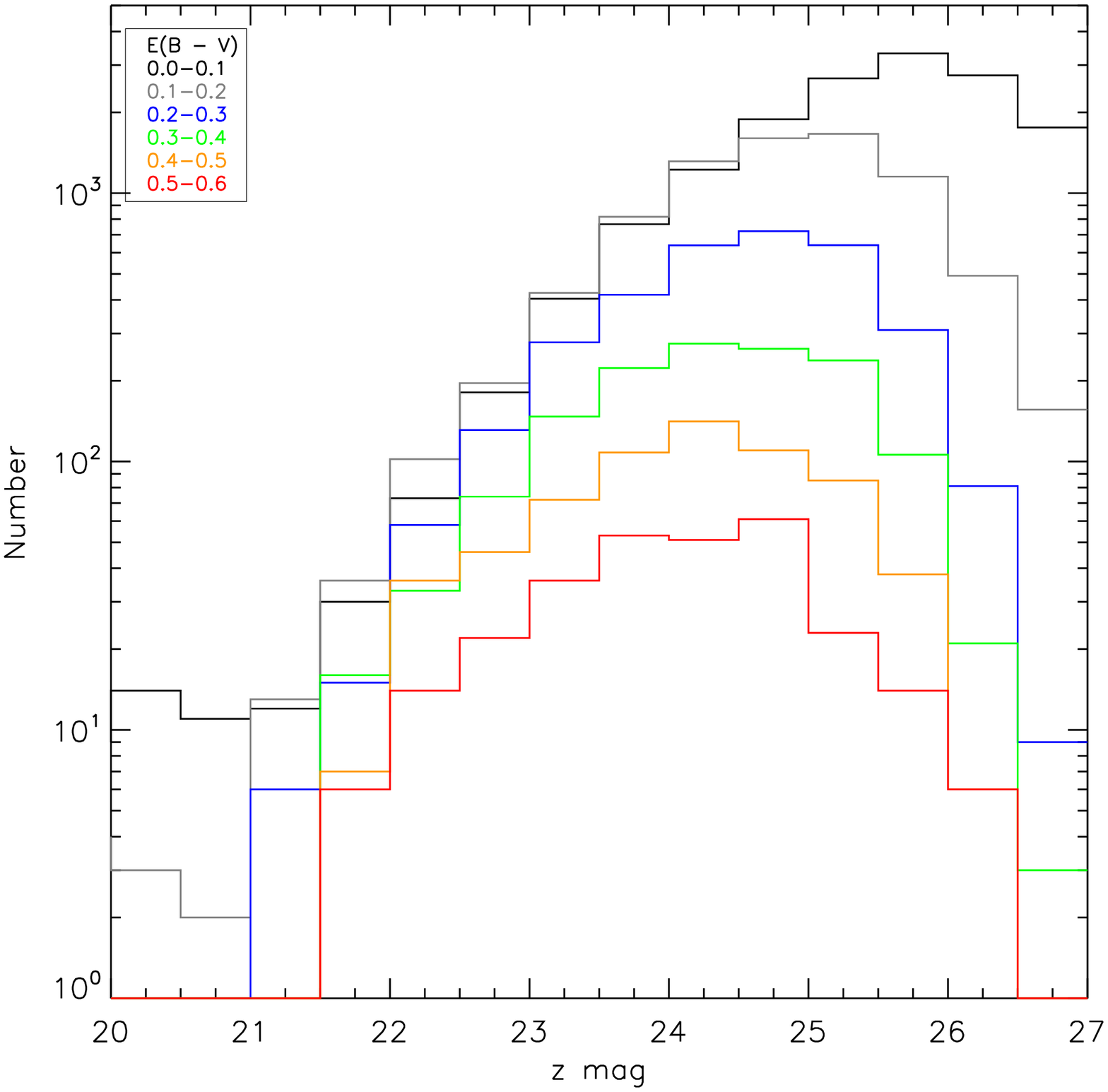}
  \caption{Redshift (left) and $z$\arcmin-band (right) distributions for galaxies with
    different measured $\EBV$. These distributions are used as prior inputs for
    the Monte Carlo simulation. This figure illustrates that (observationally) most of the
    dusty galaxies are found at $z\lesssim2$ and at relatively brighter $z$\arcmin\ magnitudes. \tcolor}
  \label{priors}
\end{figure*}

We must adopt distributions for each modeled properties (e.g., $\EBV$, redshift, and magnitude).
First, we assume that the observed $z\sim1$--3 $\EBV$ distribution applies to all galaxies in the
simulation. This simplifying assumption is needed for a proper normalization of the number of
blue-to-red galaxies, since it will affect the overlap fraction between
\sbzk\ and BX/BM or LBGs and the fraction of the Census that is contributed by either method.
The $\EBV$ distribution shows a log-normal decline with higher dust reddening following
$\log{(N)} \propto -3.3 \times \EBV$. Then for each $\EBV$ bin, we use observational constraints
on the redshift and magnitude distributions. The observed distributions show that the more
reddened galaxies are found at $z\lesssim2$ and have brighter $z$\arcmin\ magnitudes
(see Figure~\ref{priors}). In total, we generate 34,755 mock galaxies and the simulation
is repeated 100 times for 3,475,500 artificial galaxies.

One caveat with these assumptions is that observed trends and distributions are used, which
themselves are affected by incompleteness. For example, we could be underestimating the
intrinsic number of extremely dusty and/or old galaxies. However, neither we nor previous
investigators have sufficient information to attempt a further correction for this
possible incompleteness.

\indent{\it Predictions from simulations.}
With the Monte Carlo realization of our data, we find that the $\chi^2$ method begins to miss
5\% (50\%) of galaxies at $V\approx26.25$ ($V\approx27.25$) and $\Rcf\approx25.75$
($\Rcf\approx26.5$). This simulation shows that the $\chi^2$ method is highly complete down to magnitudes,
$V=25.4$ and $\Rcf=25.5$, at which we select \nuv-dropouts and BX/BM galaxies using optical colors.
\begin{figure}
  \epsscale{1.1} 
  \plotone{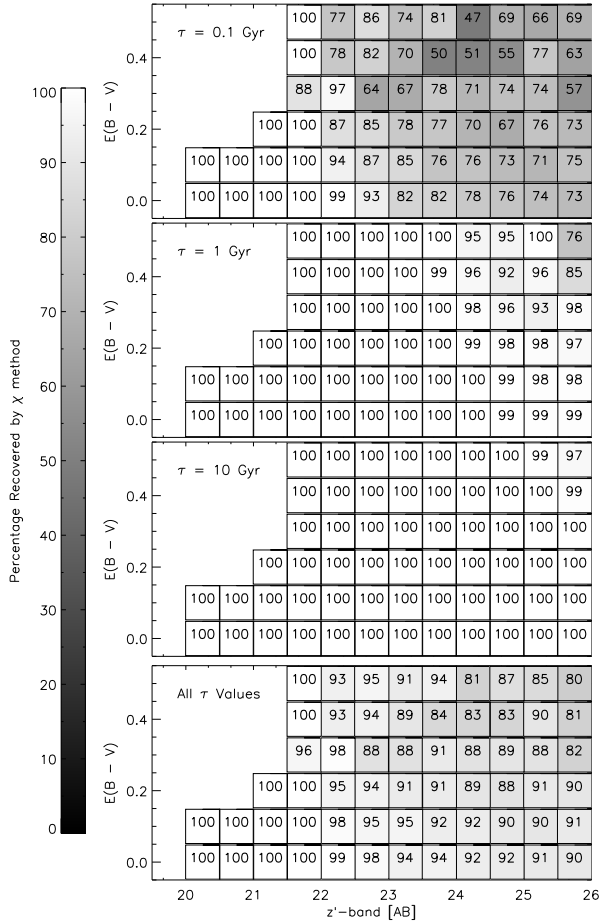}
  \caption{Recovered fraction of mock galaxies as a function of the assumed star formation
    history, dust reddening, and magnitude. Each panel plots $\EBV$ as a function of $z$\arcmin.
    Results for $\tau =0.1$, 1, and 10 Gyr and all three combined are shown from top to bottom.
    The gray scale shows the completeness from 100\% (white) to 0\% (black). Faint magnitude
    bins with high reddening have large statistical fluctuations due to input small numbers.}
  \label{EBV_comp}
\end{figure}

In Figure~\ref{EBV_comp}, we illustrate the fraction recovered by the $\chi$ method as a
function of dust reddening and the $z$\arcmin-band magnitude for the different $\tau$ models.
This figure illustrates high ($\gtrsim$80\%) completeness for the $\tau=1$ and 10 Gyr models
regardless of brightness and $\EBV$. As expected, the model with the
shortest timescale for star formation will suffer more incompleteness since many of these
galaxies are evolved and dust reddening further hampers their inclusion in the $\chi^2$ image.
By coming all three $\tau$ models, the incompleteness does not exceed 20\%.

\begin{figure*}
  \epsscale{0.9} 
  \plotone{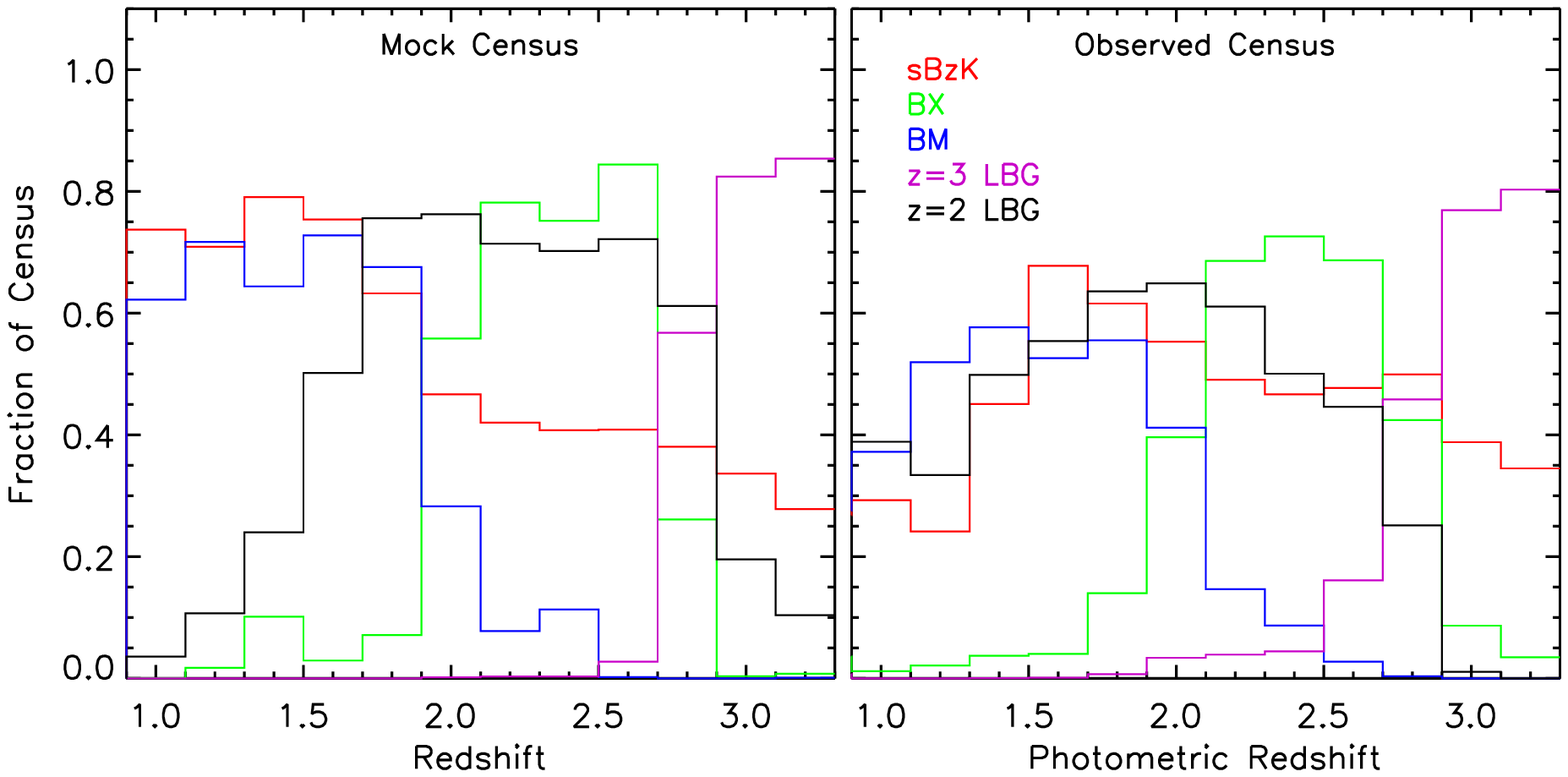}
  \caption{Comparison between the Mock Census (left) and the observed Census (right). The vertical
    axes show the fraction of the Census consisting of LBGs (black), BX (green), BM (blue), and
    \sbzk\ (red) galaxies. These plots are a reproduction of Figure~\ref{num_sfr}. Our Mock Census
    is able to reproduce the redshift peak of each galaxy population and the shape of the redshift
    distributions above $z\sim1.5$. These comparisons are further discussed in Section~\ref{6}. \tcolor}
  \label{compare1}
\end{figure*}

\indent{\it Comparisons with observations.}
In addition to the above completeness estimates, our Monte Carlo simulation provides
a few commensurable predictions.
For example, in Figure~\ref{compare1} we illustrate the fraction of the Mock and Observed
Census that is probed by each color selection technique. This is similar to Figure~\ref{num_sfr}
but with fixed redshift bins. There are several similarities that are apparent. The simulated
\Udrop\ and BX distributions are fairly consistent with observations in the terms of the
location and strength of the peak and the shape of the distributions. The \sbzk\ distributions
peak at $z\sim1.5$ at 80\% (mock) and 70\% (observed) and declines to 30\%--40\% at
$z\gtrsim3$. Above $z=1.7$, the $z\sim2$ LBG distributions are similar peaking at $z\sim2$ and
extending out to $z\sim2.8$. These similarities strongly suggest that the overlap determined
between each population is a fairly reliable result, and it is indeed true that no technique
yields 100\% of the Census at any redshift.
The greatest discrepancy, is at $z\lesssim1.5$ where the mock simulation indicates higher
completeness. The causes for such discrepancies is not known. However, these differences do
not affect much of our main results which concern $z\gtrsim1.5$.

\begin{figure}
  \epsscale{1.1} 
  \plotone{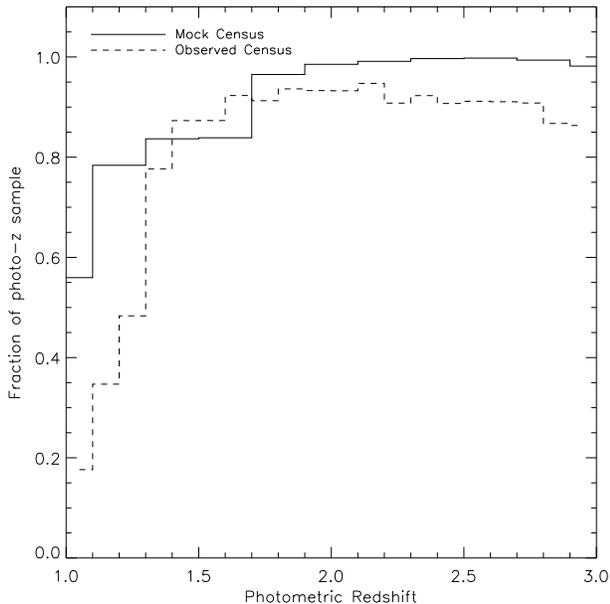}
  \caption{Comparison between the Mock Census (solid line) and the observed Census (dashed line).
    The vertical axis shows the fraction of the photo-$z$ sample that consists of galaxies that
    satisfy either one of the criteria for being BX, BM, \sbzk\, and LBGs at $z\sim2$ and $z\sim3$.
    This is a reproduction of Figure~\ref{photoz_pop} (right). The Mock Census yields a higher
    fraction than observed; however, it is consistent to within 2$\sigma$ of Poisson statistics.
    These comparisons are further discussed in Section~\ref{6}.}
  \label{compare2}
\end{figure}

Another prediction from the simulation is the fraction of the photo-$z$ census that is
recovered by the Census (see Figure~\ref{compare2}). As we have shown, the observed census
yields 90\% of galaxies above $z=1.5$, and this appears to hold out to $z\sim3$. The
Mock Census shows a similar behavior, but indicates nearly 100\%. These differences are
small since Poisson statistics is at the $\sim$5\% level.

Finally, another set of positive comparisons is the $V$-, \Rc-, and $K$-band luminosity functions
for $z\sim2$ LBGs. We find good to reasonable agreement over 5 magnitudes in the optical bands and
3 magnitudes in $K$. Other color selection techniques find agreements over a smaller range of
magnitudes. The causes of poorer agreement are still unknown.

\section{Conclusions}\label{7}
We have conducted a large photometric survey of galaxies at $z=1$--3 by synthesizing measurements from
20 broad-, intermediate-, and narrow-band filters covering observed wavelengths of 1500 \AA--4.5 \mm. This
survey is unique due to its size (0.25 deg$^2$; $5.6\times10^6$ Mpc$^3$ for $z=1$--3), depth, and reliable
photo-$z$'s derived for $\approx$23,000 $K$-band selected sources ($\approx$65,000 including
optically selected sources). Compared to a spectroscopic sample, we find that our photo-$z$'s from 20-bands
are accurate to $\accur \times (1+z)$ (1$\sigma$) for $z\lesssim1.8$. 
The accuracy of our photo-$z$ stems from coverage of both the Lyman continuum and Balmer/4000 \AA\ breaks at
$z\sim2$ with the multiple bands that we have.
Most photo-$z$ surveys use 10 photometric bands or less, are limited to accurate redshifts below $z\lesssim1$
and $z\gtrsim3$, and
suffer from a common photo-$z$ problem of ambiguity between the Balmer/4000 \AA\ break at $z\sim0.3$
and Lyman break at $z\sim3$.

We have also examined the \zphot\ for
$\approx$4500 narrow-band excess emitters at $z\approx0.25$--1.5, and found excellent to adequate agreement
between our derived \zphot, and the predictions from using simple multiple rest-frame optical colors
\citep[see][]{ly07}. This evidence further indicates that the photo-$z$ that we have generated are reliable.

In addition, the combination of deep optical and relatively deep NIR data provides a large and representative
sample of $z=1$--3 galaxies covering a wide range in stellar mass, dust content, and SFR. With
\nUaround\ (\nUazround\ with \zphot) galaxies, we are able to study statistically the overlap of
different galaxy populations (BX/BM, LBG, and BzK) obtained from each technique. We also
investigate the completeness of these techniques against a sample derived purely from photo-$z$.
The main results for this paper are:\\
\indent{\it Redshift distribution.} We determined (with accurate photo-$z$) that \zBX\ of BXs, \zBM\ of BMs,
\zU\ of \Udrops, \zsbzk\ of sBzKs, and \zNUV\ of \nuv-dropouts have $\zphotf=1$--3.5. We find that the BX, BM,
and \Udrop\ methods suffer from $\zphotf<0.5$ contamination at the \nBXif\%, \nBMif\%, and \nUif\% levels,
respectively. This problem is largely attributed to the Balmer/4000 \AA\ break occurring between the $U$ and
$B$ filters, rather than a Lyman break. We also find that the \nuv-dropout technique has \zNUVint\ contamination
from ``interlopers'' at $z=1.0$--1.5, which eliminates
a previous concern that contamination fraction estimates in \cite{ly09} were too optimistic. Also, the technique
does not suffer from $z<0.5$ interlopers, since the Balmer break is redward of the NUV.

{\it Selection bias and completeness of techniques.}
We confirmed that each color selection is biased towards some subset of galaxies
not fully representative of the entire galaxy population. By modeling the
SED of thousands of photometrically-selected galaxies, we find that the BzK method targets more massive and dusty
galaxies, while the BX, BM, and \Udrop\ methods identify galaxies of lower stellar mass with low reddening.
Compared to a combined census of color-selected galaxies (BX/BM, LBG, and BzK), we find that any UV or IR
technique finds at most \peakval\ at the peak of their redshift sensitivity. However, these techniques
are optimized for a certain redshift range, so relative to a $z=1$--3 color-selected census, the BX/BM,
\nuv-dropout and \sbzk\ techniques obtain \allz. This strongly indicates that combining multiple techniques is
favored for a complete census of high-redshift galaxies and their SFRs. We estimate that the BX/BM method is
capable of identifying (by number) \lsfrBXBM\ of $\zphotf=1$--3 census galaxies with SFR between a few and 30
\Msun\ \iyr, and this represents \lsfr\ of the $\zphotf=1$--3 census.

{\it Sample overlap between galaxy populations.}
Among the star-forming BzK-selected galaxies, the overlap fraction with the BX/BM/\Udrop\
increases at fainter $K$ luminosities, consistent with \citetalias{reddy05}. However, with a
larger sample of massive \sbzk\ galaxies, we determined that the overlap fraction is
$\sim$5 times smaller than what was reported in \citetalias{reddy05} at bright $K$ magnitudes.
We argue that this is due to the smaller FoV and sample size that they used.
Our study also shows that \statcc\ of $z\geq1.5$ \sbzk\ galaxies have a strong Lyman break
to meet the \nuv-dropout criteria. Unlike the BX/BM method, this overlap fraction is
independent of $K$ mag, indicating that the Lyman break technique is relatively more
sensitive to massive galaxies. Likewise, we find that \statca\ of $z\geq1.5$ BX and
\statcb\ of $z\geq1.5$ BM galaxies are $z\sim2$ LBGs.
These results indicate that the 912 \AA\ break is ubiquitous in $z\sim2$ galaxies. Thus, the \nuv-dropout
selection differs from the BX/BM selection by including more massive and reddened galaxies.

{\it SFR density from a census of UV- and NIR-selected galaxies.} We determined that the comoving SFR
density at $\zphotf=1.5$--2.6 is \sfrK\ from a joint census of BzK galaxies, BX/BM galaxies, and
LBGs. This census was limited to galaxies with $K\lesssim24$ (corresponds to $\approx10^{10}$ \Msun).
We find that the UV selection methods (Lyman break and BX/BM) obtain 81\% of the SFR
density from galaxies with $\EBV< \EBVlimit$ mag. These low reddening galaxies represent about \NEBVb\
of our census by number. However, there is a prominent population of galaxies with high
dust extinction ($\EBV > \EBVlimit$ mag). For example, 
we examined the contribution to the SFR density relative to the dust properties of our galaxies, and
find that galaxies below and above $\EBV = \EBVlimit$ mag contribute \pEBVl\ and \pEBV\ to the SFR
density, respectively.

{\it Completeness of color selection techniques.} We compare the census derived from color selection
techniques against a photo-$z$ sample, and find that color selection is (1) less efficient
at $z\lesssim1.5$, but (2) identifies \censuscomp\ of galaxies at $\zphotf=1.5$--2.5.
We also find that the \sbzk\ method is \sbzkgood\ complete when compared against a sample of
$K$-band detected sources with $\zphotf>1.5$.
This evidence, combined with the high fraction of overlap between BX/BM and \sbzk\ galaxies for $K\sim23$--24
AB mag, and the likelihood that faint (blue) UV-selected galaxies will be detected with deeper $K$-band
imaging, strongly suggest that future $z\approx1$--3 studies should focus on a very deep $K$-band imaging
(e.g., the Ultra-VISTA survey). But until such a deep ($K\sim25$--26 AB) survey with multi-wavelength data
is available over large areas, the UV selection techniques (either the $\Un G\Rs$ or \nuv-dropout method)
are currently the only efficient way to identify less massive $z\sim2$ galaxies.

\acknowledgements
The Keck Observatory was made possible by the generous financial support of the W.M. Keck Foundation.
The authors recognize and acknowledge the very significant cultural role and reverence that
the summit of Mauna Kea has always had within the indigenous Hawaiian community. We are most fortunate
to have the opportunity to conduct observations from this mountain.
We gratefully acknowledge NASA's support for construction, operation, and science analysis for the
\galex\ mission.
This work is based in part on observations made with the {\it Spitzer Space Telescope}, which is operated by
the Jet Propulsion Laboratory, California Institute of Technology under a contract with NASA. Support
for this work was provided by NASA through an award issued by JPL/Caltech.
C.L. thanks T. Velusamy for providing the HiRes code and helpful discussions on the topic, G. Brammer for
providing more recent SED templates to use in EAZY, and R. Quadri for providing additional unpublished
information about the MUSYC survey.
C.L. is supported by NASA grant NNX08AW14H through their Graduate Student Researcher Program.
We thank the anonymous referee for his/her comments that improved the paper.

{\it Facilities:} \facility{\galex}, \facility{Mayall (MOSAIC, NEWFIRM)}, \facility{Subaru (Suprime-Cam)},
\facility{UKIRT (WFCAM)}, \facility{\spitzer\ (IRAC)}, \facility{Keck:I (LRIS)}, \facility{Keck:II (DEIMOS)},
\facility{MMT (Hectospec)}

\begin{appendix}
\section{Unusual Sources from Color Selection}\label{extreme}
About half of the Shallow and Faint samples (restricted to the deep $K$-band area and excluding galaxies
with $\zphotf<0.5$) consists of sources that are uniquely selected by only one color selection. These sources
are of great interest, since they indicate that a particular technique culls a galaxy sample that is
unavailable by any other techniques. We summarize the uniqueness below.\\
\indent{\it Passive BzK galaxies.} We find that the entire pBzK sample does not overlap with any of the other
techniques. These galaxies at $z\approx1$--2 have the oldest ages, $\log{(t_{\rm age}/{\rm yr})\approx 9.0}$,
the highest stellar masses, $\log{(M/M_{\sun})\approx 11.0}$, and span $E(B-V)=0.0$--0.3. Their ages
imply that they formed $\Delta z = 0.6$ earlier (median formation redshift is $z=2.25$), and could be
the descendants of luminous infrared galaxies or massive LBGs with SFRs $\gtrsim 100$ \Msun\ \iyr. These galaxies
are important for a stellar mass census survey, but contribute little to the SFR density since their SFRs
are on average 0.02 \Msun\ \iyr.\\
\indent{\it Star-forming BzK galaxies.} We find that purely sBzKs, compared to UV-selected
samples, have similar SFRs ($\sim 5$ \Msun\ \iyr), but are more massive ($\sim10^{9.5}$--$10^{11}$ \Msun), cover a
broader range in dust extinction ($E[B-V]\approx0.05$--0.4), and are 0.5 dex older.\\ 
\indent{\it BX, BM, and \Udrop\ galaxies.} The UV selection techniques of Steidel et al. identify a unique set of
galaxies with relatively younger stellar ages (typically $2\times10^8$ yr),
lower stellar mass ($\sim10^9$ \Msun), and $\EBV \approx 0.1$. We find that the SFRs are systematically lower by
0.2--0.3 dex for BM and BX galaxies when compared to \Udrops, although this is likely a manifestation of redshift
dependence of the sample. The relatively poor \Udrop\ overlap with NIR techniques can be attributed to the
redshift of the sample, as well as, the lack of deeper NIR data. It would be prudent to identify a sample of
higher redshift BzK galaxies using a set of different filters \citep[e.g., $R$, $J$, and $L$;][]{daddi04} to
investigate the overlap of LBGs at higher redshift with 3.6 \mm-selected galaxies.\\
\indent{\it Lyman break galaxies at $z\sim2$.} The method that we developed to identify LBGs at $z\approx1.5$--2.5 
has significant overlap ($\approx$80\%) with the BX/BM and/or BzK techniques.
It appears that the remaining 20\% consists of galaxies with stellar masses of $\approx10^{10\pm1}$ \Msun, cover
the same stellar ages as other techniques, have an average SFR of $\sim$20 \Msun\ \iyr, and span $\EBV = 0.1$--0.4.
The main reason for why these galaxies were uniquely identified is their redshift: about two-thirds of them are
below $z=1.5$ where the \sbzk\ and BM techniques are less sensitive. We note that the higher SFRs for
\nuv-dropouts are likely due to the limit adopted for the $V$-band (our limit is comparable brighter than
$\Rcf=\LyRcutn$ for BX/BM). The fainter ($V=26.0$) limit yields an average SFR that is systematically lower by
0.2 dex, due to the higher contribution from lower-SFR LBGs.

\end{appendix}

\end{document}